\documentclass[fleqn,usenatbib]{mnras}

\usepackage{newtxtext}
\usepackage{xcolor}
\usepackage{graphicx}

\usepackage[T1]{fontenc}

\DeclareRobustCommand{\VAN}[3]{#2}
\let\VANthebibliography\thebibliography
\def\thebibliography{\DeclareRobustCommand{\VAN}[3]{##3}\VANthebibliography}

\newcommand{\Nnew}{78 } 
\newcommand{\N}{79 } 
\newcommand{\agreement}{$1\sigma$}

\usepackage{graphicx}	
\usepackage{amsmath}	
\usepackage{amssymb}	

\title[DAVOS: ZTF]{Dwarf AGNs from Variability for the Origins of Seeds (DAVOS): Optical Variability of Broad-line Dwarf AGNs from the Zwicky Transient Facility}

\author[Z. Franklin Wang et al.]{
Z. Franklin Wang,$^{1,2,3}$\thanks{E-mail: zfwang2@tamu.edu (ZFW)}
Colin J. Burke,$^{1}$\thanks{E-mail: colinjb2@illinois.edu (CJB)}
Xin Liu,$^{1,4}$
Yue Shen$^{1,4}$
\\
$^{1}$Department of Astronomy, University of Illinois at Urbana-Champaign, 1002 West Green Street, Urbana, IL 61801, USA \\
$^{2}$Department of Physics and Astronomy, Texas A\&M University, College Station, TX 77843-4242, USA \\
$^{3}$George P. and Cynthia Woods Mitchell Institute for Fundamental Physics and Astronomy, Texas A\&M University, College Station, TX 77843-4242, USA \\
$^{4}$National Center for Supercomputing Applications, 1205 West Clark Street, Urbana, IL 61801, USA\\
}

\date{Accepted XXX. Received YYY; in original form ZZZ}

\pubyear{2021}

\begin{document}
\label{firstpage}
\pagerange{\pageref{firstpage}--\pageref{lastpage}}
\maketitle

\begin{abstract}
We study the optical variability of a sample of candidate low-mass (dwarf ang Seyfert) active galactic nuclei (AGNs) using Zwicky Transient Facility \emph{g}-band light curves. Our sample is compiled from broad-line AGNs in dwarf galaxies reported in the literature with single-epoch virial black hole (BH) masses in the range $M_{\rm{BH}} \sim 10^{4}$--$10^{8}\ M_{\odot}$. We measure the characteristic ``damping'' timescale of the optical variability $\tau_{\rm{DRW}}$, beyond which the power spectral density flattens, of a final sample of \N candidate low-mass AGNs with high-quality light curves. Our results provide further confirmation of the $M_{\rm{BH}} - \tau_{\rm{DRW}}$ relation from \citet{Burke2021b} within \agreement{} agreement, adding \Nnew new low-mass AGNs to the relation. The agreement suggests that the virial BH mass estimates for these AGNs are generally reasonable. We expect that the optical light curve of an accreting intermediate-mass black hole (IMBH) to vary with a rest-frame damping timescale of $\sim$ tens of hours, which could enable detection and direct mass estimation of accreting IMBHs in wide-field time-domain imaging surveys with sufficient cadence like with the Vera C. Rubin Observatory.

\end{abstract}




\section{Introduction} \label{sec:intro}

\begin{figure*}
\includegraphics[width=0.8\textwidth]{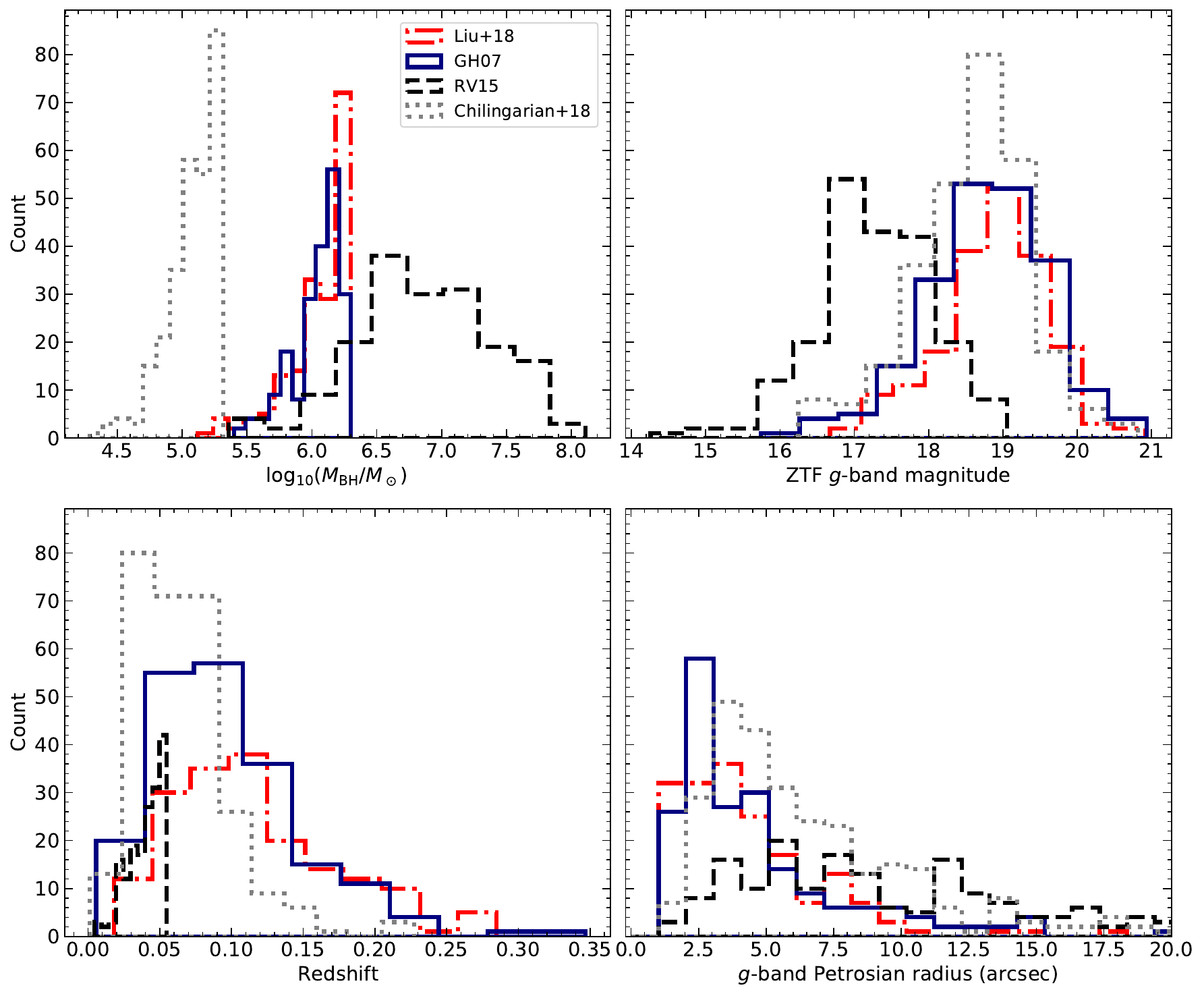}
\caption{Distributions of the virial BH mass estimates (\emph{top left}), ZTF \emph{g}-band PSF magnitude (\emph{top right}), spectroscopic redshift (\emph{bottom left}), and \emph{g}-band Petrosian radius (\emph{bottom right}) for our parent sample of low-mass AGNs selected from \citet{Greene2007} ("GH07"), \citet{Reines2015} ("RV15"), \citet{Chilingarian2018} ("Chinlingarian+18"), and \citet{Liu2018} ("Liu+18").  \label{fig:histogram}}
\end{figure*}

Intermediate-mass black holes (IMBHs; $M_{\rm{BH}} = 10^2 - 10^5\ M_{\odot}$) are thought to be the missing link between stellar-mass and supermassive black holes (SMBHs). The demographics of active galactic nuclei (AGN) in dwarf galaxies hosting IMBHs traces the formation mechanisms of SMBH seeds which are believed to have formed in nascent galaxies at high redshift \citep[e.g.,][]{Volonteri2010,Reines2016,Greene2020,Inayoshi2020}. Perhaps the best observational evidence for IMBHs is the gravitational wave event GW190521 from the merger of two black holes (with a remnant mass of $142^{+28}_{-16}$ $M_{\odot}$; \citealt{LIGOVirgo2020}). Accreting IMBHs remain difficult to identify, with only a few exceptional candidates such as the X-ray tidal disruption event 3XMM J215022.4-055108 \citep{Lin2018}; and the $M_{\rm{BH}} \sim 10^4\ M_{\odot}$ hyper-luminous X-ray source ESO 243-49 HLX-1 \citep{Farrell2009}.

Recent works demonstrated success in identifying dwarf AGNs with BH masses on the boundary between IMBHs and SMBHs with $M_{\rm{BH}} \sim 10^4 - 10^6\ M_{\odot}$ using optical spectroscopic signatures (e.g., \citealt{Greene2005a,Reines2013,Baldassare2015,Baldassare2016,Cann2020}) or variability (e.g., \citealt{Baldassare2018,Baldassare2020,Martinez-Palomera2020,Guo2020,Ward2021b,Burke2021b}). We have compiled a parent sample of dwarf AGN candidates in the literature with optical signatures of accretion activity (variability, narrow emission line diagnostics, and/or broad emission lines). Then, we restricted our sample to broad-line AGNs and those with virial BH mass estimates. These virial BH masses are in the range $M_{\rm{BH}} \sim 10^{4} - 10^{6}\ M_{\odot}$ \citep{Greene2007,Chilingarian2018,Liu2018}. We add the sample of low-mass (dwarf and Seyfert) AGNs with $M_{\rm{BH}} \sim 10^{5} - 10^{8}\ M_{\odot}$ from \citealt{Reines2015} used to calibrate their local BH - host galaxy stellar mass relation, because we will attempt to benchmark our variability timescale -derived BH masses against the \citealt{Reines2015} relation. These low-mass AGNs offer the best cases currently to test our understanding of accretion physics close to the IMBH regime.

Optical variability in AGN light curves can be used to probe accretion disk structure and the dominant physical processes at the radius of optical emission. The power spectral density (PSD) of optical AGN light curves are well-described by a damped random walk (DRW) model from hours to years timescales, where the PSD transitions from white noise to red noise ($f^{-2}$) at a characteristic frequency ($f_{\rm DRW}=(2\pi\tau_{\rm DRW})^{-1}$) \citep[e.g.,][]{Kelly2009,MacLeod2010}. The corresponding  rest-frame ``damping'' timescale $\tau_{\rm{DRW}}$ correlates strongly with $M_{\rm{BH}}$ and is likely associated with the thermal timescale in the accretion disk at the UV-emitting radius \citep{Burke2021}. \citet{Ward2021b} found evidence that their sample of variability-selected dwarf AGNs have more rapid variability compared to a control sample of massive AGNs using ZTF light curves.

The $M_{\rm{BH}} - \tau_{\rm{DRW}}$ relation of \citet{Burke2021} predicts a $\tau_{\rm{DRW}}$ of tens of days for a $10^6\ M_{\odot}$ BH as opposed to several hundred days for a $10^9\ M_{\odot}$ SMBH. Therefore, the variability timescales for dwarf AGNs are well-sampled by Zwicky Transient Facility (ZTF; \citealt{Masci2019,Bellm2019}) light curves with a typical baseline and cadence of $\sim 1000$ days and $\sim 3$ days respectively. We will test the $M_{\rm{BH}} - \tau_{\rm{DRW}}$ relation in the dwarf AGN regime, motivated by the possibility of extrapolating the relation to even lower masses to enable not only IMBH discovery, but also host-independent BH mass estimation, which may not be feasible with traditional spectroscopy-based methods due to the lack of broad emission lines \citep{Chakravorty2014}.

In this work, we report our time series analysis of ZTF light curves of \N low-mass AGNs with virial BH mass estimates of $10^{5.69}\leq M_{\rm{BH}} (M_{\odot}) \leq 10^{7.80}$. We focus on spectroscopically-confirmed broad-line dwarf AGNs. This includes \Nnew new objects which we place on the $M_{\rm{BH}} - \tau_{\rm{DRW}}$ relation compared to the previous \citet{Burke2021} sample with $\sim 4$ years long light curves, sufficient to constrain the damping timescale for dwarf AGNs. These objects were not included in the \citet{Burke2021} sample, which only considered objects from \citet{Reines2013} and used a previous ZTF data release with shorter light curve lengths of $\sim 2$ years. We describe the AGN parent sample and data in \S\ref{sec:data}, present our results in \S\ref{sec:results}, discuss the possible implications in \S\ref{sec:discussion}, and conclude in \S\ref{sec:conclusion}. All uncertainties and plotted error bars are $1\sigma$ unless otherwise specified.

\section{Observations and Data Analysis} \label{sec:data}

\subsection{Dwarf AGN Parent Sample}

We searched the literature for dwarf AGN candidates with broad-line measurements that enabled single-epoch virial BH mass estimates \citep{Shen2013}. Our parent sample of low-mass AGNs come from the following sources: \citet{Greene2007} ("GH07"), \citet{Reines2015} ("RV15"), \citet{Chilingarian2018} ("Chinlingarian+18"), and \citet{Liu2018} ("Liu+18"). These AGNs were identified using one of two methods. First, dwarf galaxies were identified with narrow emission line ratios consistent with an AGN based on the classical BPT diagram \citep{Baldwin1981,Veilleux1987,Kewley2006} with broad emission lines detected afterwards. Second, AGNs were selected via detection of the broad emission line itself from Sloan Digital Sky Survey (SDSS) spectra \citep{York2000,Abazajian2009} even if the line ratios are consistent with star-formation. In the later case, the broad emission lines can be due to a supernova or stellar transient, requiring follow-up observations years later \citep{Baldassare2016}. However, long-duration variability can also be used to test the AGN scenario. In this work, we only included sources with published BH masses from broad emission line detections with persistent, AGN-like variability. We show the virial BH mass estimate, $g$-band PSF magnitude, spectroscopic redshift, and host-galaxy \emph{g}-band Petrosian radii \citep{Petrosian1976,Graham2005} distributions for the parent sample sources in Figure~\ref{fig:histogram}. We checked for duplicate sources between the samples, and used the $M_{\rm{BH}}$ value from the more recent publication in those cases.

\begin{figure*}
\includegraphics[width=0.98\textwidth]{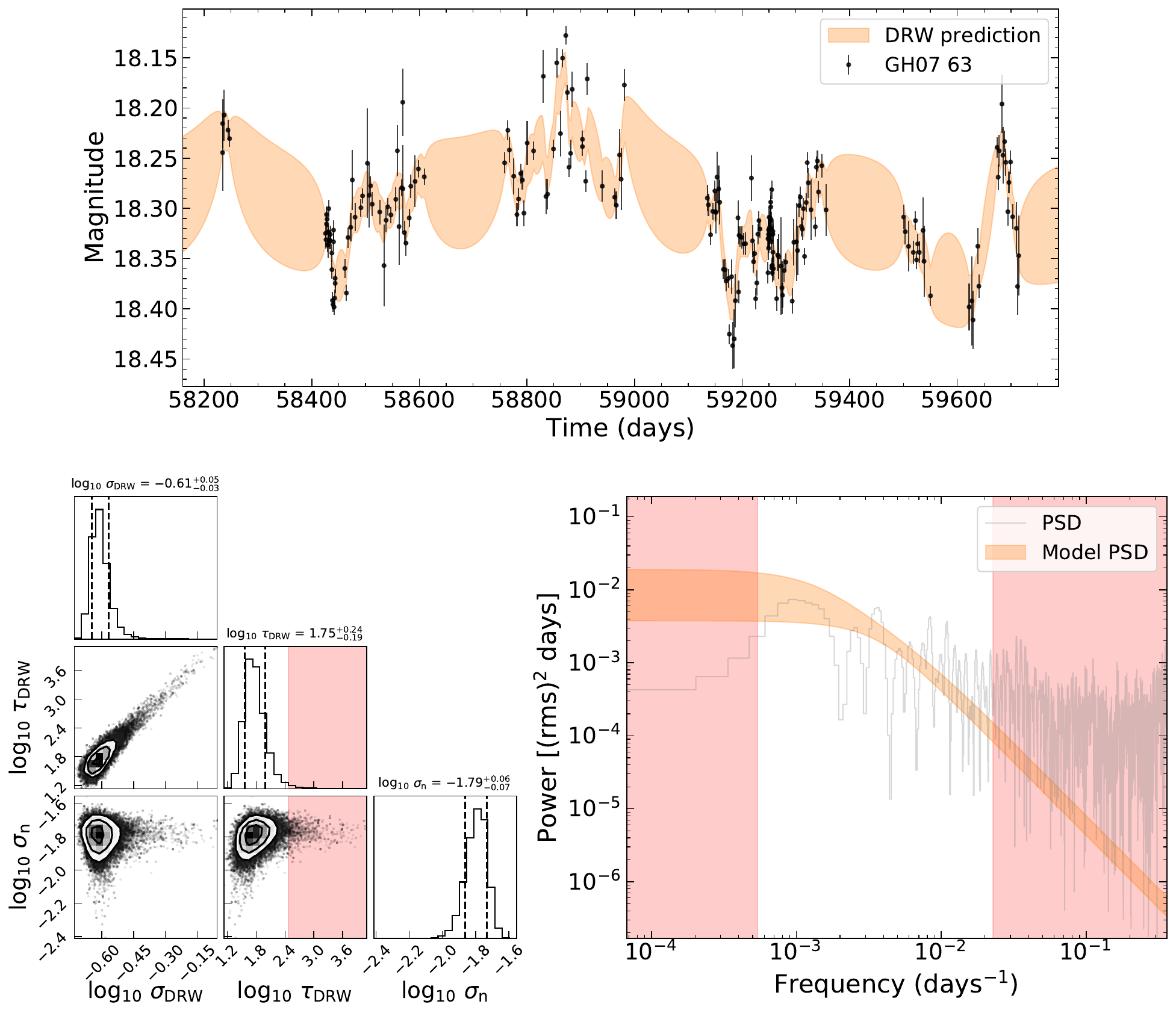}
\caption{Example DRW modeling of SDSS J090320.97+045738.0 with $M_{\rm{BH}} = 10^{6.1\, \pm\, 0.3} M_{\odot}$ and $z = 0.0567$ \citep{Greene2007}. We show the $g$-band difference imaging light curve and the best-fit DRW model with 1$\sigma$ uncertainty (orange shaded area) (\emph{top panel}). We show the posterior probability distributions for the fitted DRW parameters and their covariance (Equation~\ref{eq:drw}). In the covariance panels, the contours trace the 1, 2, 3$\sigma$ levels overplotted on the sample density map (black being higher density) with individual samples in the lowest-density regions shown as black points (\emph{bottom left panel}). Finally, we show the normalized Lomb-Scargle PSD in gray and the 1$\sigma$ range of the DRW PSD from the posterior prediction as the orange shaded area (\emph{bottom right panel}). Note the Lomb-Scargle PSD includes a white noise floor at high frequency that is not included in the DRW model.  The red shaded regions correspond to periods greater than 20\% the light curve length and less than the mean cadence, where the PSD is not well sampled. \label{fig:example}}
\end{figure*}

\subsection{Light Curve Data and Cleaning}

ZTF is a time-domain optical (primarily \emph{g} and \emph{r} with some \emph{i}-band) imaging survey using the Samuel Oschin 48-inch Schmidt telescope at Palomar Observatory. ZTF light curves began in 2018 with a nominal cadence of 2 or 3 days (with seasonal gaps), covering declinations above $-30$ degrees with a single-epoch imaging depth of 20.5 mag ($5\sigma$, \emph{r} band). The ZTF data release light curves are PSF-based, and will introduce additional measurement error for resolved sources given the median $g$-band PSF full width at half maximum (FWHM) of $2\farcs1$ \citep{Bellm2019}. For this reason, we use difference imaging-based photometry as described below.

We downloaded difference images from the ZTF data release 13 (March 2018 $-$ July 2022) in the \emph{g} band from the NASA/IPAC Infrared Science Archive\footnote{\url{https://irsa.ipac.caltech.edu/Missions/ztf.html}} which had coverage of any one of our sample targets. After collecting the images, we extracted photometry using a $2\farcs5$ diameter aperture, which is well-matched to the ZTF seeing, using the \textsc{Photutils} code \citep{Bradley2022}. After constructing the light curves, we performed outlier rejection at the $3\sigma$ level using a moving window size of 1 year. Given AGNs tend to be more variable in at bluer wavelengths, we restrict our analysis to the \emph{g} band. This effect may be even stronger for dwarf AGNs, as the disk temperature of an IMBH shifts the SED into the UV (e.g., \citealt{Cann2018}).



\subsection{Light Curve Modeling}

It has become increasingly popular to model AGN light curves in the time domain (as opposed to the frequency domain) with a Continuous Auto-Regressive Moving Average (CARMA) model \citep[e.g.,][]{Kelly2009,Kelly2014,Simm2016,Suberlak2021}. The CARMA models are a set of flexible Gaussian process models \citep{Aigrain2022} to describe stochastic time series, with two parameters, $p$ and $q$, describing the orders of the auto-regression (AR) part and the moving average (MA) part. The DRW model is the lowest order CARMA model with $p=1$ and $q=0$, or a CAR(1) process. In this work we use the DRW model to constrain the characteristic damping timescale of an AGN light curve. Deviations from the simple DRW model for SMBHs have been reported on short timescales with space- or ground-based photometry \citep[e.g.,][]{Mushotzky2011,Kasliwal2015,Stone2022}. To check that the DRW model is sufficient for our data, we visually confirmed that the ACF of the $\chi^2$ (model$-$data) residuals are consistent with Gaussian white noise using Bartlett's formula. Given that the best-fitting DRW models have sufficiently small residuals, higher-order CARMA models are not justified given the data quality. \citet{Burke2021} provided a more detailed discussion on the advantages and caveats of the DRW model compared with the more complicated higher-order CARMA models.

The major advantage of the Gaussian process models is that they can provide an accurate description of irregularly-sampled stochastic light curves and the underlying PSD while taking into account measurement uncertainties. Because the modeling is performed in the time domain rather than in the frequency domain, the inferred PSD is more robust against windowing effects (such as aliasing and red noise leakage) than traditional power spectral analysis. Fitting a broken power law to the PSD of light curves can produce consistent results with time domain DRW fitting, albeit with larger intrinsic scatter and greater demands on the light curve sampling and quality \citep{Burke2021}. We compute the PSD using the Lomb$-$Scargle periodogram \citep{Lomb1976,Scargle1982} as a consistency check. A typical example of our time series analysis and DRW fitting is shown in Figure~\ref{fig:example}. The light curves and their best-fit DRW models of our final sample are shown in Appendix~\ref{appx:lc}.

The covariance function for a DRW plus extra white noise model is,
\begin{equation}
    k(t_{nm}) = 2 \sigma_{\rm{DRW}}^2 \exp{(-t_{nm} / \tau_{\rm{DRW}})} + \sigma_{n}^2\delta_{nm}\
    \label{eq:drw}
\end{equation}
where $t_{nm}=|t_{n}-t_{m}|$ is the time lag between measurements $m$ and $n$, $\sigma_{\rm{DRW}}$ is the amplitude term, $\tau_{\rm{DRW}}$ is the damping timescale, $\sigma_{\rm{n}}$ is the white noise amplitude, and $\delta_{nm}$ is the Kronecker $\delta$ function. We include the white noise term to account for excess measurement noise in the ZTF light curves due to e.g., seeing variations or under-estimated uncertainties from our photometric extraction procedure, which does not include systematic sources of uncertainty in addition to the statistical uncertainties. This model has an exact correspondence to the structure function (SF) definition,
\begin{align}
\begin{split}
    {\rm{SF}}^2 & = \rm{SF}_{\infty}^2(1 - {\rm{ACF}}(t_{nm})) \\
    & = 2 \sigma_{\rm{DRW}}^2 (1 - \exp{(-t_{nm} / \tau_{\rm{DRW}}}))\
    \label{eq:sf}
\end{split}
\end{align}
where the asymptotic variability amplitude $\rm{SF}_{\infty} = \sqrt{2} \sigma_{\rm{DRW}}$ and the autocorrelation function ${\rm{ACF}}(t_{nm}) = \exp{(-t_{nm} / \tau_{\rm{DRW}}})$. Therefore, we have three free parameters to fit in the model: $\sigma_{\rm{DRW}}$, $\tau_{\rm{DRW}}$, and $\sigma_{\rm{n}}$. We use the \textsc{celerite} package \citep{Foreman-Mackey2017} to perform the model fitting. We use Markov chain Monte Carlo (MCMC) implemented in the \textsc{emcee} package \citep{Foreman-Mackey2013} to sample the joint posterior probability density, with uniform priors for all parameters. We take the 16th and 84th percentiles of the marginalized posterior distributions for each parameter to estimate the $1\sigma$ uncertainties.

\subsection{Recoverability of Input Damping Timescale}
\label{sec:simulations}

\begin{figure*}
\includegraphics[width=0.98\textwidth]{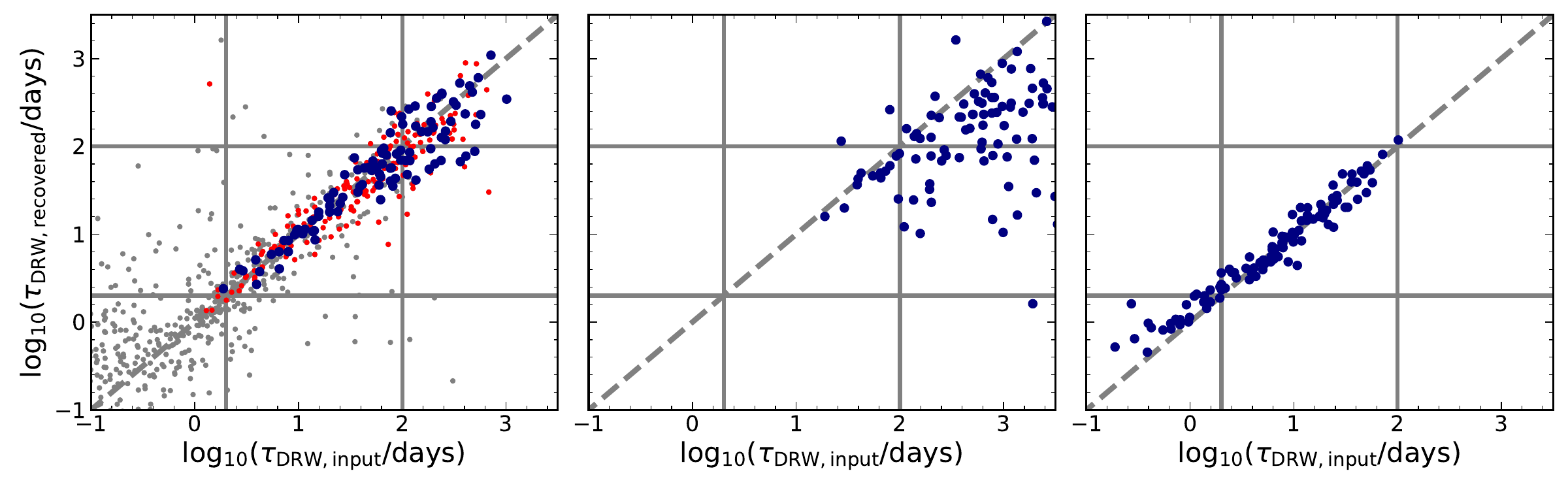}
\caption{Input versus recovered damping timescale for simulated realistic DRW light curves. \emph{Left panel}: Recoverability using the input $\tau_{\rm{DRW}}$ from our parent sample with no quality constraints (grey points), with constrains (ii) ($\tau_{\rm{DRW}} > {\rm{cadence}}$) and (iv) ($\sigma_{\rm{LB}}>3$) (red points), and with constraints (i) ($\rm{SNR}>1$), (ii), and (iii) (blue circles). \emph{Center panel}: Recoverability using input $\tau_{\rm{DRW}}$ of $10$ times the result from our parent sample after applying constraints (i), (ii), and (iii) on the original light curves. The recovered $\tau_{\rm{DRW}}$ are biased and saturate at values near $20-30\%$ the light curve length when the input $\tau_{\rm{DRW}}$ is greater than about $10\%$ the light curve length. \emph{Right panel}: Recoverability using input $\tau_{\rm{DRW}}$ of $0.1$ times the result from our parent sample after applying constraints (i), (ii), and (iii) on the original light curves. Light curves with input $\tau_{\rm{DRW}}$ smaller than the mean cadence typically have unreliable recovered $\tau_{\rm{DRW}}$. No duration constraints were applied in our simulations to demonstrate the bias for light curves of insufficient duration \citep{Kozowski2017}. The pair of (vertical or horizontal) gray lines corresponds to the cadence and $0.1 \times$ the light curve length of a typical ZTF light curve in our sample. The dashed gray line shows the $y=x$ line. \label{fig:simulations}}
\end{figure*}

Following \citet{Burke2021}, we impose the following minimal quality constraints on our light curves for our final sample:

\begin{enumerate}
  \item \textbf{Length:} $\tau_{\rm{DRW}} < 0.1 \times {\rm{baseline}}$. 
  \item \textbf{Sampling:} $\tau_{\rm{DRW}} > 6\ {\rm{days}}$. 
  \item \textbf{Signal-to-noise ratio (SNR):} ${\rm{SNR}} > 1$, where ${\rm{SNR}} = \sigma_{\rm{DRW}} / \sqrt{\sigma_{\rm{n}}^2 + \overline{dy}^2}$ where $\overline{dy}$ is the mean uncertainty of the light curve measurements.
  \item \textbf{AGN-like variability:} $\sigma_{\rm{LB}} > 3$, where $\sigma_{\rm{LB}}$ is the Ljung$-$Box test \citep{LjungBox1978} significance under the null hypothesis that the light curve is white noise.
\end{enumerate}

Constraint (i) is required to avoid a bias in the value of $\tau_{\rm{DRW}}$ if the light curve is of insufficient length. If the light curve is too short, the most likely value of the recovered $\tau_{\rm{DRW}}$ is $20-30\%$ the light curve length with a large scatter \citep{Kozowski2017}. Constraint (ii) is required so that the measured damping timescale does not artificially converge near the cadence. Given the uncertainty of a factor of $0.3$ dex, we require $\tau_{\rm{DRW}}$ to be a factor of $\sim2$ greater than the typical ZTF cadence of 3 days (i.e., $\tau_{\rm{DRW}}\lesssim6$ days could be consistent with fitting noise). Constraint (iii) ensures the fitted model amplitude is reasonable compared to the light curve measurement errors and the DRW SNR is sufficient for a reliable fit. Constraints (iii) and (iv) are both required, because \textsc{celerite} will attempt to fit a DRW to a light curve even if it is dominated by measurement noise, resulting in ${\rm{SNR}} > 1$. Both these constraints assure the DRW is properly fitted to a light curve with AGN-like (correlated) variability.

To justify each of the constraints above, we perform DRW simulations to test the recoverability of input $\tau_{\rm{DRW}}$. For each of the light curves in our \emph{g}-band parent sample, we create a simulated DRW light with the same sampling/cadence, length, and noise as the real data. For the first test, we use the same amplitude $\sigma_{\rm{DRW}}$ and damping timescale $\tau_{\rm{DRW}}$ as the real light curve. Then, we performed a set of simulations in the same fashion with input $\tau_{\rm{DRW}}$ $10$ times and $0.1$ times the values from the real light curves. Figure~\ref{fig:simulations} shows the results of these simulations. When all quality and duration constraints are applied, the recovered $\tau_{\rm{DRW}}$ are shown to be reliable (blue circles in the central box in the left panel of Figure~\ref{fig:simulations}).

In our final sample of \N low-mass AGN candidates that pass both our initial criteria (compactness and variability) as well as the DRW model fitting quality checks above (criteria i-iii), we found one object in common with the final \citet{Burke2021} sample. The low-mass end of the \citet{Burke2021} relation is dominated by a few dwarf AGNs from \citet{Reines2013} with shorter light curve durations from a previous ZTF data release and a light curve of NGC 4395 from the Transiting Exoplanet Survey Satellite \citep{Burke2020}. Of the five ZTF sources in the final sample from \citet{Burke2021}, we excluded four from this work for the following reasons: Mrk 1044 (SNR < 1.0), NGC 4395 (insufficient cadence/sampling), RGG 127 (SNR < 1.0), and RGG 123 (SNR < 1.0). This does not indicate the \citet{Burke2021} measurements for these sources were incorrect, because in this work we enforced stricter quality controls starting from a larger sample of ZTF light curves. Compared to the previous work, more of our ZTF light curves would pass the length requirement given the longer ZTF light curves from the more recent data release. However, we found that many of those light curves which were too short for inclusion in the final \citet{Burke2021} sample are noisy from e.g., variations in weather across the light curve. Given the much worse average quality of parent AGN sample of ZTF data release light curves, we required stricter quality cuts in this work at a cost of sample completeness. Indeed, visual inspection of the model fitting results for these four sources indicates that the fits may be reasonable and their damping timescale are within $1\sigma$ agreement. A summary table of our DRW fitting results for the final sample is given in Table~\ref{tab:1}.

\section{Results} \label{sec:results}

\begin{figure}
\includegraphics[width=0.48\textwidth]{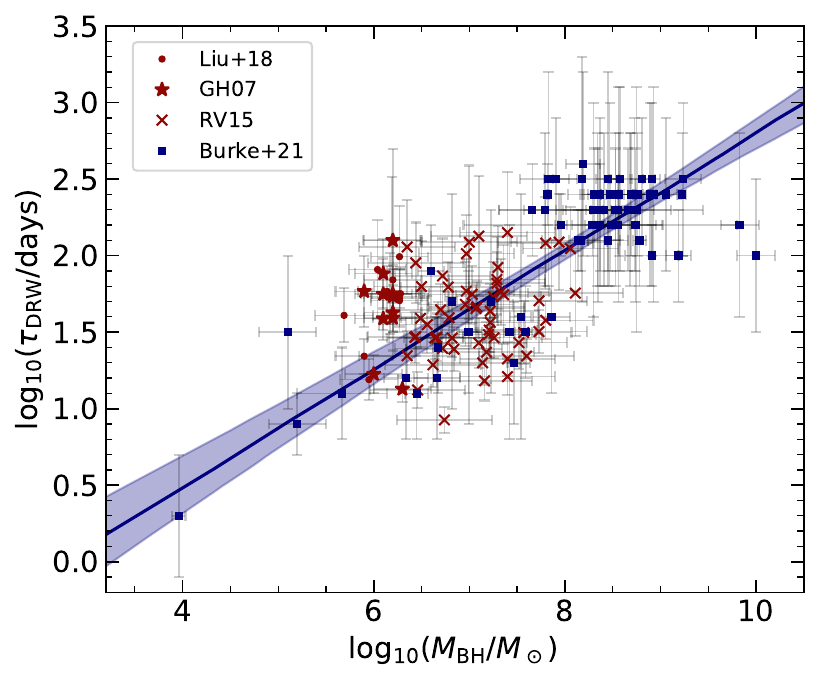}
\caption{Rest-frame damping timescale versus black hole mass for our final sample (red symbols) with $1\sigma$ uncertainty measured using \emph{g}-band ZTF light curves. We also show the data (blue square symbols) and best-fit relation (blue line) with $1\sigma$ uncertainty (light blue band) from \citet{Burke2021}. Our best-fit relation is consistent with the relation of \citet{Burke2021} within \agreement{}. \label{fig:tau_mass}}
\end{figure}

We perform linear regression between $\tau_{\rm DRW}$ (rest-frame) and $M_{\rm{BH}}$ incorporating measurement uncertainties of both quantities. We do not correct for differences between observed wavelength at different redshifts, because the observed scaling of $\tau_{\rm DRW} \propto \lambda^{0.17}$ \citep{MacLeod2010,Suberlak2021,Stone2022} is generally smaller than the measurement uncertainties given the moderate redshift range of our AGNs. These measurement uncertainties are slightly asymmetric in general, therefore we symmetrize them by taking the mean. We use a hierarchical Bayesian model for fitting a line to data with measurement uncertainties \citep{Kelly2007}. The best-fit relation and $1\sigma$ uncertainties for our parent sample of \N low-mass AGNs follow the simple linear relation,
\begin{equation}\label{eqn:tau_mass}
    \tau_{\rm{DRW}} = \alpha\ {\rm{days}}\  \left(\frac{M_{\rm{BH}}}{10^6\ M_{\odot}}\right)^\beta\
\end{equation}
where $\alpha = 22.19^{+2.66}_{-1.68}$, $\beta = 0.31^{+0.03}_{-0.03}$ with a $1\sigma$ intrinsic scatter of $0.20^{+0.03}_{-0.03}$ dex in $\tau_{\rm{DRW}}$. The data and best-fit relation are shown in Figure~\ref{fig:tau_mass}. Our best-fitting linear model is consistent with the \citet{Burke2021} results within \agreement{}. However, the dynamic range in BH mass for the sample of dwarf AGNs is small, which results in a much shallower slope with larger uncertainties than the best-fitting relation in \citet{Burke2021}. Given the small dynamic range in mass of dwarf AGNs, it is not surprising that a correlation is not statistically significant using our sample alone.

\label{sec:tau_mass_noduration}

\subsection{Extension to Stellar Mass Accretion Disks}

\begin{figure}
\includegraphics[width=0.48\textwidth]{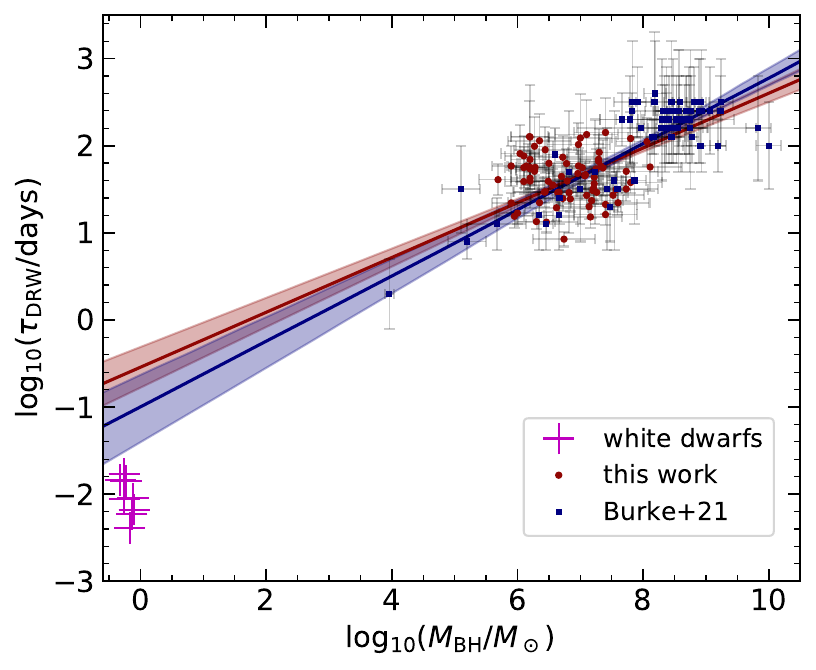}
\caption{Rest-frame damping timescale versus black hole mass for our final \emph{g}-band ZTF sample combined with the final sample from \citet{Burke2021} (black symbols). We show the best-fit relation (black line) with $1\sigma$ uncertainty (gray band) measured using the combined sample data with duplicate sources removed (see text). We also show the data and best-fit relation (blue line) with $1\sigma$ uncertainty (light blue band) from \citet{Burke2021}. For comparison, we show the values for accreting white dwarfs \citep{Scaringi2015}, scaled to the typical AGN Eddington luminosity ratio of $0.15$, in magenta. The typical uncertainties on $M_{\rm WD}$ and the white dwarf damping timescale are 0.2 dex and 0.01 days respectively \citep{Scaringi2015}. 
\label{fig:tau_mass_full}}
\end{figure}

Next, we re-fit the combined final sample of \emph{g}-band ZTF dwarf AGNs with the final sample from \citet{Burke2021}. The results are shown in Figure~\ref{fig:tau_mass_full}. We removed the one \emph{g}-band duplicate source from the \citet{Burke2021} sample which was included in our sample with longer ZTF light curves and updated BH masses from the relation of \citet{Reines2015}. For comparison to stellar mass accretion disks, we also show accreting white dwarfs of the nova-like class from \citet{Scaringi2015}. However, we do not include the white dwarfs in the fitting. Although our combined sample results appear consistent with previous work, we caution that the ZTF light curves are generally of poorer quality and the virial BH masses of dwarf AGNs may have greater uncertainties, results in a larger scatter for these dwarf AGNs. Nevertheless, this provides an important confirmation to the \citet{Burke2021} results after better populating the AGN mass range of $M_{\rm{BH}} \sim 10^6 - 10^8\ M_{\odot}$.

\begin{table*}
\centering
\caption{Final sample parameters and variability timing results. We list the J2000 right ascension (RA), declination (DEC), spectroscopic SDSS redshift ($z$), black hole mass, fitted damping timescale and amplitude, and light curve baseline, Ljung–Box test significance ($\sigma_{\rm{LB}}$), and signal to noise ratio (SNR) as defined in \S\ref{sec:simulations} in the text. $\sigma_{\rm{LB}}$ values of ``inf'' indicates a high significance beyond the numerical precision limits of our code. The reference refers to the source of the broad-line BH mass estimate we adopted. Logarithms are 10 based.}
\label{tab:1}
\scriptsize
\begin{tabular}{ccccccccccc}
\hline
Used Name & RA & DEC & z & $\log(M_{\rm{BH}}/M_\odot)$ & $\log(\tau_{\rm{DRW}} / {\rm{days}})$ & $\log(\sigma_{\rm{DRW}} / {\rm{mag}})$ & baseline & $\sigma_{\rm{LB}}$ & SNR & reference \\
 & (deg) & (deg) &  & (dex) & (dex) & (dex) & (days) &  &   & \\
\hline
Liu+18 5 & 122.29468 & 11.10531 & 0.0526 & 5.95 ± 0.3 & 1.2 ± 0.1 & -1.5 ± 0.4 & 1505 & 17.0 & 1.4 & \citet{Liu2018} \\
Liu+18 33 & 143.53584 & 17.94556 & 0.0271 & 5.69 ± 0.3 & 1.6 ± 0.2 & -1.4 ± 0.4 & 1512 & 26.3 & 1.5 & \citet{Liu2018} \\
Liu+18 34 & 143.77839 & 14.97909 & 0.0831 & 6.04 ± 0.3 & 1.9 ± 0.3 & -1.6 ± 0.5 & 1512 & 20.1 & 1.4 & \citet{Liu2018} \\
Liu+18 100 & 188.65214 & 16.53123 & 0.0736 & 6.14 ± 0.3 & 1.8 ± 0.2 & -1.2 ± 0.4 & 1553 & inf & 2.9 & \citet{Liu2018} \\
Liu+18 116 & 196.20814 & 20.64553 & 0.0644 & 6.17 ± 0.3 & 1.8 ± 0.3 & -1.5 ± 0.5 & 1561 & 33.6 & 1.6 & \citet{Liu2018} \\
Liu+18 117 & 196.23734 & 39.92492 & 0.0274 & 5.90 ± 0.3 & 1.3 ± 0.1 & -1.7 ± 0.4 & 1563 & inf & 1.1 & \citet{Liu2018} \\
Liu+18 125 & 200.32909 & 30.07328 & 0.0916 & 6.28 ± 0.3 & 1.8 ± 0.2 & -1.4 ± 0.4 & 1564 & inf & 1.5 & \citet{Liu2018} \\
Liu+18 126 & 201.16139 & 29.17003 & 0.0727 & 6.27 ± 0.3 & 1.7 ± 0.2 & -1.4 ± 0.5 & 1563 & inf & 1.6 & \citet{Liu2018} \\
Liu+18 157 & 218.54809 & 16.84901 & 0.0936 & 6.20 ± 0.3 & 1.8 ± 0.3 & -1.5 ± 0.5 & 1564 & inf & 1.3 & \citet{Liu2018} \\
Liu+18 173 & 226.26005 & 13.03378 & 0.2045 & 6.19 ± 0.3 & 2.1 ± 0.4 & -1.7 ± 0.5 & 1562 & 31.5 & 1.1 & \citet{Liu2018} \\
Liu+18 186 & 238.10368 & 4.29062 & 0.0459 & 6.27 ± 0.3 & 2.0 ± 0.3 & -1.5 ± 0.5 & 1563 & 31.8 & 1.2 & \citet{Liu2018} \\
Liu+18 199 & 248.75609 & 30.90337 & 0.0543 & 6.27 ± 0.3 & 1.7 ± 0.2 & -1.5 ± 0.4 & 1562 & inf & 2.5 & \citet{Liu2018} \\
GH07 60 & 132.96933 & 52.47582 & 0.0644 & 6.20 ± 0.3 & 1.6 ± 0.2 & -1.8 ± 0.4 & 1515 & inf & 1.1 & \citet{Greene2007} \\
GH07 63 & 135.83736 & 4.96059 & 0.0567 & 6.10 ± 0.3 & 1.7 ± 0.2 & -1.4 ± 0.5 & 1512 & 19.5 & 1.9 & \citet{Greene2007} \\
GH07 105 & 165.74471 & 46.63652 & 0.1490 & 6.10 ± 0.3 & 1.9 ± 0.2 & -1.7 ± 0.5 & 1538 & 37.7 & 1.3 & \citet{Greene2007} \\
GH07 124 & 174.15992 & 42.76473 & 0.0710 & 6.00 ± 0.3 & 1.2 ± 0.1 & -1.7 ± 0.4 & 1540 & inf & 1.1 & \citet{Greene2007} \\
GH07 160 & 199.86049 & 10.93638 & 0.0644 & 5.90 ± 0.3 & 1.8 ± 0.2 & -1.6 ± 0.5 & 1563 & 29.3 & 1.4 & \citet{Greene2007} \\
GH07 171 & 209.35217 & 65.41832 & 0.1060 & 6.20 ± 0.3 & 1.8 ± 0.2 & -1.4 ± 0.5 & 1563 & inf & 2.2 & \citet{Greene2007} \\
GH07 177 & 215.32969 & 63.21673 & 0.1340 & 6.20 ± 0.3 & 1.7 ± 0.2 & -1.6 ± 0.5 & 1563 & inf & 1.5 & \citet{Greene2007} \\
GH07 203 & 239.79010 & 35.02988 & 0.0310 & 6.20 ± 0.3 & 1.6 ± 0.1 & -1.1 ± 0.4 & 1564 & inf & 3.2 & \citet{Greene2007} \\
GH07 204 & 240.18746 & 50.87044 & 0.1000 & 6.10 ± 0.3 & 1.6 ± 0.1 & -1.8 ± 0.4 & 1562 & inf & 1.1 & \citet{Greene2007} \\
GH07 215 & 254.15411 & 37.24434 & 0.0626 & 6.30 ± 0.3 & 1.1 ± 0.1 & -1.6 ± 0.4 & 1562 & inf & 2.3 & \citet{Greene2007} \\
GH07 219 & 314.59226 & -6.83455 & 0.0737 & 6.20 ± 0.3 & 2.1 ± 0.6 & -1.4 ± 0.5 & 1506 & 23.8 & 1.3 & \citet{Greene2007} \\
RV15 4 & 32.54789 & -9.05987 & 0.0419 & 8.11 ± 0.5 & 1.8 ± 0.3 & -1.9 ± 0.5 & 1303 & 22.8 & 1.1 & \citet{Reines2015} \\
RV15 6 & 46.07405 & 0.47423 & 0.0447 & 6.50 ± 0.5 & 1.8 ± 0.3 & -1.4 ± 0.5 & 1315 & inf & 1.4 & \citet{Reines2015} \\
RV15 7 & 50.02634 & 40.36628 & 0.0479 & 6.72 ± 0.5 & 1.4 ± 0.1 & -1.7 ± 0.4 & 1553 & 34.8 & 1.1 & \citet{Reines2015} \\
RV15 12 & 118.85534 & 39.18606 & 0.0333 & 6.74 ± 0.5 & 0.9 ± 0.1 & -1.3 ± 0.4 & 1508 & inf & 2.4 & \citet{Reines2015} \\
RV15 14 & 119.70951 & 9.58909 & 0.0455 & 7.25 ± 0.5 & 1.7 ± 0.2 & -1.4 ± 0.5 & 1505 & inf & 1.3 & \citet{Reines2015} \\
RV15 17 & 120.68084 & 31.06759 & 0.0410 & 7.80 ± 0.5 & 2.1 ± 0.3 & -1.1 ± 0.5 & 1506 & inf & 1.9 & \citet{Reines2015} \\
RV15 21 & 122.00730 & 38.32648 & 0.0413 & 6.46 ± 0.5 & 1.1 ± 0.1 & -1.8 ± 0.4 & 1508 & 31.2 & 1.0 & \citet{Reines2015} \\
RV15 23 & 123.33051 & 46.14712 & 0.0539 & 7.08 ± 0.5 & 1.7 ± 0.2 & -1.2 ± 0.4 & 1496 & inf & 3.6 & \citet{Reines2015} \\
RV15 26 & 124.36005 & 10.20281 & 0.0455 & 7.40 ± 0.5 & 2.2 ± 0.4 & -0.9 ± 0.5 & 1505 & inf & 1.4 & \citet{Reines2015} \\
RV15 37 & 131.72539 & 25.37009 & 0.0506 & 7.73 ± 0.5 & 1.7 ± 0.2 & -1.0 ± 0.5 & 1507 & 30.5 & 1.8 & \citet{Reines2015} \\
RV15 45 & 139.14301 & 17.91898 & 0.0268 & 7.40 ± 0.5 & 1.2 ± 0.1 & -1.7 ± 0.4 & 1512 & 21.6 & 1.8 & \citet{Reines2015} \\
RV15 47 & 140.31484 & 10.29473 & 0.0390 & 7.36 ± 0.5 & 1.7 ± 0.3 & -1.6 ± 0.5 & 1512 & 4.9 & 2.8 & \citet{Reines2015} \\
RV15 56 & 145.51997 & 23.68526 & 0.0211 & 7.10 ± 0.5 & 2.1 ± 0.4 & -1.3 ± 0.5 & 1512 & inf & 1.5 & \citet{Reines2015} \\
RV15 58 & 146.37234 & 9.60290 & 0.0137 & 6.35 ± 0.5 & 1.3 ± 0.2 & -1.6 ± 0.4 & 1512 & 22.5 & 2.3 & \citet{Reines2015} \\
RV15 68 & 150.52930 & 3.05767 & 0.0228 & 7.22 ± 0.5 & 1.6 ± 0.2 & -1.4 ± 0.4 & 1513 & 31.3 & 4.5 & \citet{Reines2015} \\
RV15 69 & 150.57826 & 26.80159 & 0.0518 & 7.52 ± 0.5 & 1.4 ± 0.2 & -1.6 ± 0.4 & 1515 & inf & 2.1 & \citet{Reines2015} \\
RV15 73 & 153.01080 & 30.21751 & 0.0497 & 7.29 ± 0.5 & 1.8 ± 0.2 & -1.3 ± 0.5 & 1523 & inf & 1.3 & \citet{Reines2015} \\
RV15 83 & 159.46589 & 33.81392 & 0.0510 & 6.43 ± 0.5 & 1.5 ± 0.2 & -1.7 ± 0.4 & 1523 & 28.0 & 1.1 & \citet{Reines2015} \\
RV15 85 & 160.72055 & 4.24476 & 0.0519 & 7.29 ± 0.5 & 1.8 ± 0.3 & -1.4 ± 0.5 & 1529 & 26.9 & 1.8 & \citet{Reines2015} \\
RV15 92 & 165.06680 & 46.27092 & 0.0315 & 6.44 ± 0.5 & 1.5 ± 0.1 & -1.5 ± 0.4 & 1538 & inf & 2.4 & \citet{Reines2015} \\
RV15 98 & 167.69151 & 11.61153 & 0.0417 & 7.18 ± 0.5 & 1.4 ± 0.2 & -1.5 ± 0.4 & 1527 & 15.7 & 1.1 & \citet{Reines2015} \\
RV15 102 & 169.74039 & 58.05665 & 0.0282 & 7.10 ± 0.5 & 1.4 ± 0.2 & -1.5 ± 0.4 & 1540 & inf & 1.8 & \citet{Reines2015} \\
RV15 106 & 171.61964 & 47.41495 & 0.0331 & 6.35 ± 0.5 & 2.1 ± 0.3 & -1.6 ± 0.5 & 1540 & inf & 1.3 & \citet{Reines2015} \\
RV15 113 & 175.05809 & 24.69706 & 0.0127 & 7.00 ± 0.5 & 2.1 ± 0.5 & -1.9 ± 0.5 & 1538 & 33.1 & 1.6 & \citet{Reines2015} \\
RV15 119 & 176.43826 & 55.79998 & 0.0540 & 7.16 ± 0.5 & 1.2 ± 0.1 & -2.0 ± 0.4 & 1547 & inf & 1.3 & \citet{Reines2015} \\
RV15 122 & 179.29451 & 22.29617 & 0.0520 & 7.30 ± 0.5 & 1.9 ± 0.3 & -1.5 ± 0.5 & 1547 & 27.4 & 2.3 & \citet{Reines2015} \\
RV15 123 & 180.30980 & -3.67809 & 0.0193 & 6.96 ± 0.5 & 1.7 ± 0.3 & -1.4 ± 0.5 & 1536 & 10.8 & 1.8 & \citet{Reines2015} \\
RV15 129 & 181.60947 & 42.74056 & 0.0522 & 7.33 ± 0.5 & 1.8 ± 0.2 & -1.1 ± 0.5 & 1549 & inf & 3.7 & \citet{Reines2015} \\
RV15 130 & 181.86655 & 25.60303 & 0.0481 & 6.64 ± 0.5 & 1.5 ± 0.1 & -1.6 ± 0.4 & 1545 & inf & 1.5 & \citet{Reines2015} \\
RV15 133 & 184.02951 & 50.82512 & 0.0313 & 8.06 ± 0.5 & 2.0 ± 0.3 & -0.9 ± 0.5 & 1553 & inf & 2.5 & \citet{Reines2015} \\
RV15 134 & 184.61134 & -0.13059 & 0.0499 & 6.56 ± 0.5 & 1.5 ± 0.2 & -1.7 ± 0.4 & 1545 & 12.9 & 1.7 & \citet{Reines2015} \\
RV15 137 & 185.56405 & 28.35659 & 0.0277 & 6.84 ± 0.5 & 1.4 ± 0.1 & -1.4 ± 0.4 & 1551 & inf & 1.6 & \citet{Reines2015} \\
RV15 140 & 188.15618 & 66.41459 & 0.0473 & 7.40 ± 0.5 & 1.3 ± 0.1 & -1.1 ± 0.4 & 1563 & inf & 2.8 & \citet{Reines2015} \\
RV15 152 & 195.08309 & 61.65512 & 0.0526 & 7.60 ± 0.5 & 1.3 ± 0.1 & -1.2 ± 0.4 & 1561 & inf & 3.3 & \citet{Reines2015} \\
RV15 157 & 198.27422 & 1.46553 & 0.0288 & 6.44 ± 0.5 & 2.0 ± 0.3 & -1.5 ± 0.5 & 1563 & 25.0 & 2.1 & \citet{Reines2015} \\
RV15 164 & 204.41614 & 39.15456 & 0.0202 & 6.70 ± 0.5 & 1.6 ± 0.2 & -1.8 ± 0.4 & 1563 & inf & 1.4 & \citet{Reines2015} \\
RV15 167 & 206.11005 & 44.27223 & 0.0546 & 6.82 ± 0.5 & 1.5 ± 0.1 & -1.1 ± 0.4 & 1564 & inf & 2.0 & \citet{Reines2015} \\
RV15 174 & 211.31193 & -2.98367 & 0.0535 & 7.21 ± 0.5 & 1.5 ± 0.2 & -1.6 ± 0.4 & 1559 & 16.8 & 1.4 & \citet{Reines2015} \\
RV15 176 & 214.12843 & 1.61890 & 0.0536 & 7.22 ± 0.5 & 1.6 ± 0.2 & -1.4 ± 0.4 & 1564 & 32.9 & 2.7 & \citet{Reines2015} \\
RV15 177 & 214.49809 & 25.13687 & 0.0167 & 7.94 ± 0.5 & 2.1 ± 0.3 & -0.3 ± 0.5 & 1564 & inf & 1.1 & \citet{Reines2015} \\
RV15 180 & 216.03493 & 21.08848 & 0.0471 & 7.03 ± 0.5 & 1.7 ± 0.2 & -1.4 ± 0.5 & 1564 & inf & 1.9 & \citet{Reines2015} \\
RV15 186 & 222.73097 & 27.57853 & 0.0303 & 6.79 ± 0.5 & 1.6 ± 0.2 & -1.5 ± 0.4 & 1563 & 30.2 & 1.7 & \citet{Reines2015} \\
RV15 191 & 225.41164 & 10.42040 & 0.0302 & 7.07 ± 0.5 & 1.7 ± 0.2 & -1.2 ± 0.4 & 1564 & inf & 1.5 & \citet{Reines2015} \\
RV15 200 & 230.85180 & 55.31542 & 0.0394 & 6.72 ± 0.5 & 1.9 ± 0.2 & -1.9 ± 0.5 & 1564 & inf & 1.0 & \citet{Reines2015} \\
RV15 202 & 232.41909 & 30.48592 & 0.0359 & 6.97 ± 0.5 & 1.8 ± 0.2 & -1.5 ± 0.5 & 1564 & inf & 1.2 & \citet{Reines2015} \\
RV15 211 & 238.57264 & 32.64387 & 0.0486 & 7.73 ± 0.5 & 1.5 ± 0.1 & -1.3 ± 0.4 & 1564 & inf & 2.4 & \citet{Reines2015} \\
RV15 212 & 239.17043 & 12.28831 & 0.0355 & 6.66 ± 0.5 & 1.5 ± 0.2 & -1.6 ± 0.4 & 1562 & 31.9 & 1.6 & \citet{Reines2015} \\
RV15 215 & 241.26026 & 33.09584 & 0.0532 & 7.80 ± 0.5 & 1.6 ± 0.1 & -1.3 ± 0.4 & 1563 & inf & 3.9 & \citet{Reines2015} \\
RV15 220 & 242.98464 & 52.18809 & 0.0416 & 7.26 ± 0.5 & 1.5 ± 0.1 & -1.4 ± 0.4 & 1562 & inf & 2.1 & \citet{Reines2015} \\
RV15 223 & 244.68689 & 25.65212 & 0.0483 & 7.20 ± 0.5 & 1.5 ± 0.1 & -1.3 ± 0.4 & 1563 & inf & 2.0 & \citet{Reines2015} \\
RV15 224 & 244.96380 & 40.97981 & 0.0383 & 6.78 ± 0.5 & 1.8 ± 0.2 & -1.0 ± 0.5 & 1563 & inf & 4.2 & \citet{Reines2015} \\
RV15 228 & 247.53080 & -0.19345 & 0.0467 & 7.14 ± 0.5 & 1.3 ± 0.1 & -1.7 ± 0.4 & 1555 & 8.3 & 1.3 & \citet{Reines2015} \\
RV15 233 & 258.11851 & 35.88406 & 0.0269 & 6.97 ± 0.5 & 2.0 ± 0.3 & -1.5 ± 0.5 & 1562 & inf & 2.2 & \citet{Reines2015} \\
RV15 236 & 316.38934 & 0.47478 & 0.0549 & 6.49 ± 0.5 & 1.6 ± 0.2 & -1.6 ± 0.4 & 1532 & 25.2 & 1.4 & \citet{Reines2015} \\
RV15 238 & 323.43439 & -8.27101 & 0.0533 & 7.57 ± 0.5 & 1.5 ± 0.2 & -1.8 ± 0.4 & 1503 & 7.9 & 1.1 & \citet{Reines2015} \\
RV15 242 & 342.10264 & 0.15573 & 0.0544 & 6.62 ± 0.5 & 1.3 ± 0.1 & -1.6 ± 0.4 & 1496 & 20.8 & 1.2 & \citet{Reines2015} \\
\hline
\end{tabular}
\end{table*}

\section{Discussion} \label{sec:discussion}

\begin{figure*}
\includegraphics[width=0.98\textwidth]{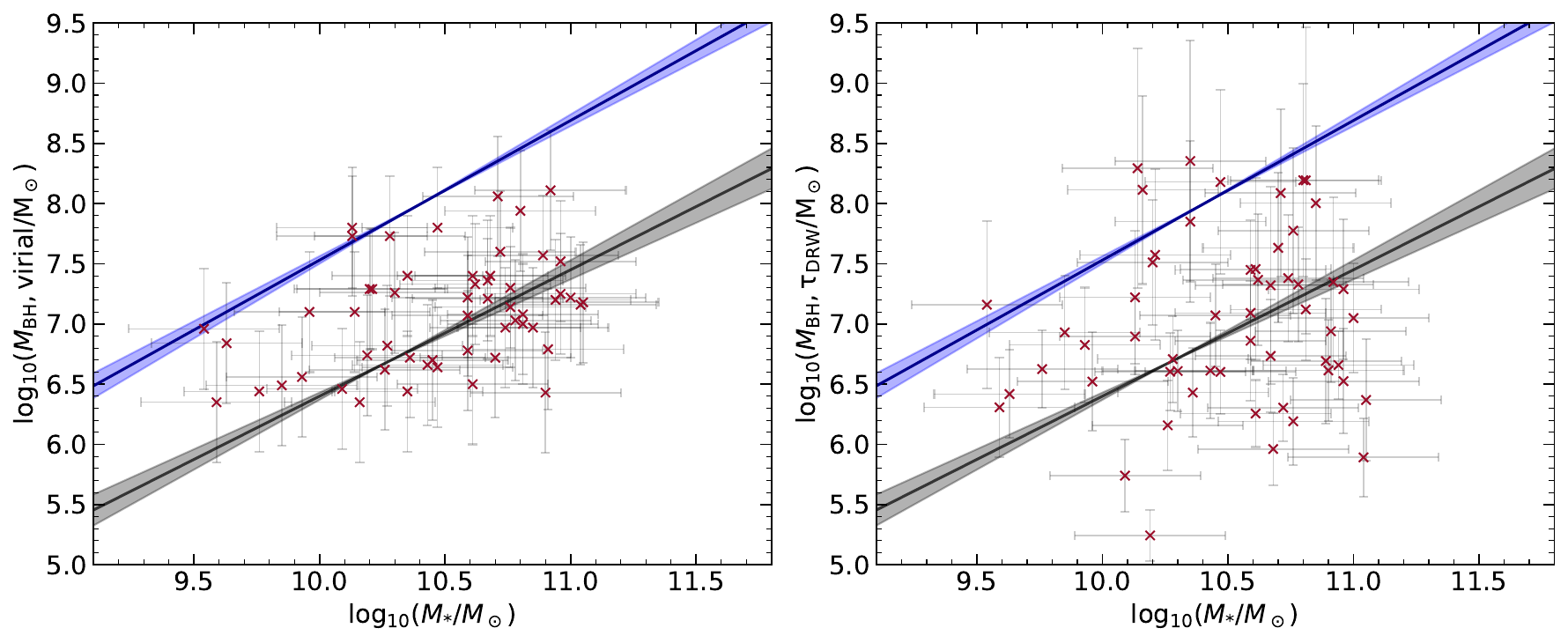}
\caption{Comparison between $M_{\rm{BH}} - M_{\ast}$ and $M_{\rm{BH}} - \tau_{\rm{DRW}}$ relations. We plot the virial BH mass $M_{\rm{BH, virial}}$ versus $M_{\ast}$ against the relations of \citet{Reines2015} and \citet{Kormendy2013} (\emph{left}). We also plot the BH mass predicted from the $M_{\rm{BH}} - \tau_{\rm{DRW}}$ relation of \citet{Burke2021} using our ZTF light curves versus the stellar masses from \citet{Reines2015} in the \emph{g} band (\emph{right}). The gray (blue) line and shaded band is the best-fit local AGN scaling relation and $1\sigma$ uncertainty from \citet{Reines2015} (\citet{Kormendy2013}) relations. The fitted intrinsic scatter of our data in each panel is, from left to right: $0.14^{+0.09}_{-0.06}$ and $0.46^{+0.09}_{-0.10}$ dex.  \label{fig:mass-tau}}
\end{figure*}

To test the usefulness of the $M_{\rm{BH}} - \tau_{\rm{DRW}}$ relation, we obtained host galaxy stellar masses $M_{\ast}$ for objects in our final sample when available from \citet{Reines2015}. These stellar masses were estimated from \cite{Zibetti2009} mass-to-light ratios as a function of their $g-i$ colors after subtracting the contribution from AGN emission using mock AGN spectra \citep{Reines2015}. We then compared the predicted BH masses from the $M_{\rm{BH}} - \tau_{\rm{DRW}}$ relation of \citet{Burke2021}, $M_{\rm{BH}, \tau_{\rm{DRW}}}$, to the expectation from the $M_{\rm{BH}} - M_{\ast}$ relation. The results are shown in Figure~\ref{fig:mass-tau}.

We find that the BH masses predicted from $M_{\rm{BH}} - \tau_{\rm{DRW}}$ relation are less precise (intrinsic scatter of $0.46^{+0.09}_{-0.10}$ dex in predicted BH mass but total rms scatter of $0.77^{+0.02}_{-0.01}$ dex) than the $M_{\rm{BH}} - M_{\ast}$ relation (intrinsic scatter of $0.14^{+0.09}_{-0.06}$ dex with total rms scatter of $0.51^{+0.01}_{-0.01}$ dex). This intrinsic scatter in predicted BH mass from the $M_{\rm{BH}} - \tau_{\rm{DRW}}$ relation is broadly consistent with the intrinsic scatter of $0.33^+{0.11}_{-0.11}$ dex found by \cite{Burke2021}, and similar to the scatter in single-epoch virial BH masses \citep{Shen2013}. The larger total rms scatter could be due to the larger measurement uncertainties in our ZTF damping timescale measurements from poorer light curve quality. The BH mass prediction is proportional to $\tau_{\rm DRW}^{2.54}$, hence a 0.2 dex measurement uncertainty in $\tau_{\rm DRW}$ translates to $\sim 0.5$ dex uncertainty in BH mass, leading to the large rms scatter seen in Figure~\ref{fig:mass-tau}. Given such large measurement uncertainties in predicted BH mass, the nominal intrinsic scatter constrained from the regression fit in this work is more difficult to constrain. Therefore, with higher quality light curves of dwarf AGNs, we expect the predicted BH masses to improve. Alternatively, the large scatter could be due to larger variations of other parameters in dwarf AGNs which were unaccounted for, such as black hole spin and accretion rate. Notably, the \citet{Greene2007} sample, which may have systematically longer damping timescales in Figure~\ref{fig:mass-tau}, have relatively high Eddington ratios of $\log(L_{\rm{bol}}/L_{\rm{Edd}}) \sim -0.5$ \citep{Greene2007}. Future studies with a sample spanning a large range of Eddington luminosity ratios could help inform this work.

We caution that the single-epoch virial BH masses in this work have a typical systematic uncertainty of $\sim0.4$ dex \citep{Shen2013}. For a dwarf AGN, the uncertainty may be even larger, because of enhanced complications from outflows or stellar continuum absorption features to measure the weak broad-line emission from the AGN. In addition, the ZTF data release light curves are generally noisier than composite Stripe 82 light curves adopted in \citet{Burke2021} or light curves from dedicated AGN monitoring programs despite the quality constraints we applied. These sources of systematic uncertainty will contribute to the intrinsic scatter in our relations.


\section{Conclusion} \label{sec:conclusion}
\label{sec:conclusion}

We have analyzed the characteristic ``damping'' timescale of a final sample of \N low-mass (dwarf and Seyfert) AGNs with virial BH mass estimates of $M_{\rm{BH}} \sim 10^6 - 10^8\ M_{\odot}$ at a redshift range of $z \sim 0 - 0.3$ using ZTF data release 13 with $\sim4$ year long light curves with sufficient quality to enable reliable DRW model fitting. Our best-fit linear model is consistent with the \citet{Burke2021} results within \agreement{} where the IMBH range is better populated by the new ZTF sample. This work demonstrates consistency of dwarf AGN variability with the $M_{\rm{BH}} - \tau_{\rm{DRW}}$ relation, which indicates their single-epoch virial masses are reasonable on average. Given their short damping timescales, we conclude that the dwarf AGN candidates in our sample may indeed be dwarf AGNs, rather than long-lived transients \citep{Burke2020} or more massive AGNs. This provides a useful and independent line of evidence for the AGN nature of these sources, even for those with narrow emission line ratios consistent with star formation.

High-quality, long-duration light curves will improve the uncertainties on individual $\tau_{\rm{DRW}}$ measurements and improve the robustness of the $M_{\rm{BH}} - \tau_{\rm{DRW}}$ relation, particularly at the high-mass end ($M_{\rm{BH}} \gtrsim 10^9\ M_{\odot}$) where exclusion of AGNs with $\tau_{\rm{DRW}}$ values greater than several hundred days with insufficient light curve baselines may result in a slope slightly shallower than the predicted slope of $0.5$ from the standard model of accretion disks \citep{Burke2021}. 

High-cadence monitoring of the lowest-mass dwarf AGNs ($M_{\rm{BH}} \lesssim 10^5\ M_{\odot}$) can be used to estimate the BH mass without the need for spectroscopy. The $M_{\rm{BH}} - \tau_{\rm{DRW}}$ of \citet{Burke2021} relation predicts a damping timescale of about 13 hours (3 days) for a $10^2\ M_{\rm{\odot}}$ ($10^4\ M_{\rm{\odot}}$) IMBH. In this case, the $M_{\rm{BH}} - \tau_{\rm{DRW}}$ relation can serve as a discovery machine for IMBHs which fail to produce a broad line region \citep{Chakravorty2014} or where the broad emission line is too weak to detect, as long as the optical variability is dominated by the accretion disk emission rather than from a companion star or reprocessed light from X-rays. This motivates optical surveys with intra-day cadence to enable discovery and mass estimation of IMBHs using variability \citep{Bellm2022}. The variable sky at these timescales is still relatively unexplored at sufficient volumes and depths to enable a systematic search for rapidly-accreting IMBHs.

\section*{Acknowledgements}

We thank Gautham Narayan for useful discussions. C.J.B. acknowledges the Illinois Graduate Survey Science Fellowship for partial support. Z.F.W. and X.L. acknowledge support from the University of Illinois Campus Research Board and NSF grant AST-2206499. Y.S. acknowledges support from NSF grant AST-2009947. This research was supported in part by the National Science Foundation under PHY-1748958. This research made use of Photutils, an Astropy package for
detection and photometry of astronomical sources \citep{Bradley2022}.

Based on observations obtained with the Samuel Oschin Telescope 48-inch and the 60-inch Telescope at the Palomar Observatory as part of the Zwicky Transient Facility project. ZTF is supported by the National Science Foundation under Grant No. AST-2034437 and a collaboration including Caltech, IPAC, the Weizmann Institute for Science, the Oskar Klein Center at Stockholm University, the University of Maryland, Deutsches Elektronen-Synchrotron and Humboldt University, the TANGO Consortium of Taiwan, the University of Wisconsin at Milwaukee, Trinity College Dublin, Lawrence Livermore National Laboratories, and IN2P3, France. Operations are conducted by COO, IPAC, and UW.

This work makes use of SDSS-I/II and SDSS-III/IV data (http://www.sdss.org/ and http://www.sdss3.org/).

{\it Facilities:} ZTF, Sloan

\section*{Data Availability}

All ZTF light curves can be downloaded at \url{https://irsa.ipac.caltech.edu/Missions/ztf.html}. Our timing analysis code is available at \url{https://github.com/burke86/taufit}.



\bibliographystyle{mnras}
\bibliography{ref} 



\appendix

\section{Final Sample Light Curves and Best-fit DRW Models}\label{appx:lc}

\begin{figure*}
\includegraphics[width=0.48\textwidth]{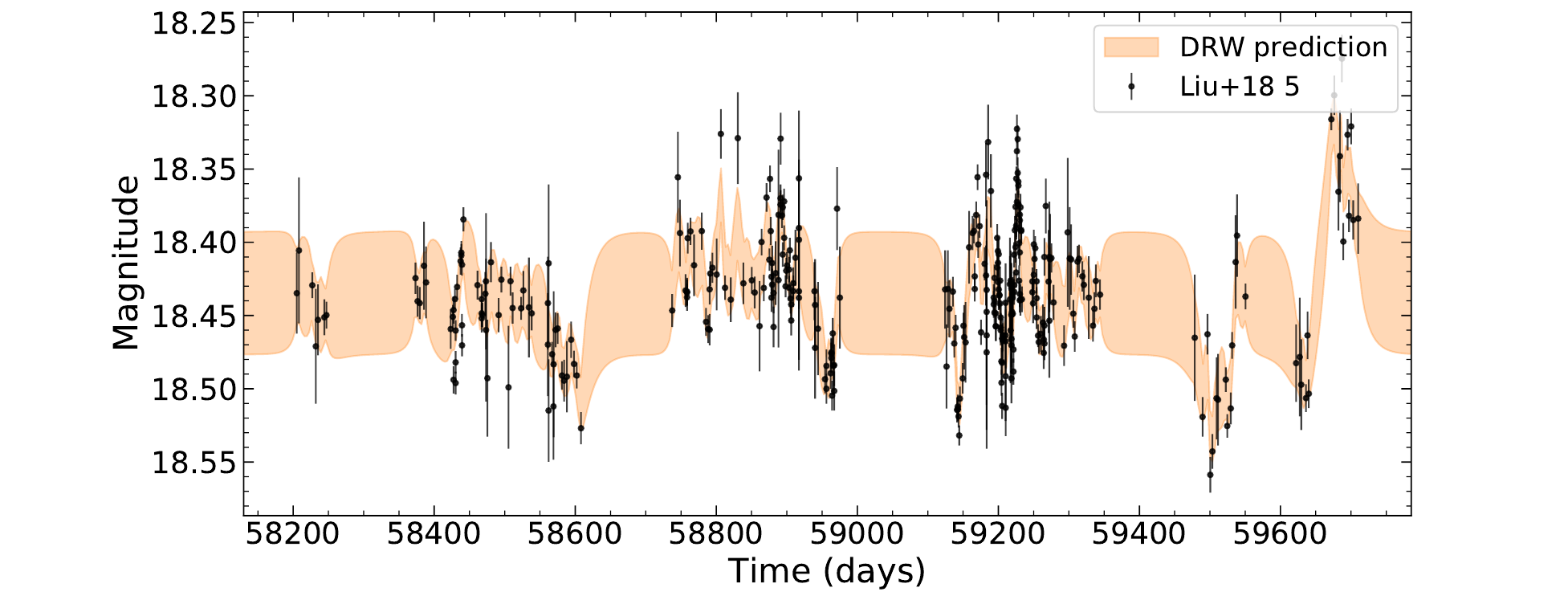}
\includegraphics[width=0.48\textwidth]{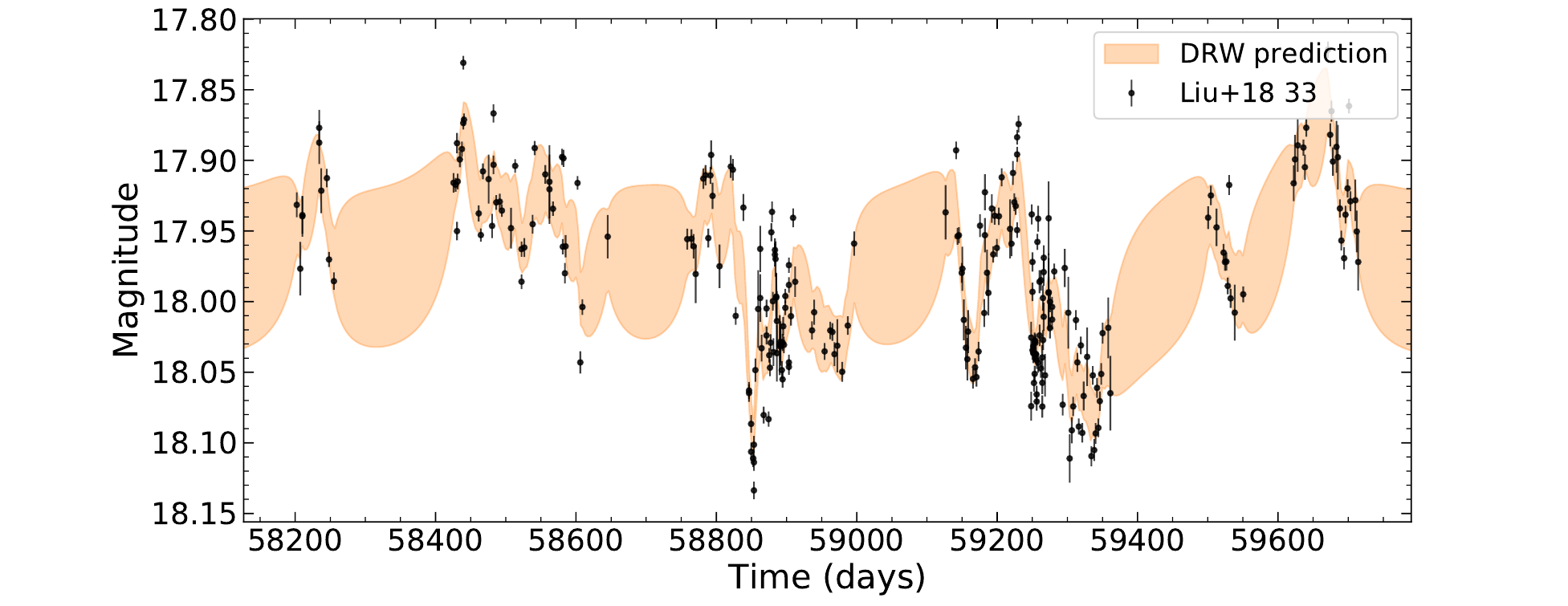}
\includegraphics[width=0.48\textwidth]{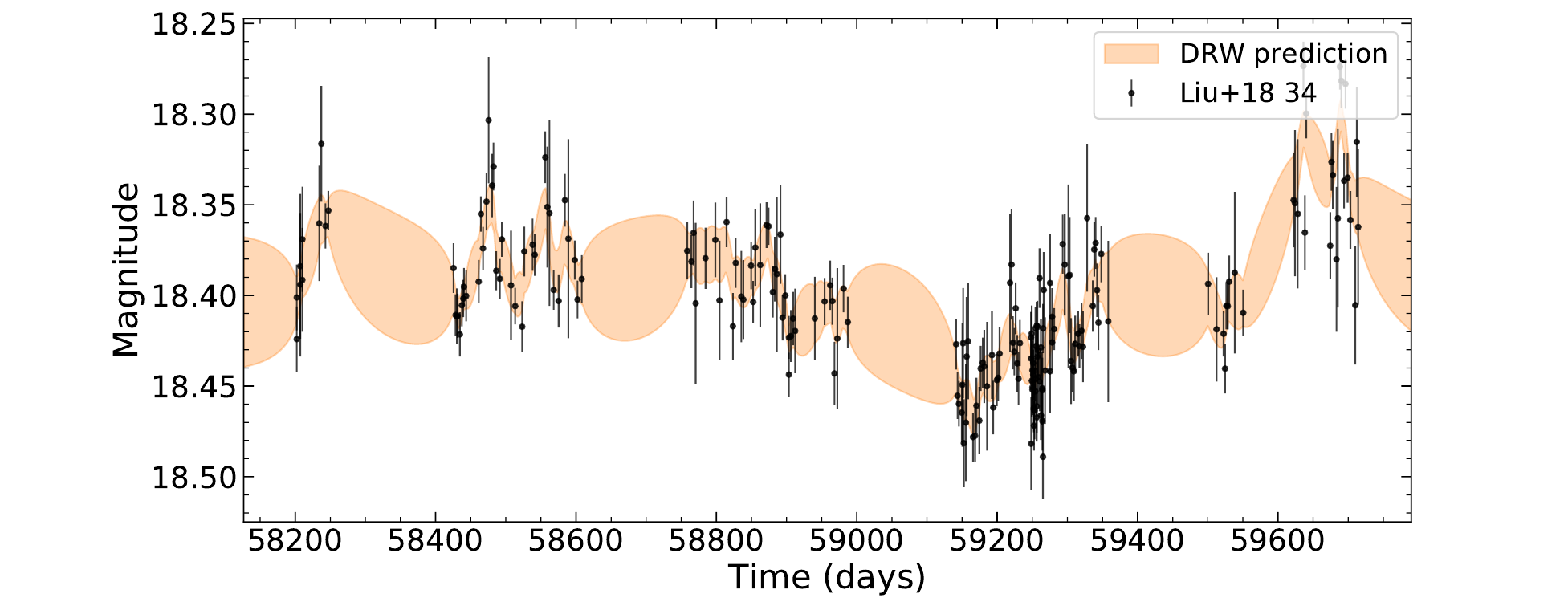}
\includegraphics[width=0.48\textwidth]{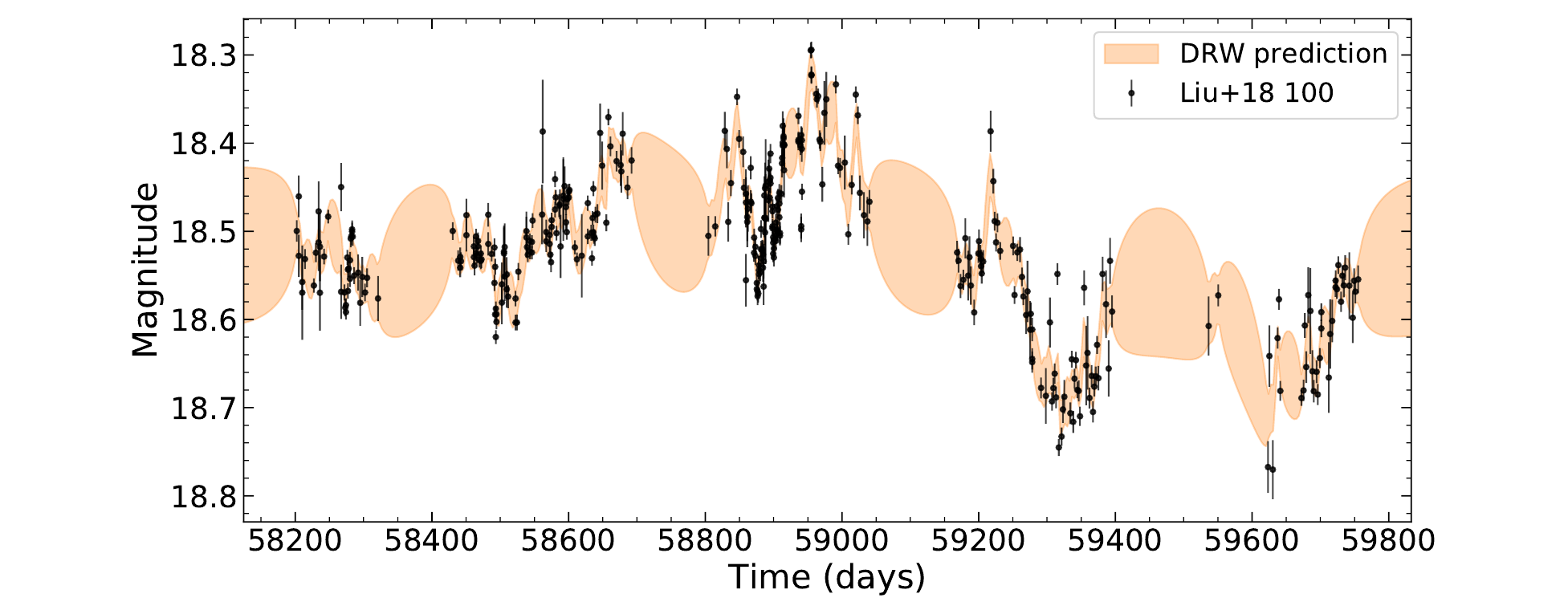}
\includegraphics[width=0.48\textwidth]{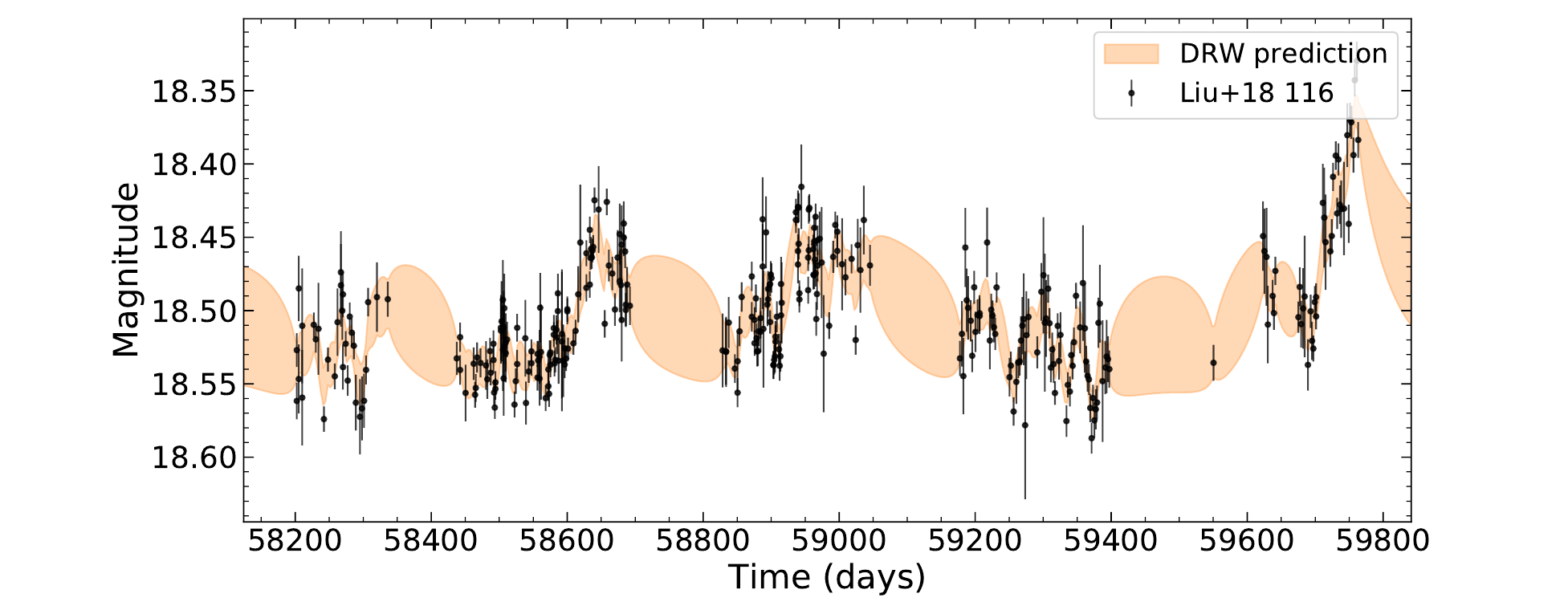}
\includegraphics[width=0.48\textwidth]{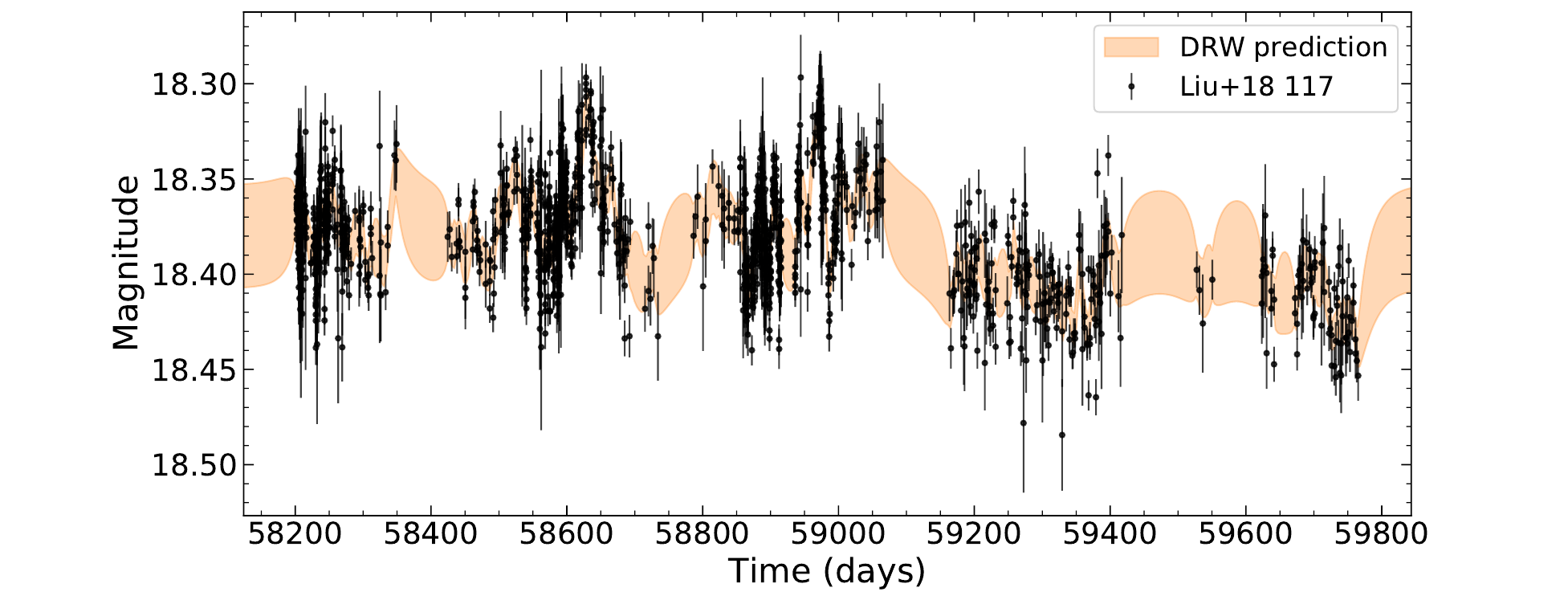}
\includegraphics[width=0.48\textwidth]{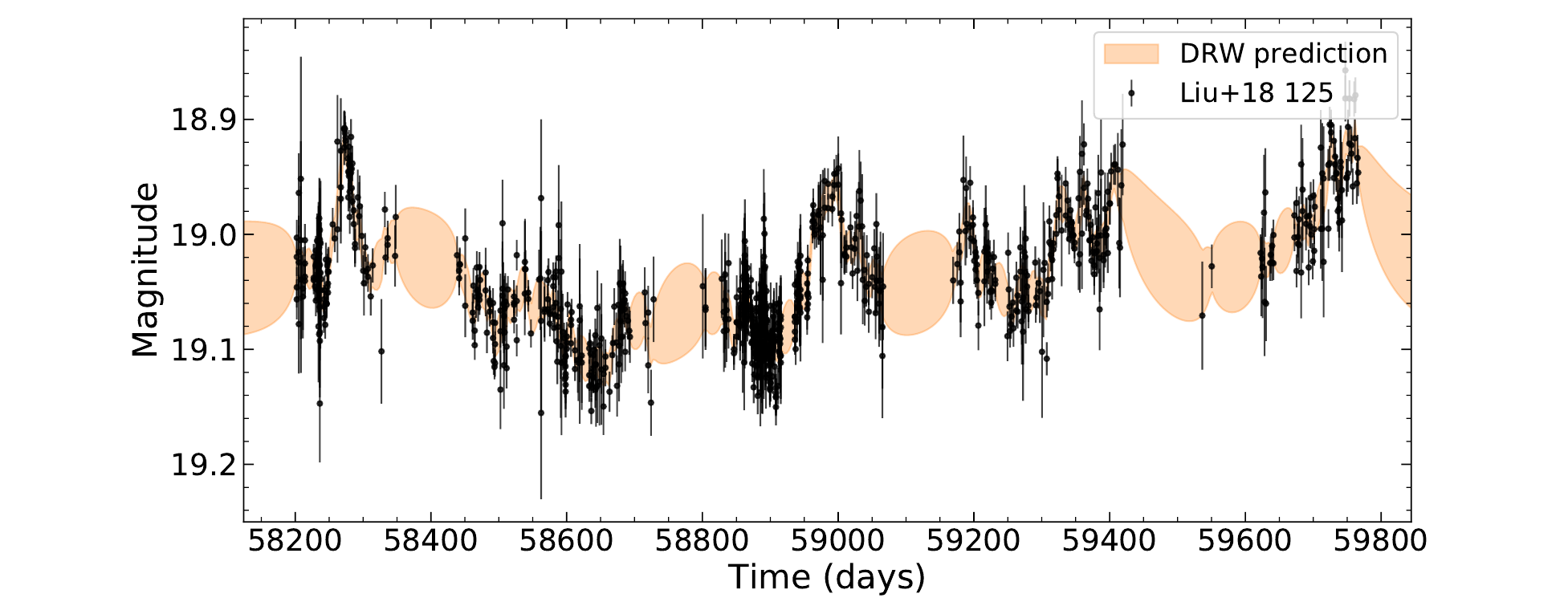}
\includegraphics[width=0.48\textwidth]{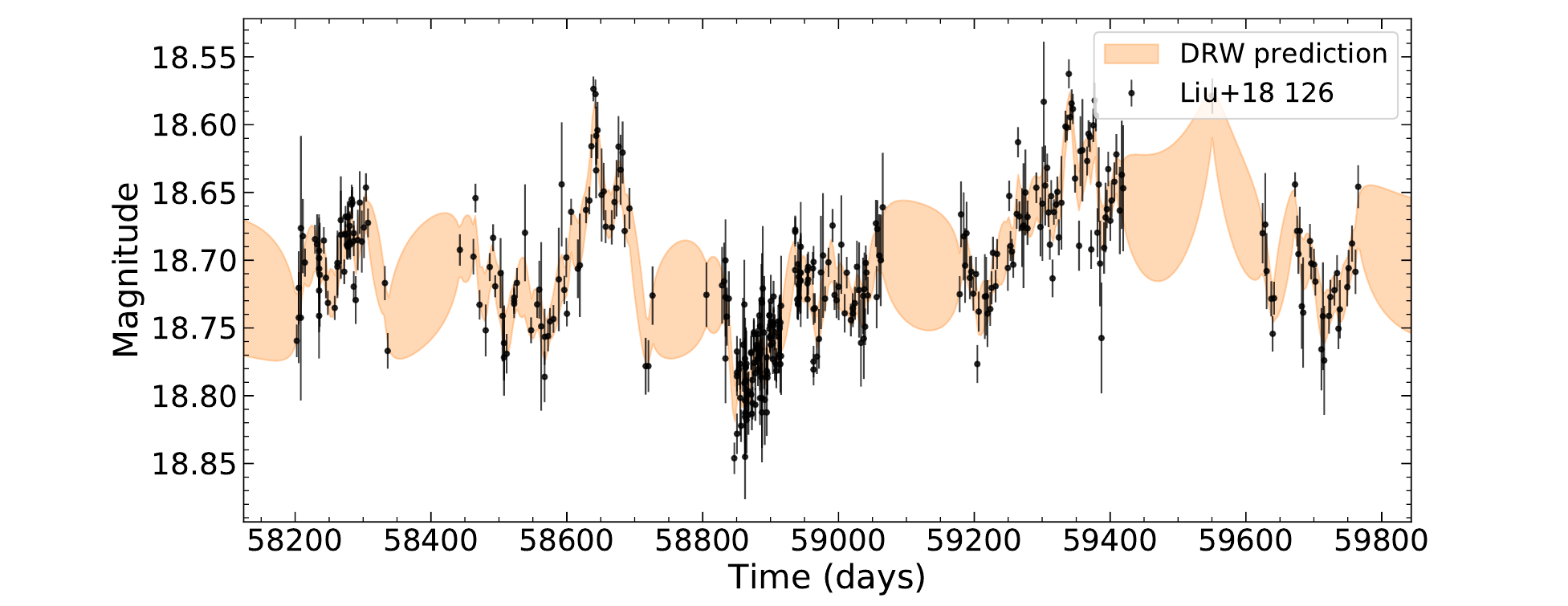}
\includegraphics[width=0.48\textwidth]{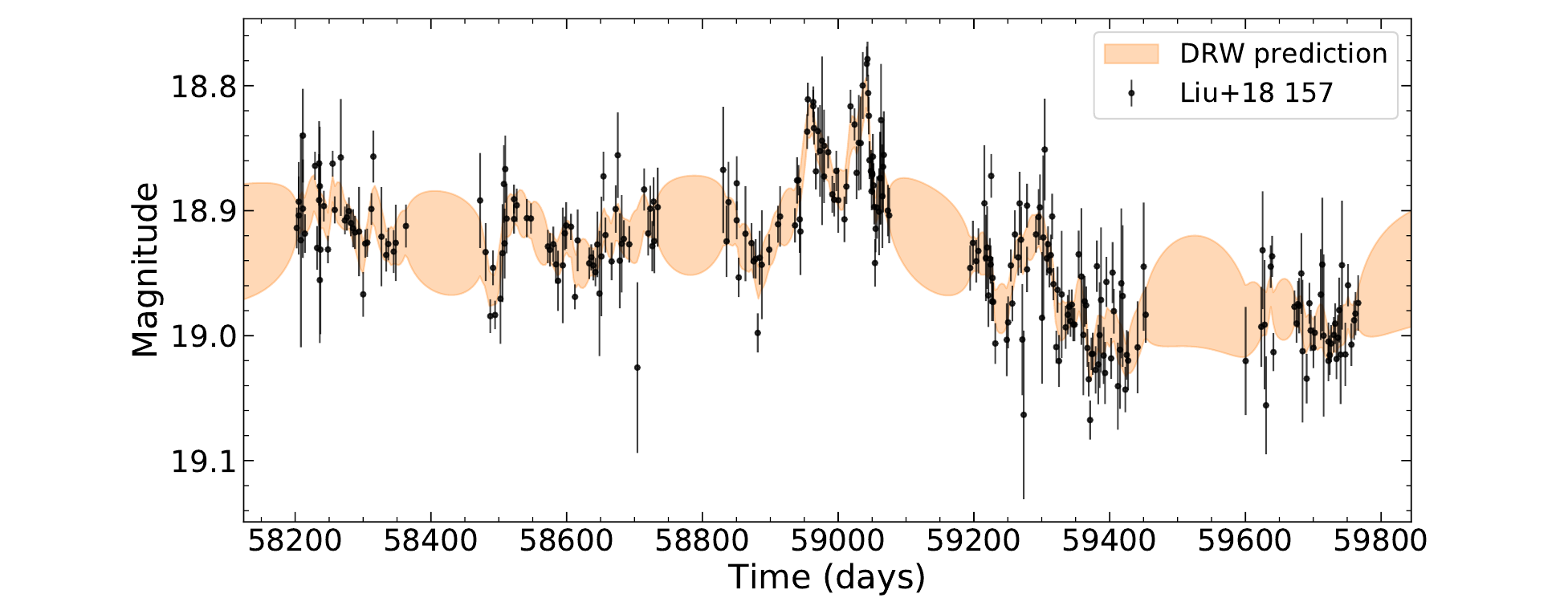}
\includegraphics[width=0.48\textwidth]{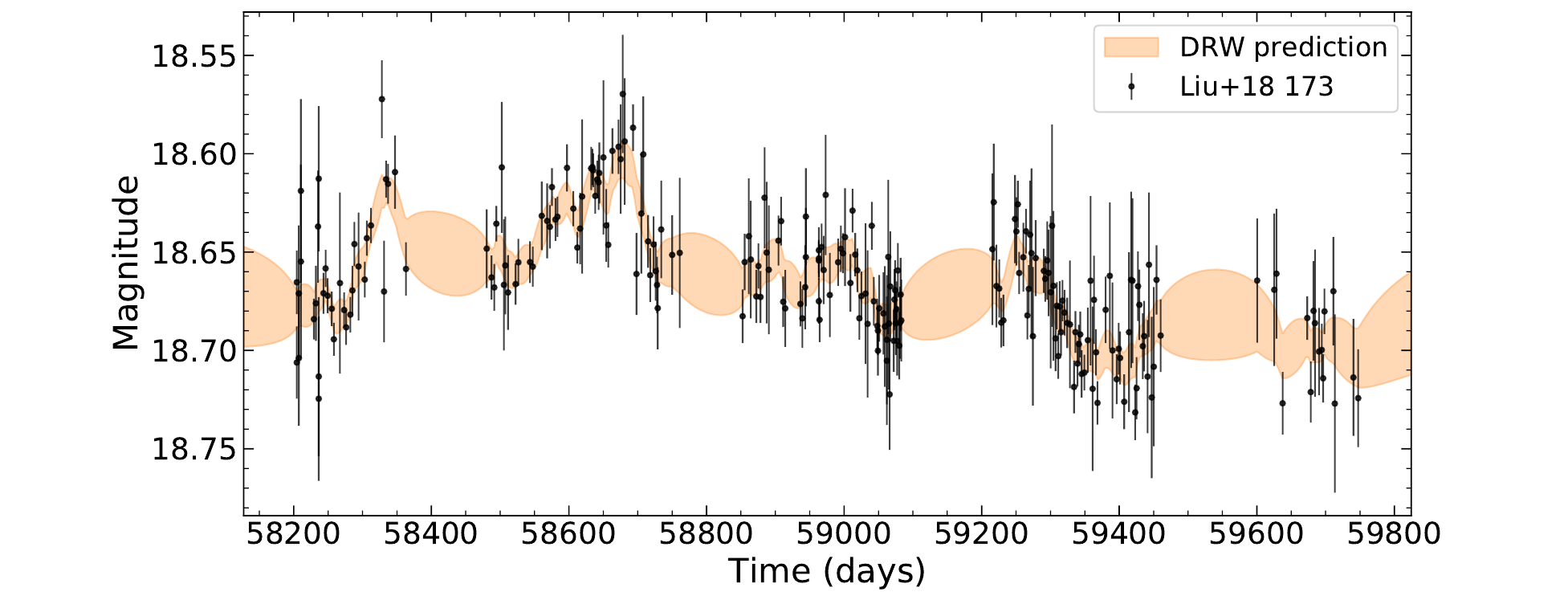}
\includegraphics[width=0.48\textwidth]{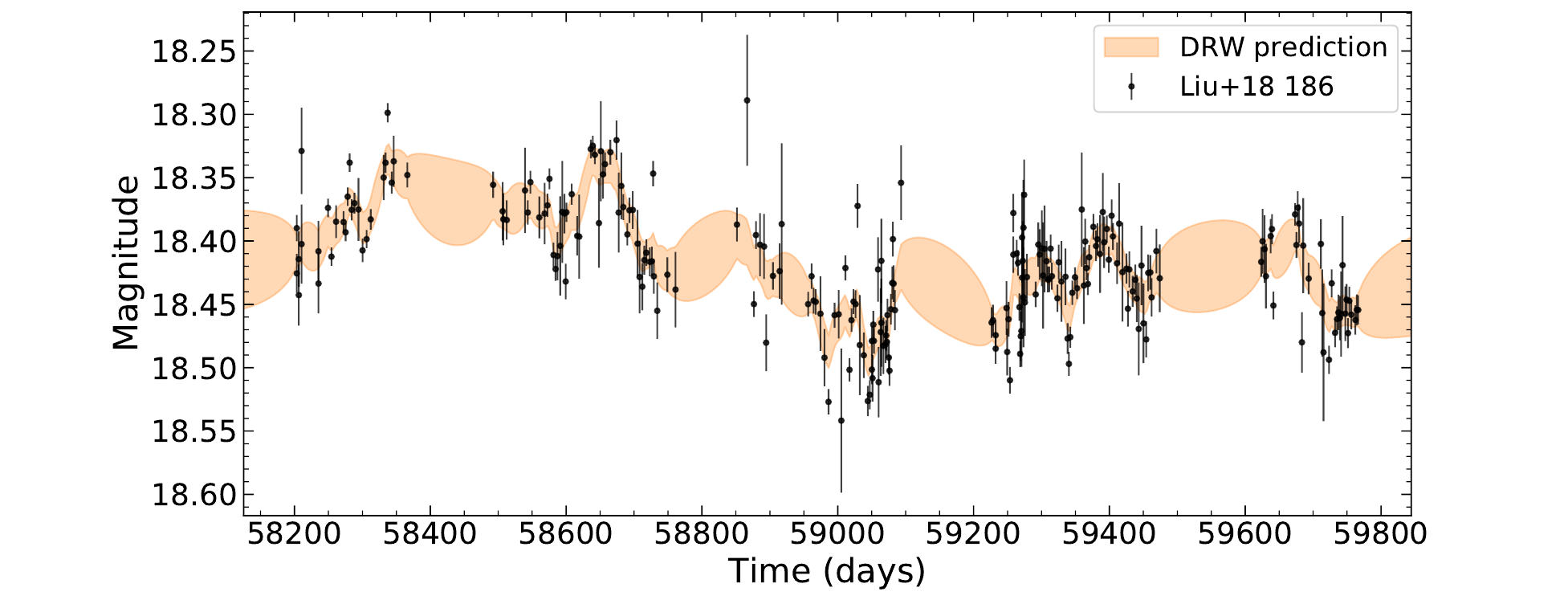}
\includegraphics[width=0.48\textwidth]{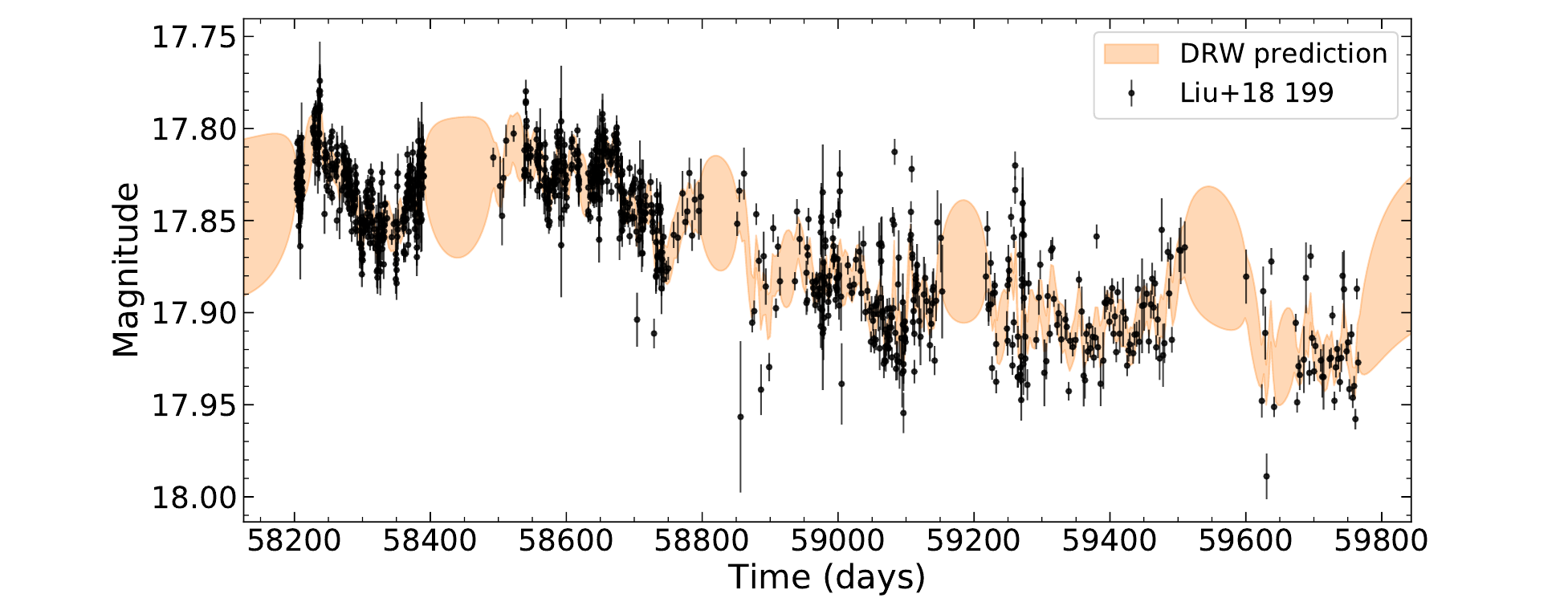}
\caption{ZTF \emph{g}-band difference imaging light curves and DRW model predictions for dwarf AGNs in our final sample (continued below).\label{fig:lc_appx}}
\end{figure*}

\begin{figure*}
\includegraphics[width=0.48\textwidth]{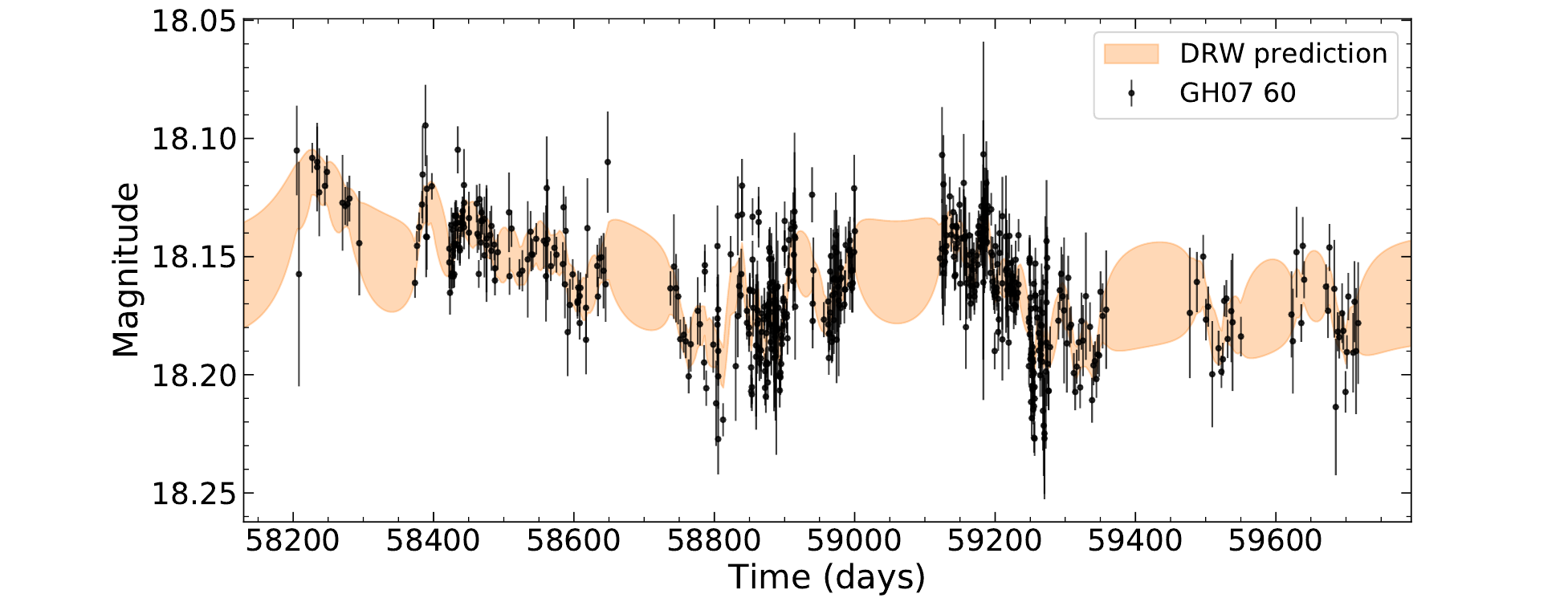}
\includegraphics[width=0.48\textwidth]{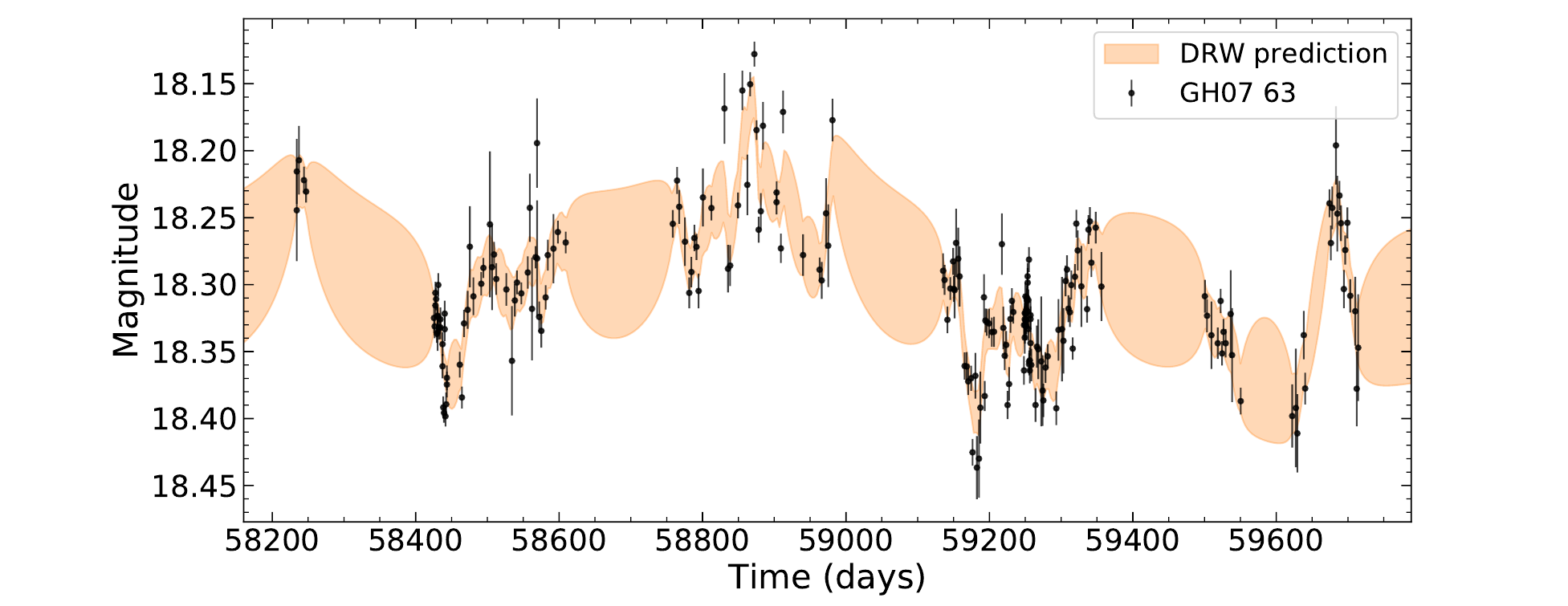}
\includegraphics[width=0.48\textwidth]{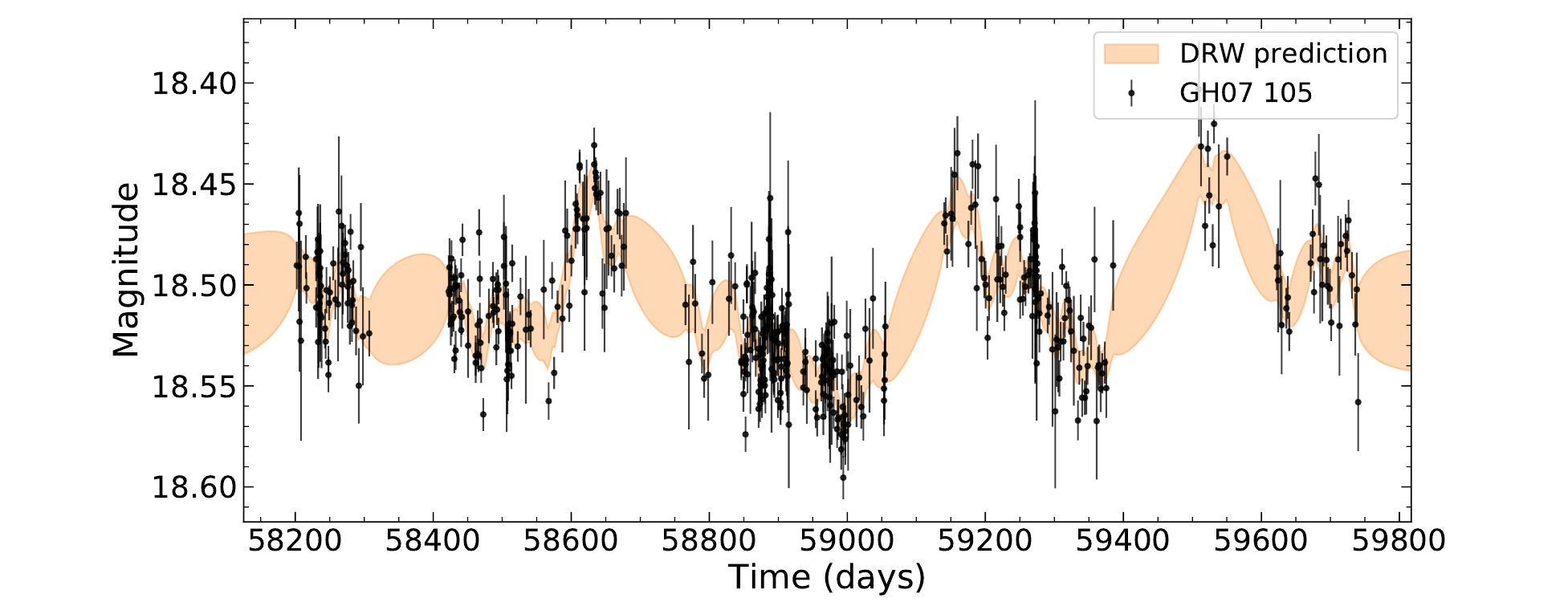}
\includegraphics[width=0.48\textwidth]{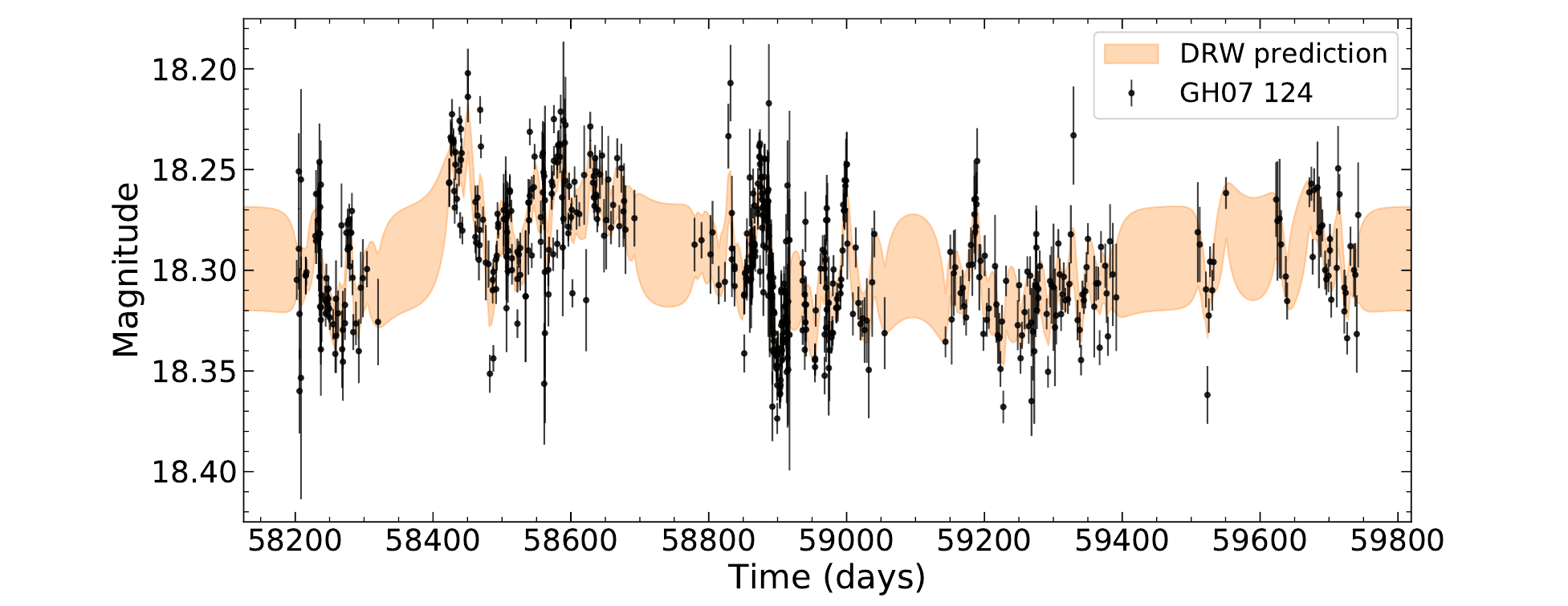}
\includegraphics[width=0.48\textwidth]{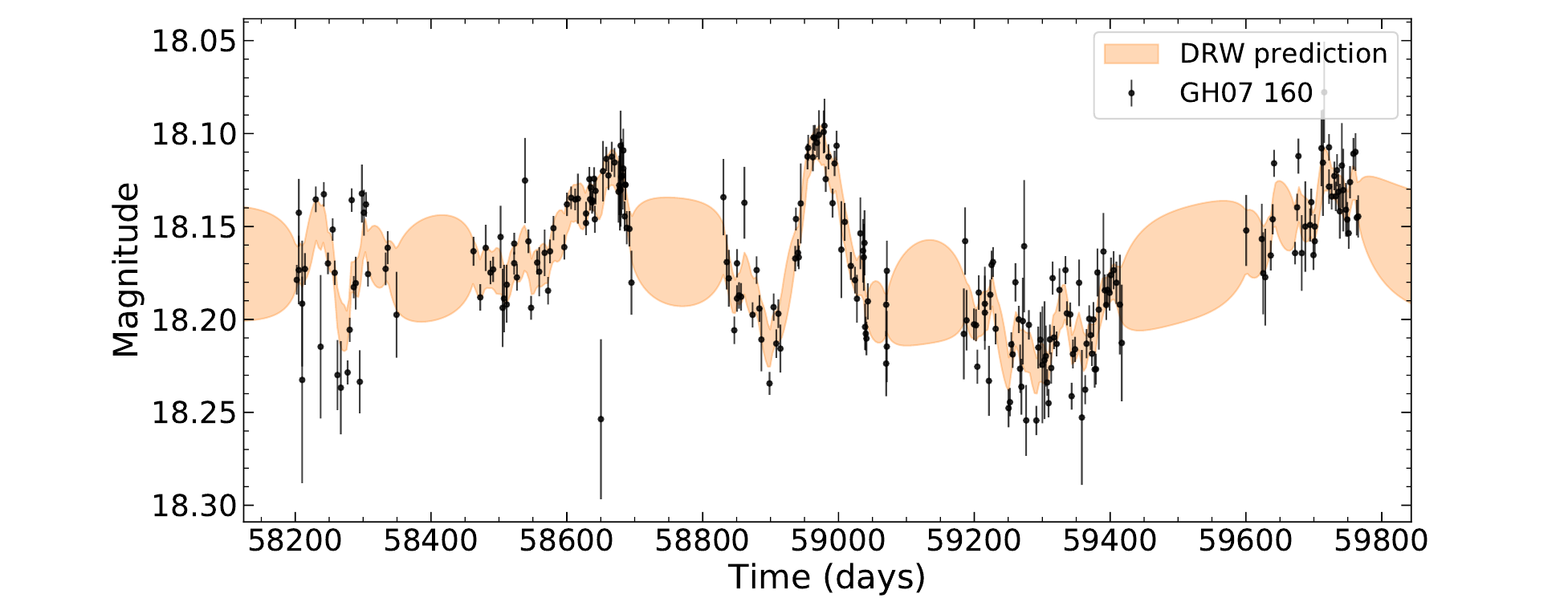}
\includegraphics[width=0.48\textwidth]{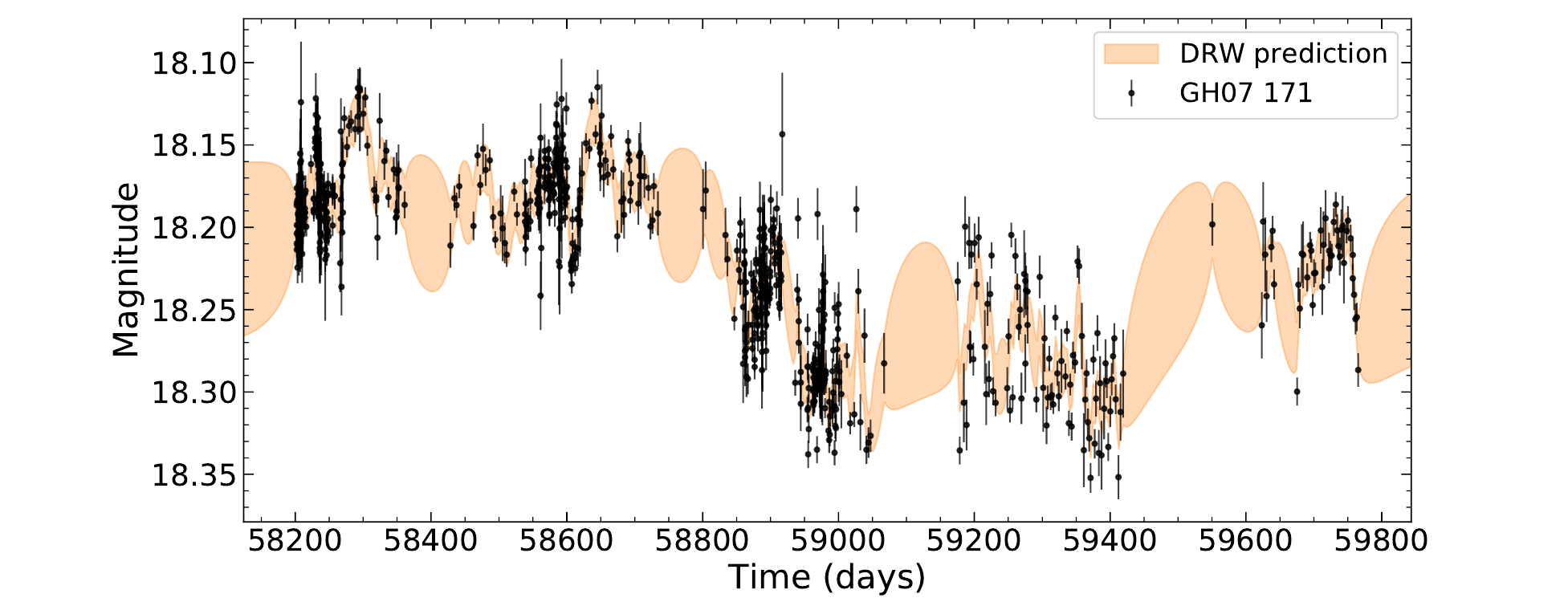}
\includegraphics[width=0.48\textwidth]{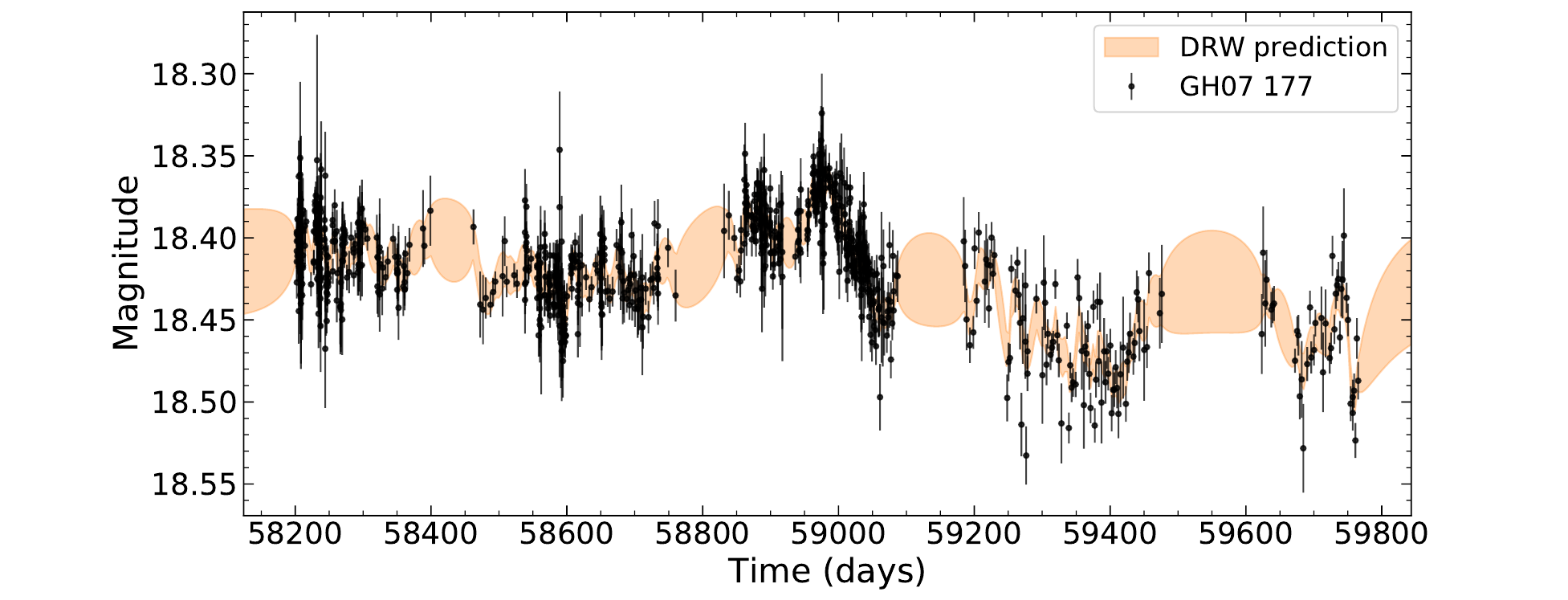}
\includegraphics[width=0.48\textwidth]{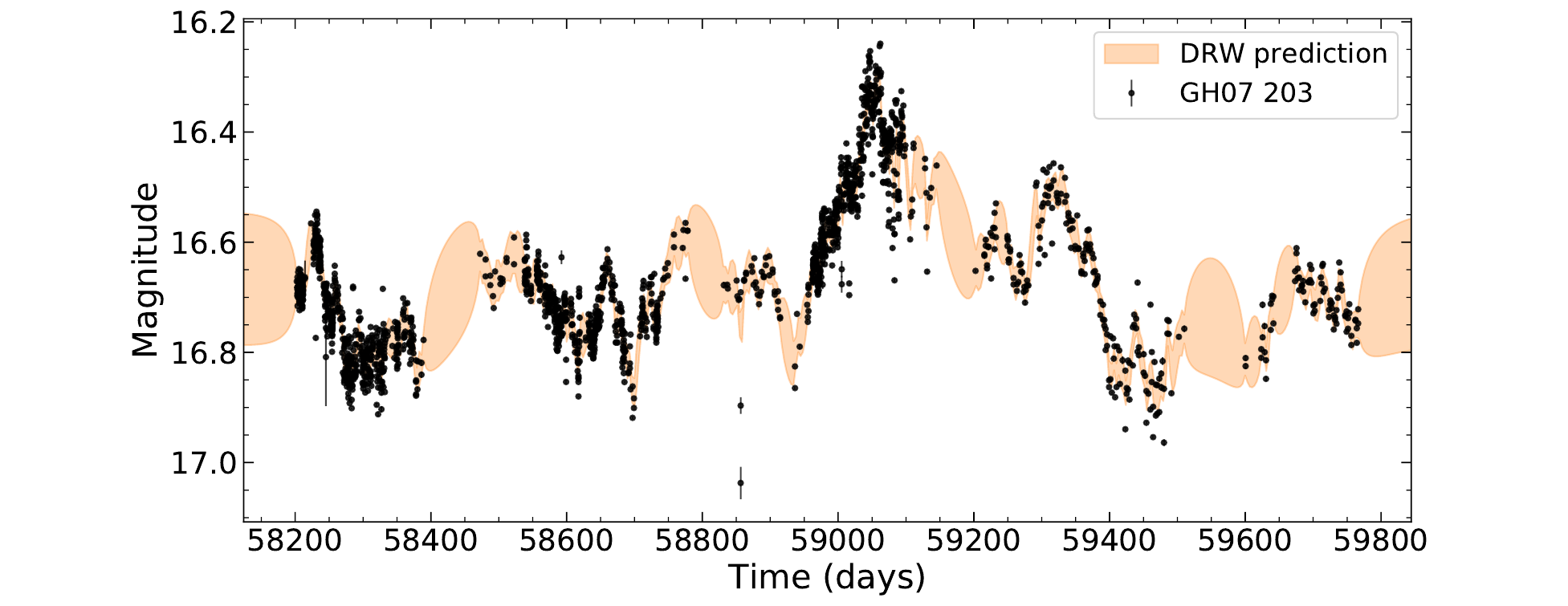}
\includegraphics[width=0.48\textwidth]{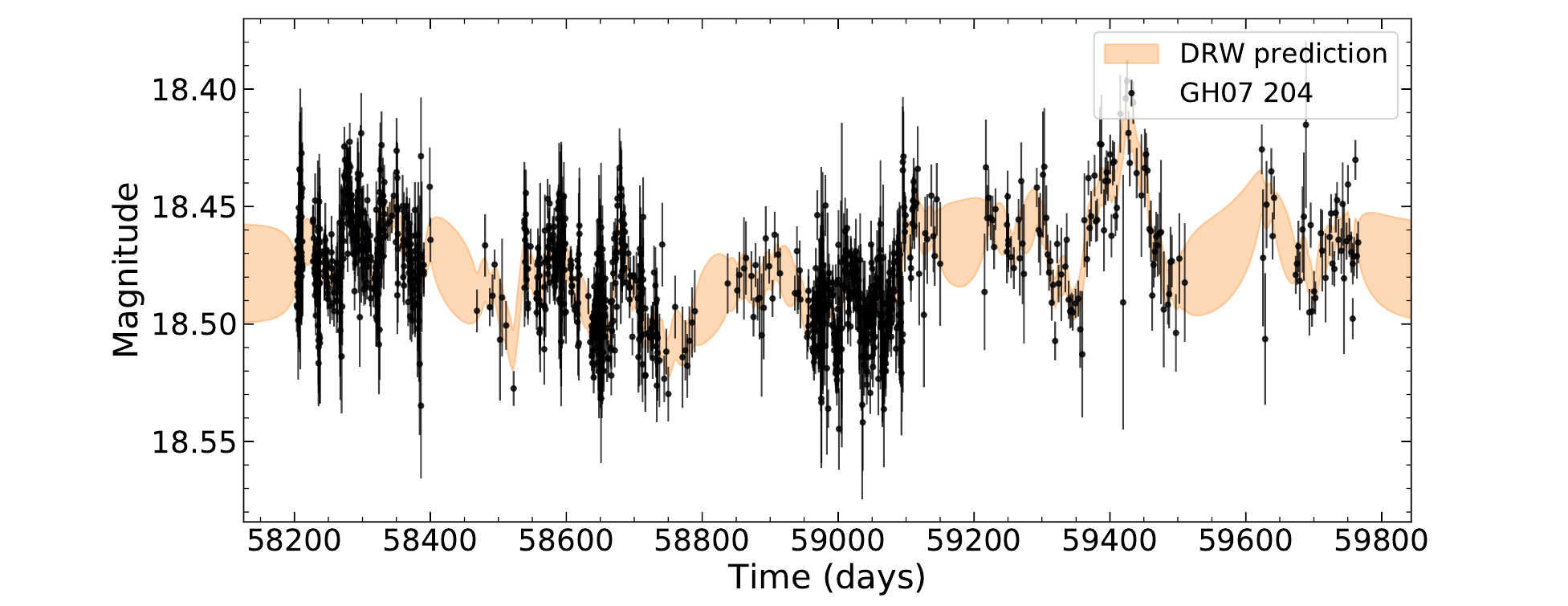}
\includegraphics[width=0.48\textwidth]{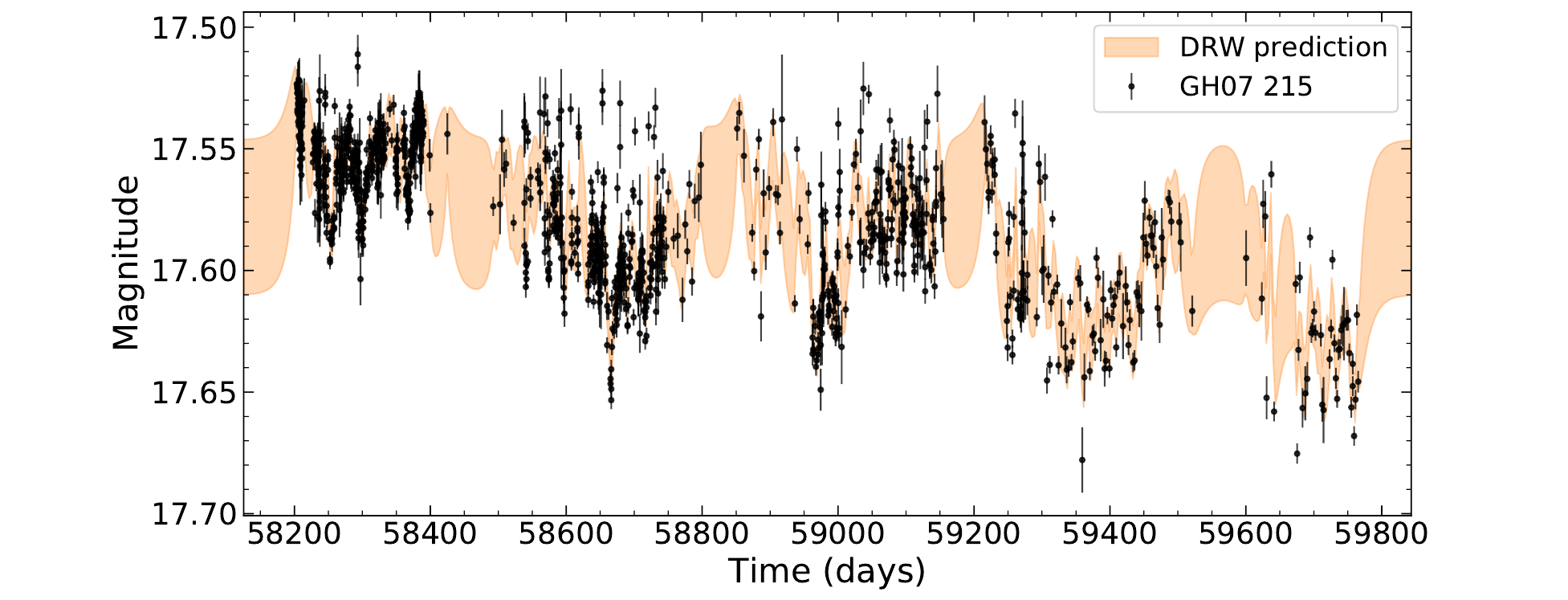}
\includegraphics[width=0.48\textwidth]{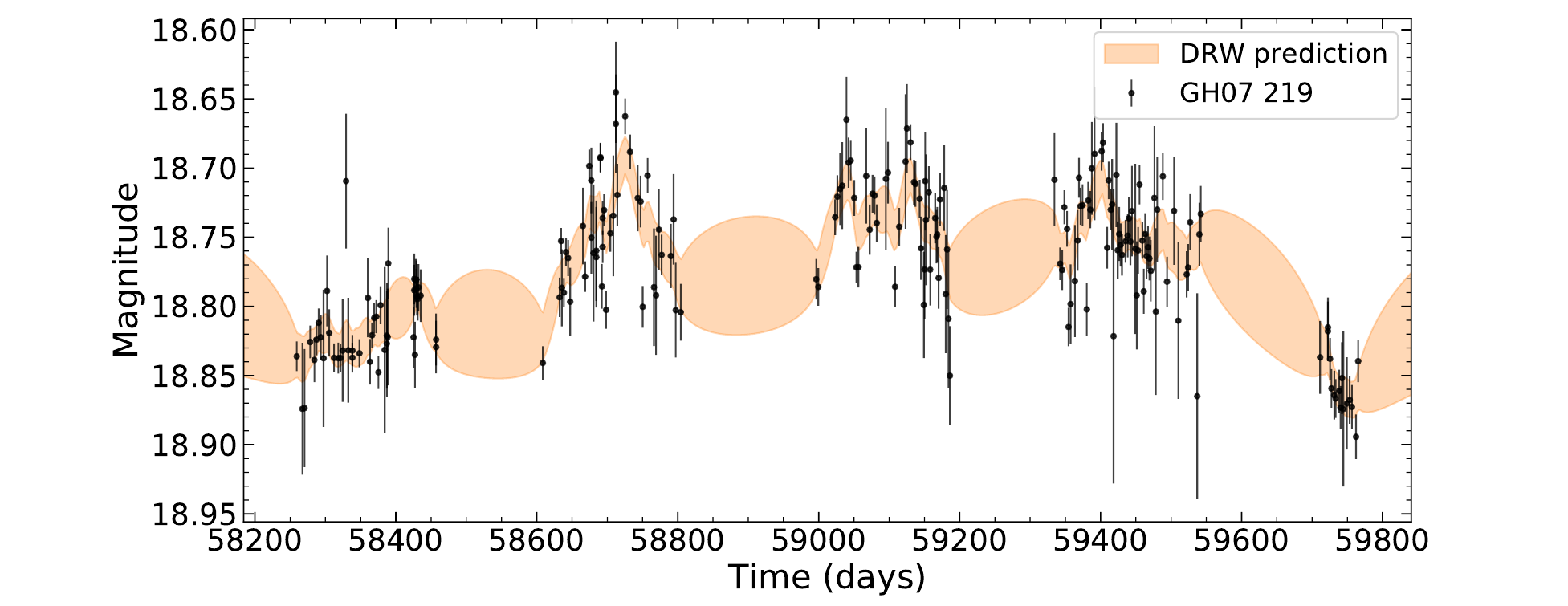}
\includegraphics[width=0.48\textwidth]{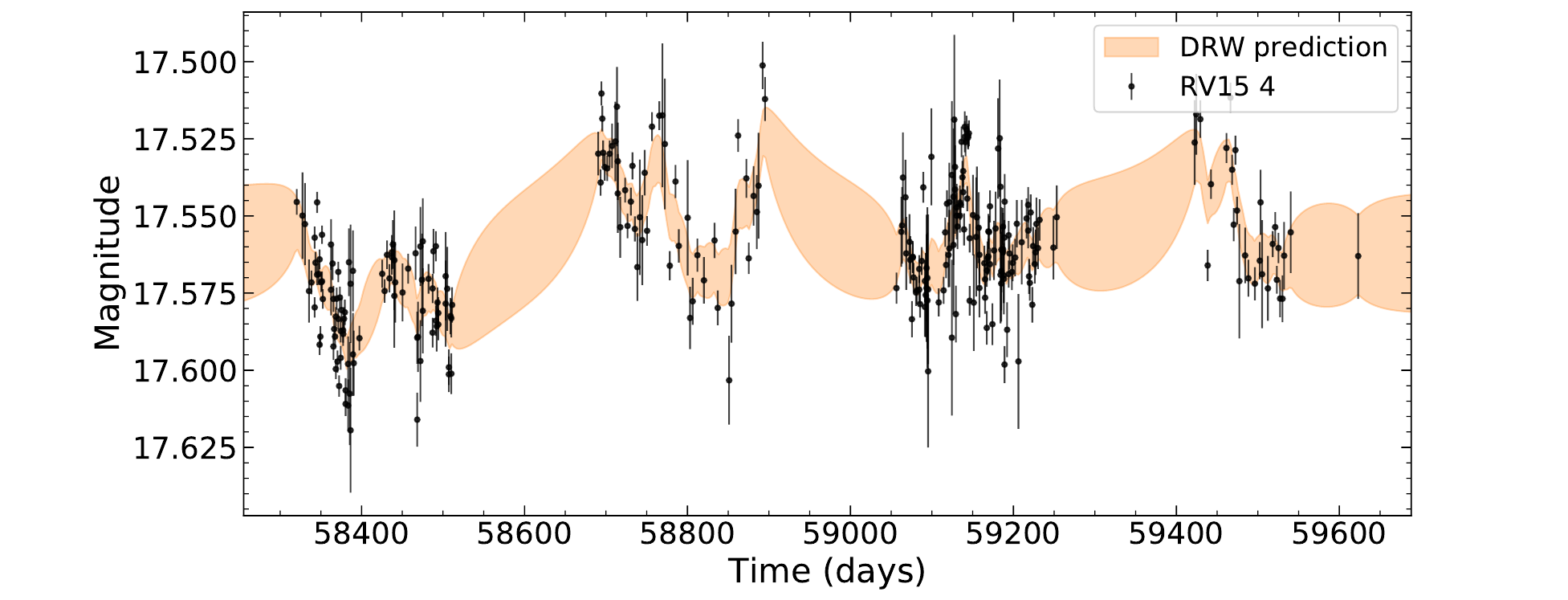}
\includegraphics[width=0.48\textwidth]{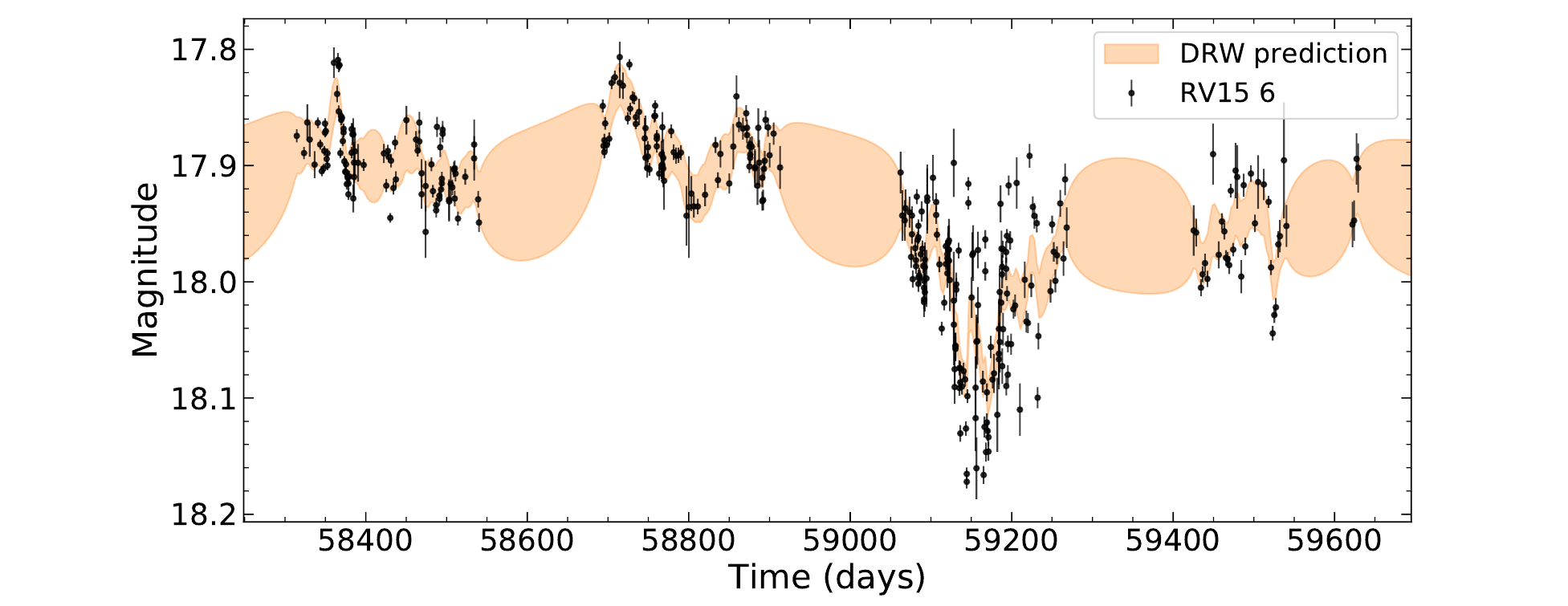}
\includegraphics[width=0.48\textwidth]{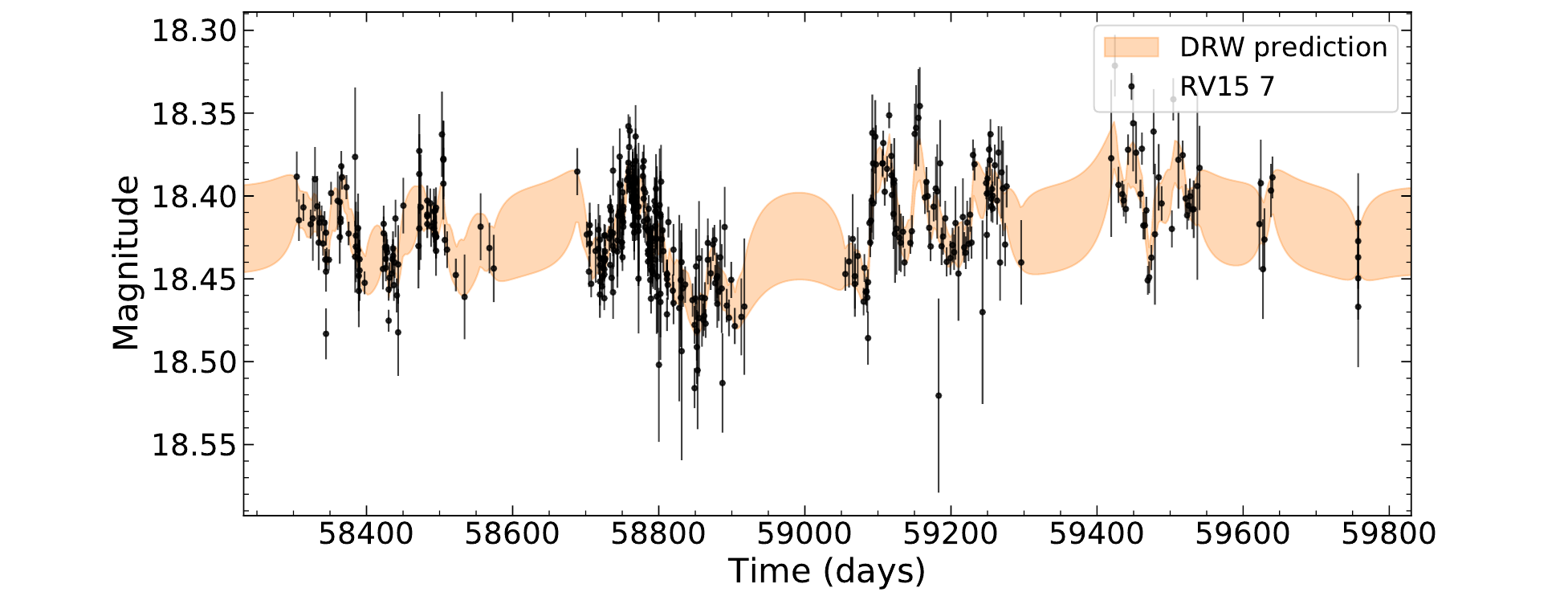}
\caption{Continuation of Figure~\ref{fig:lc_appx}.}
\end{figure*}

\begin{figure*}
\includegraphics[width=0.48\textwidth]{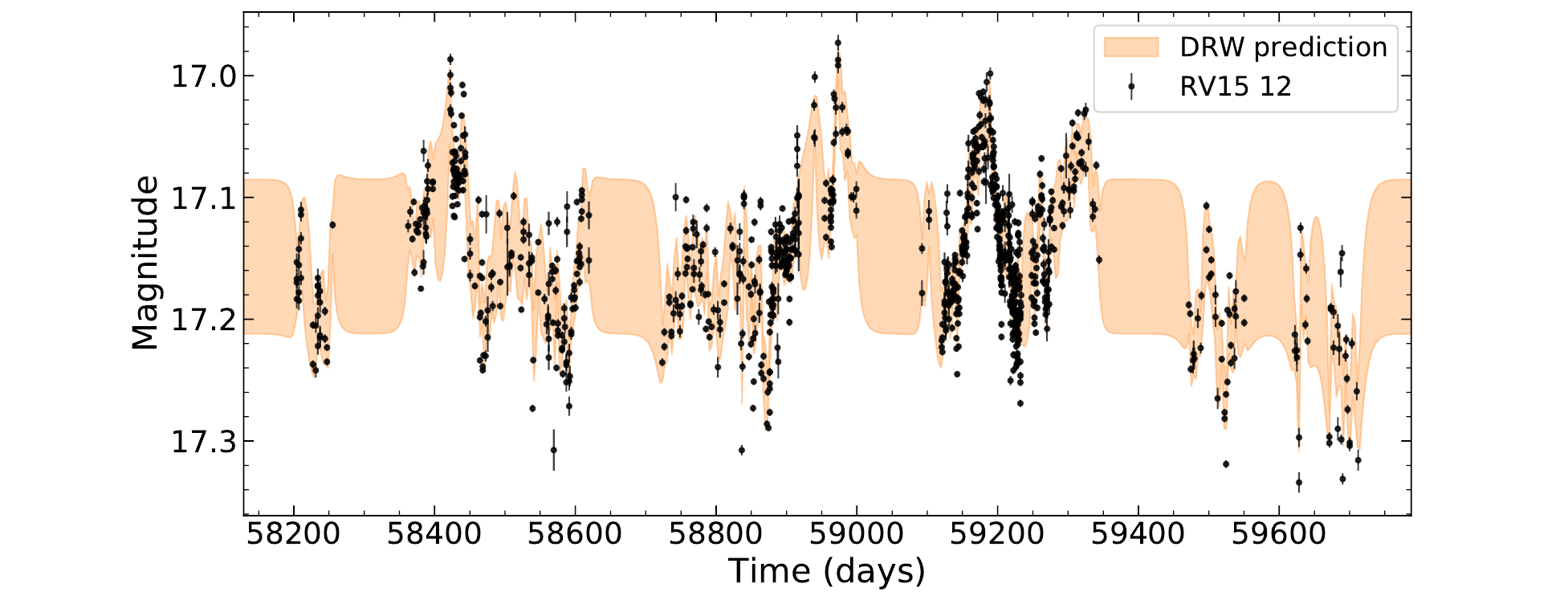}
\includegraphics[width=0.48\textwidth]{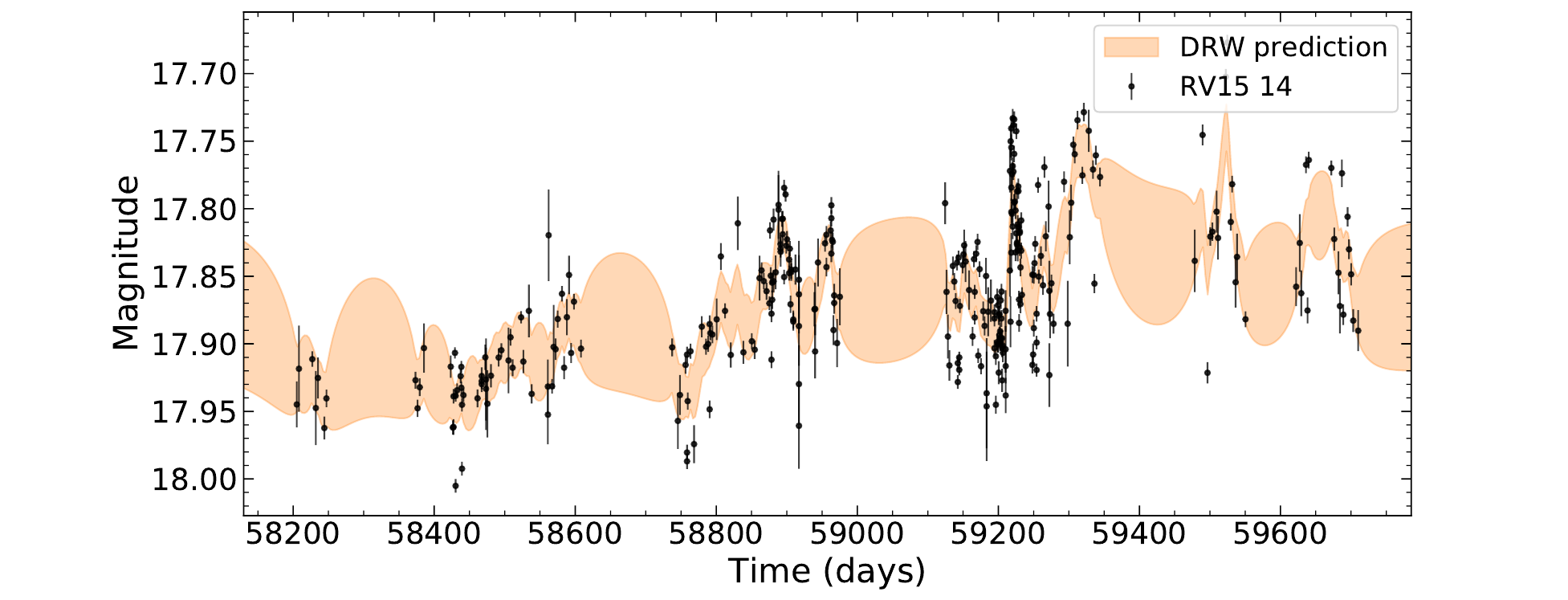}
\includegraphics[width=0.48\textwidth]{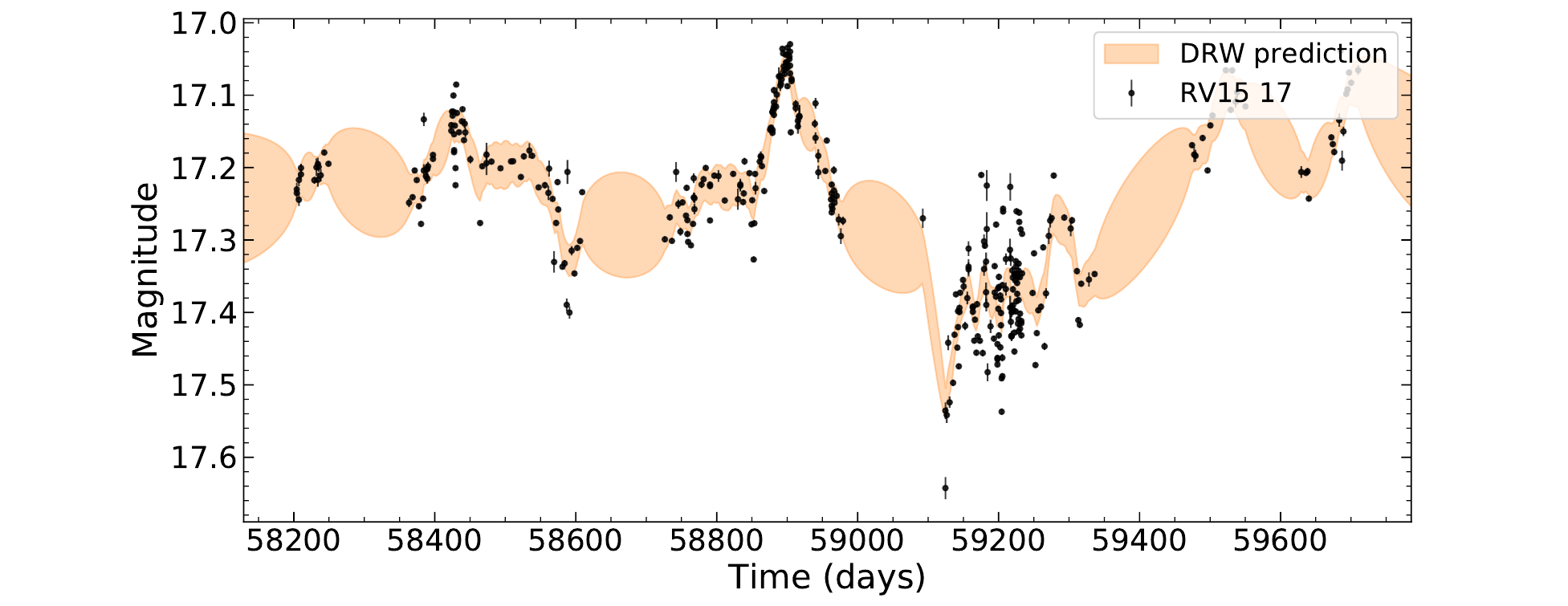}
\includegraphics[width=0.48\textwidth]{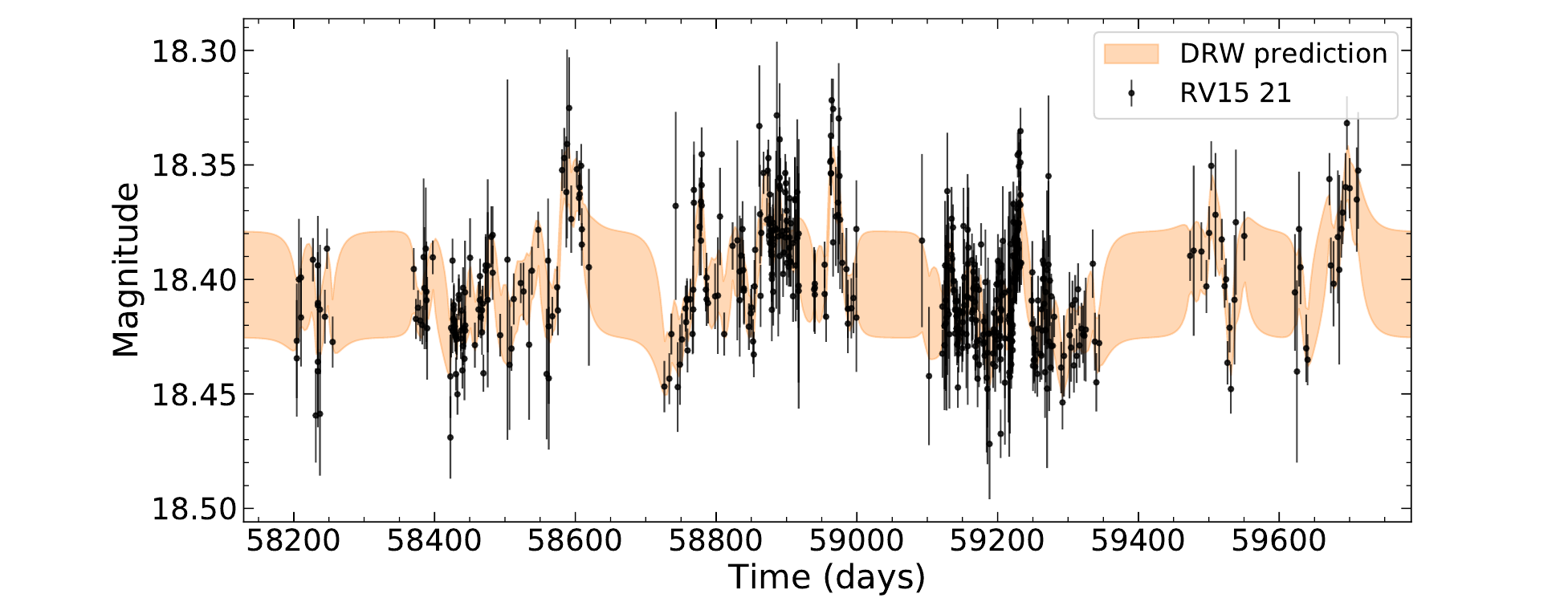}
\includegraphics[width=0.48\textwidth]{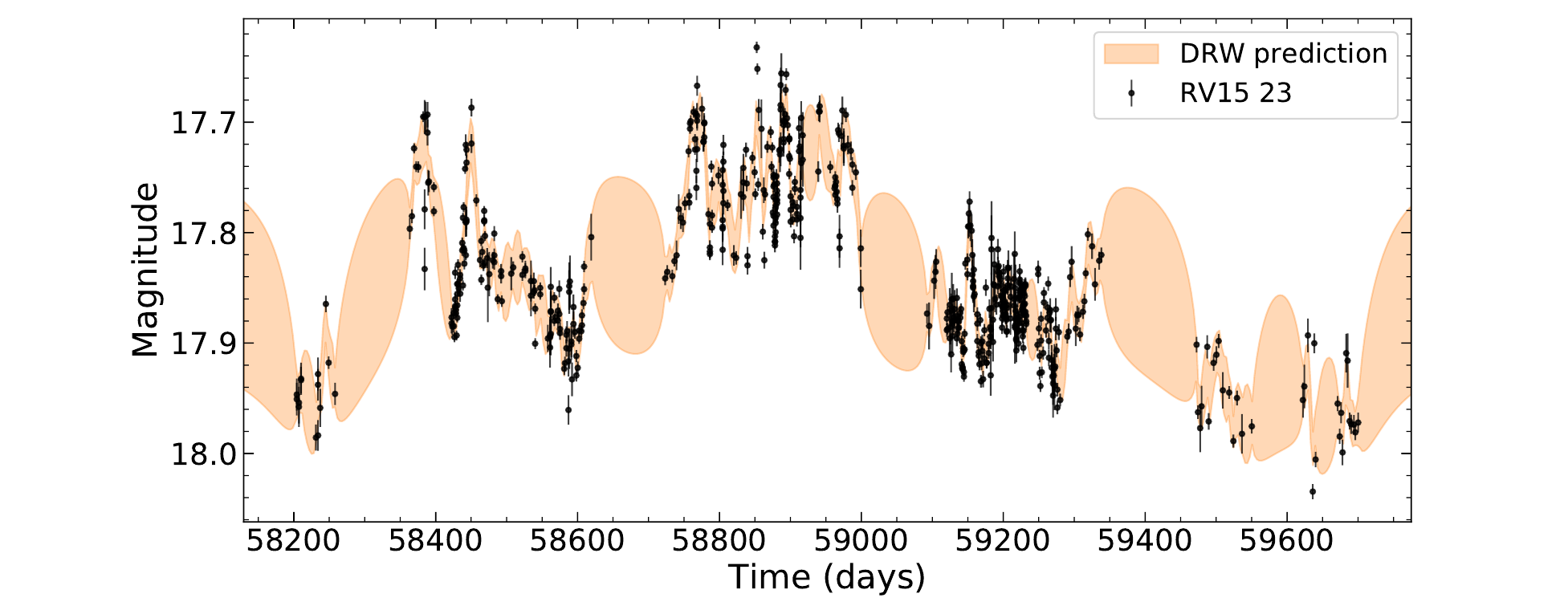}
\includegraphics[width=0.48\textwidth]{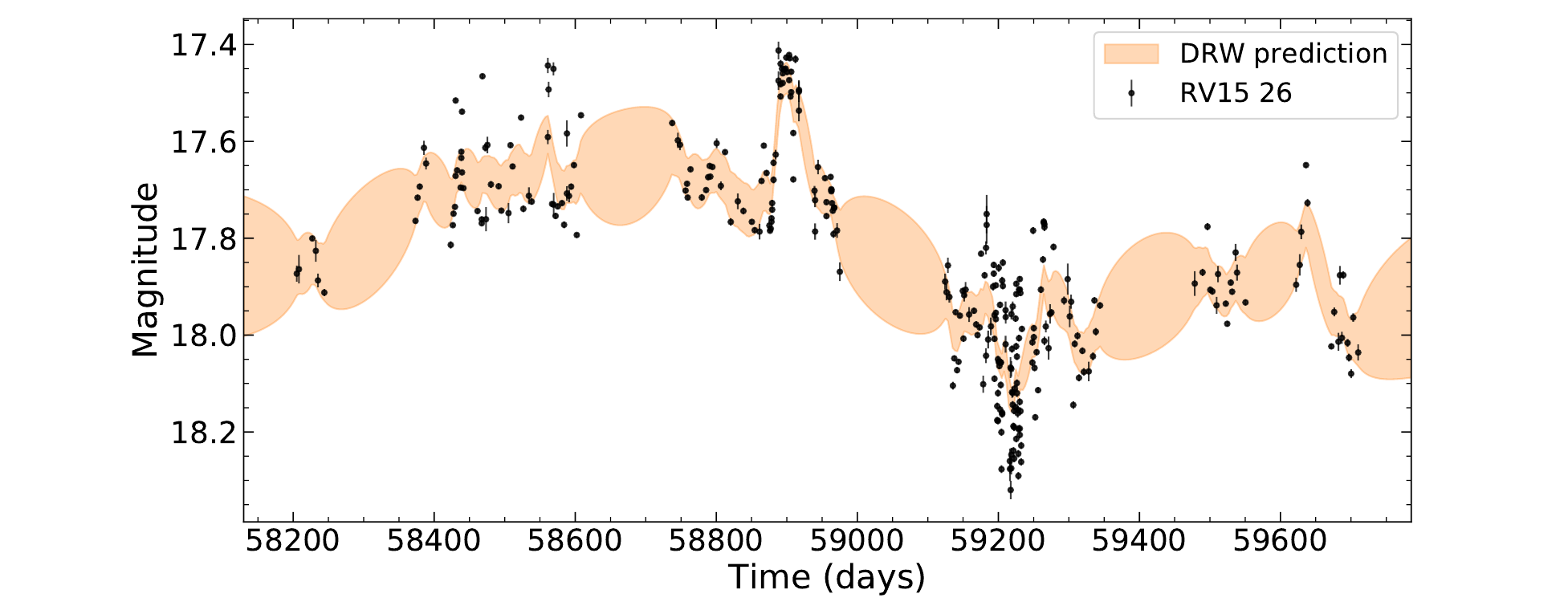}
\includegraphics[width=0.48\textwidth]{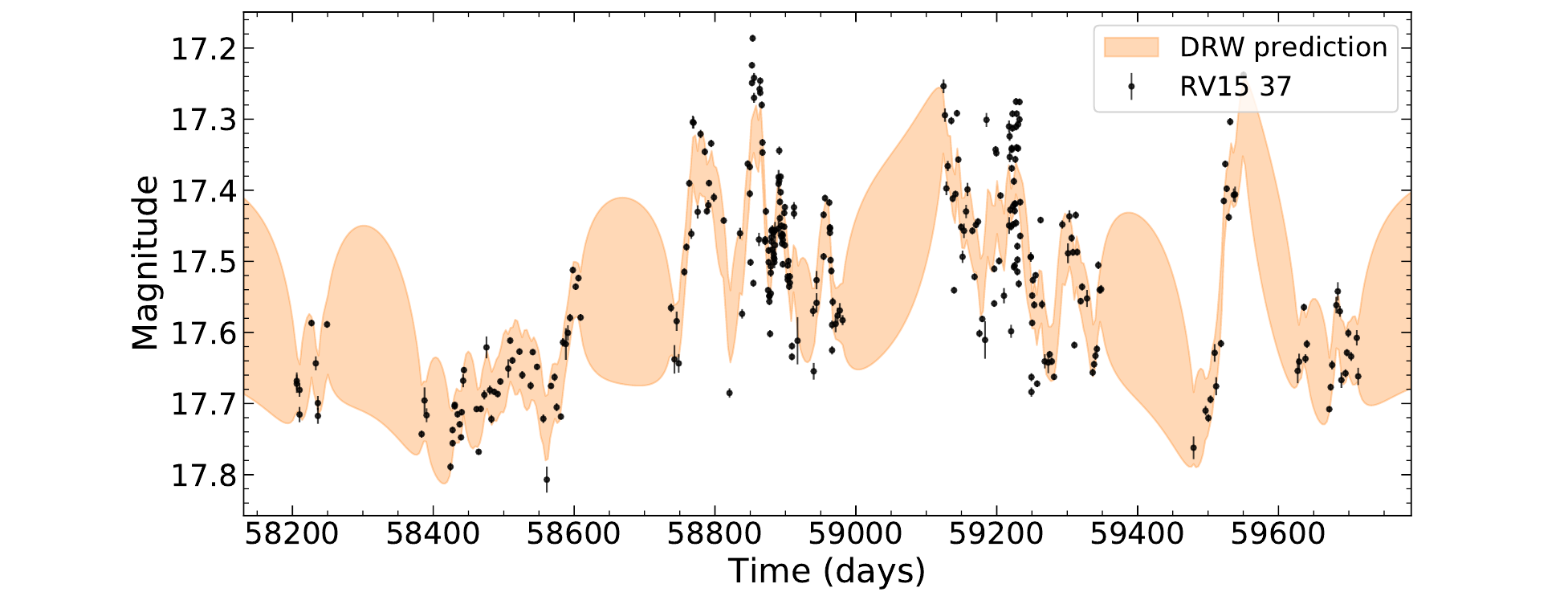}
\includegraphics[width=0.48\textwidth]{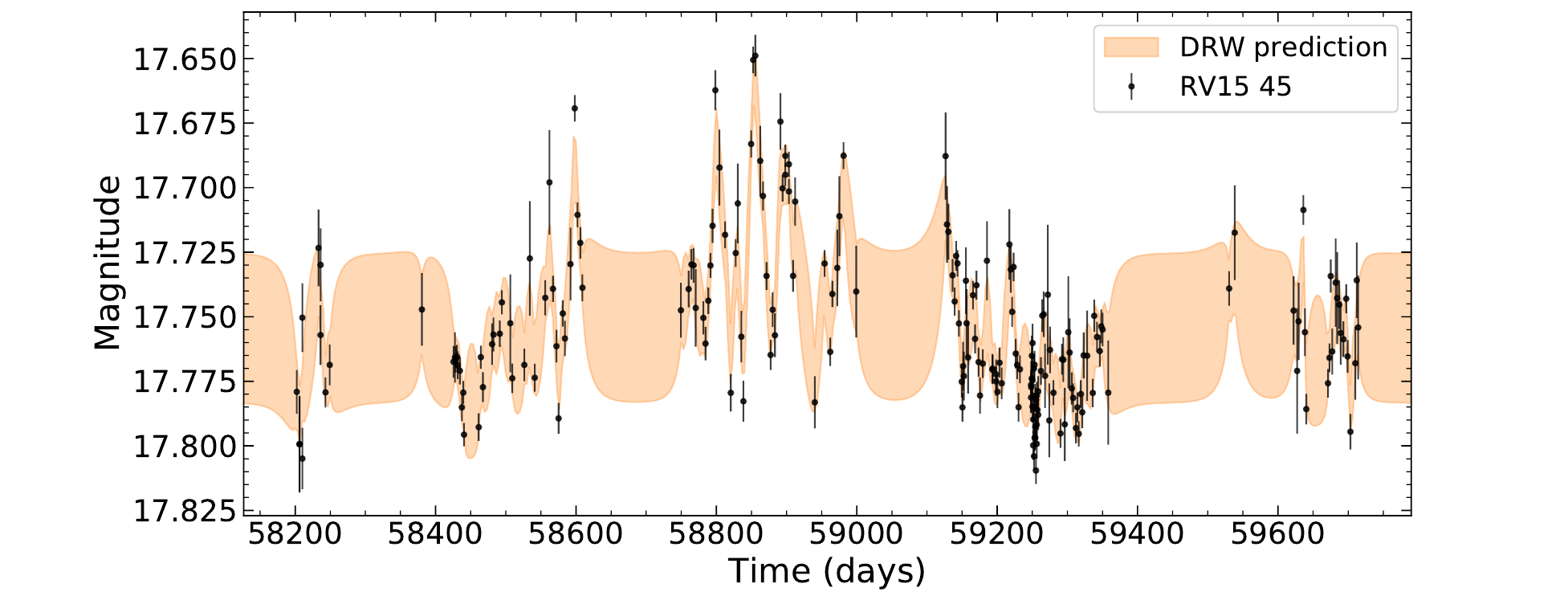}
\includegraphics[width=0.48\textwidth]{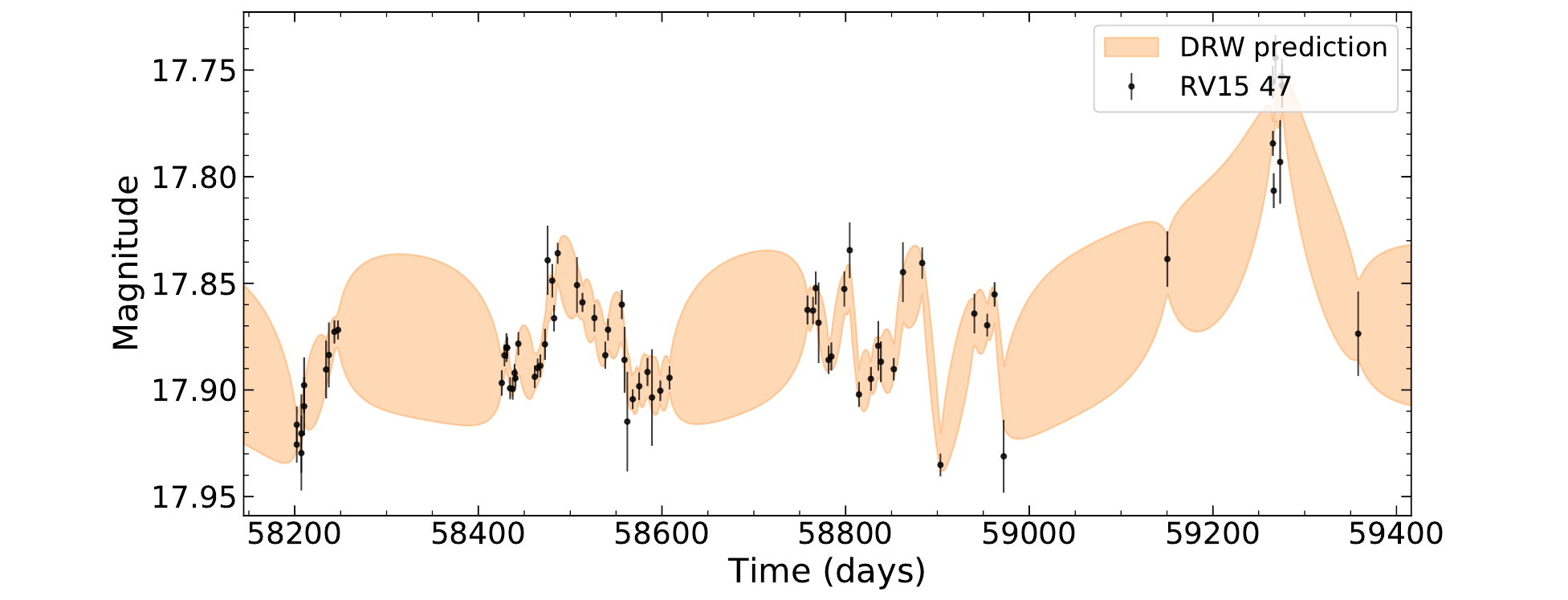}
\includegraphics[width=0.48\textwidth]{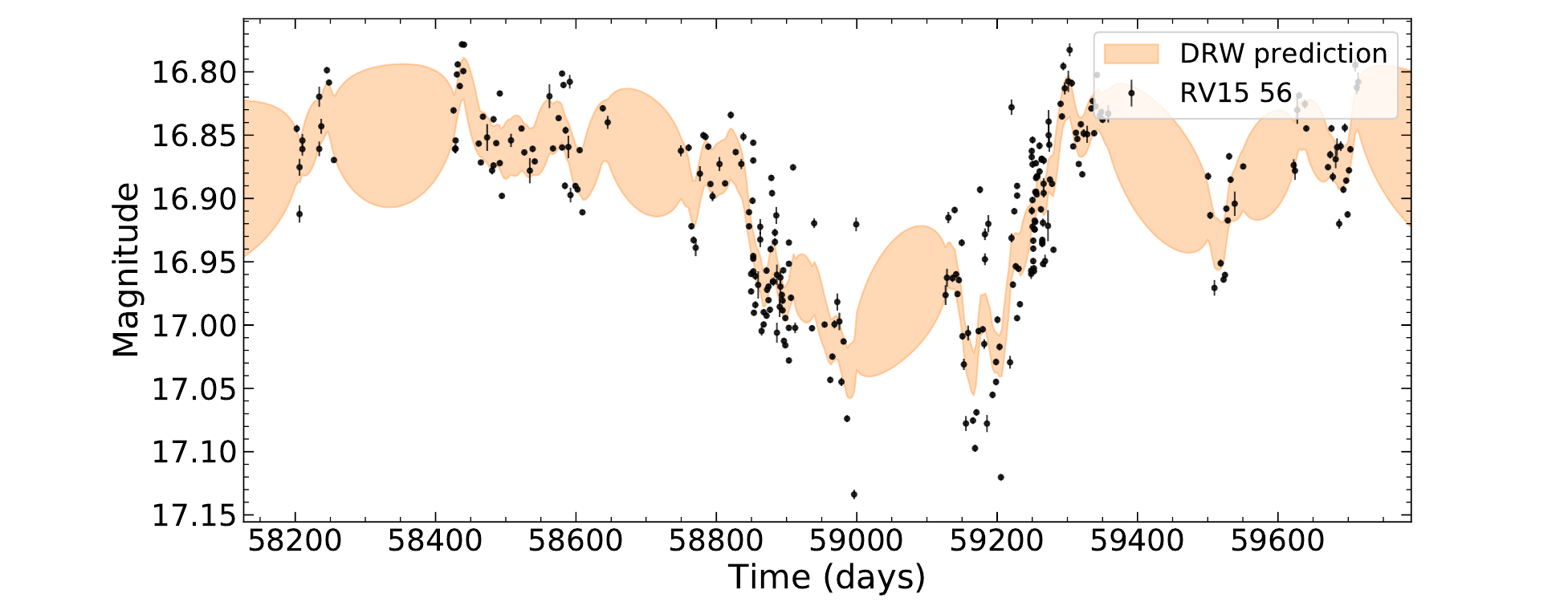}
\includegraphics[width=0.48\textwidth]{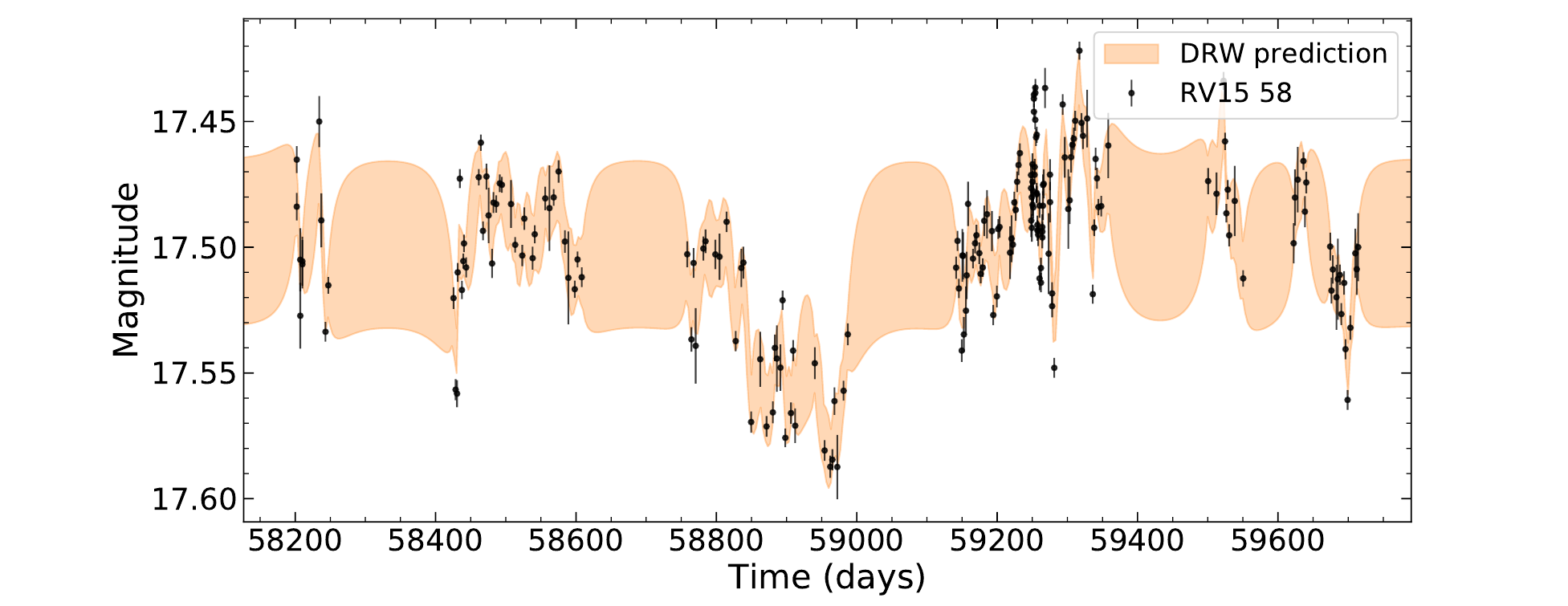}
\includegraphics[width=0.48\textwidth]{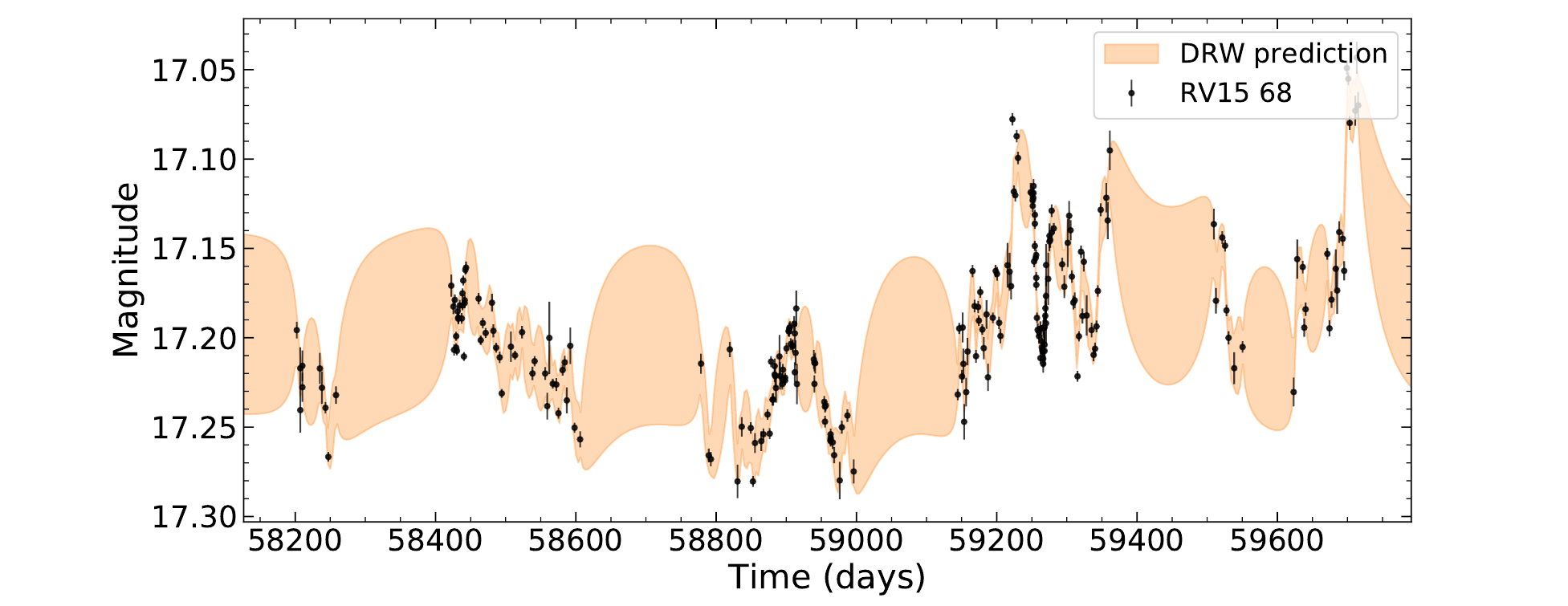}
\includegraphics[width=0.48\textwidth]{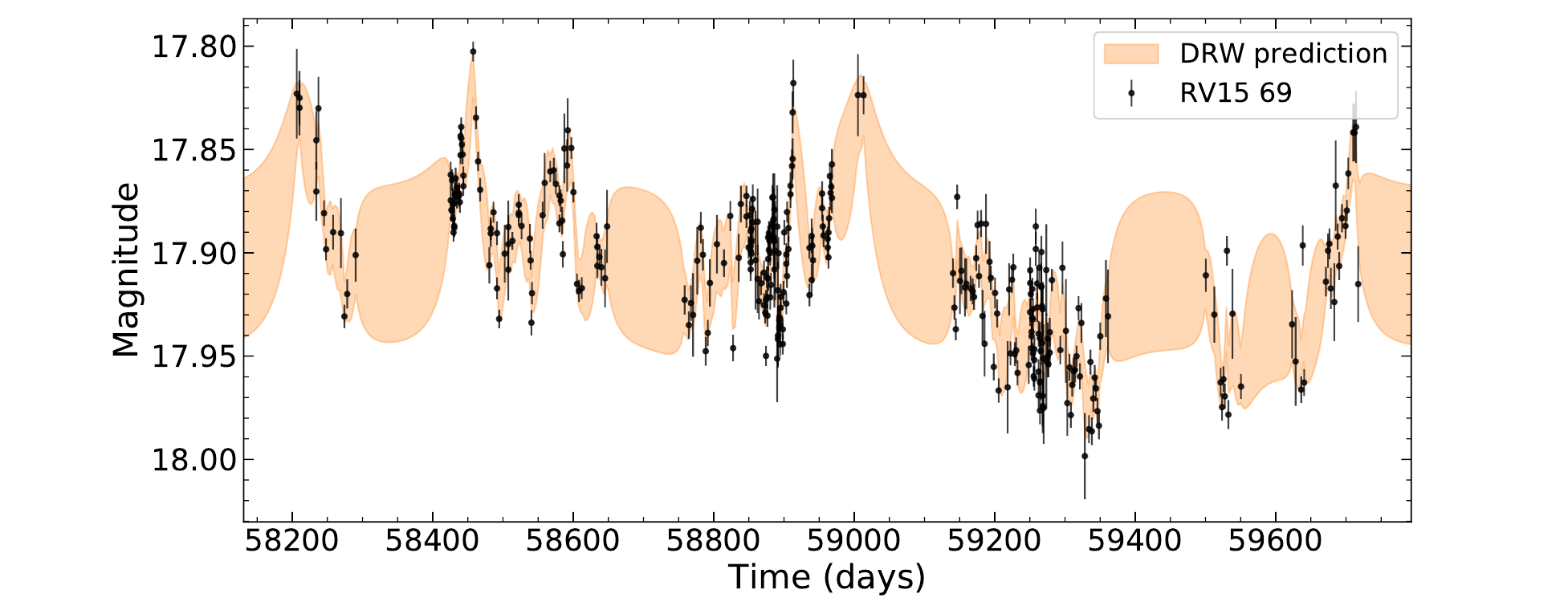}
\includegraphics[width=0.48\textwidth]{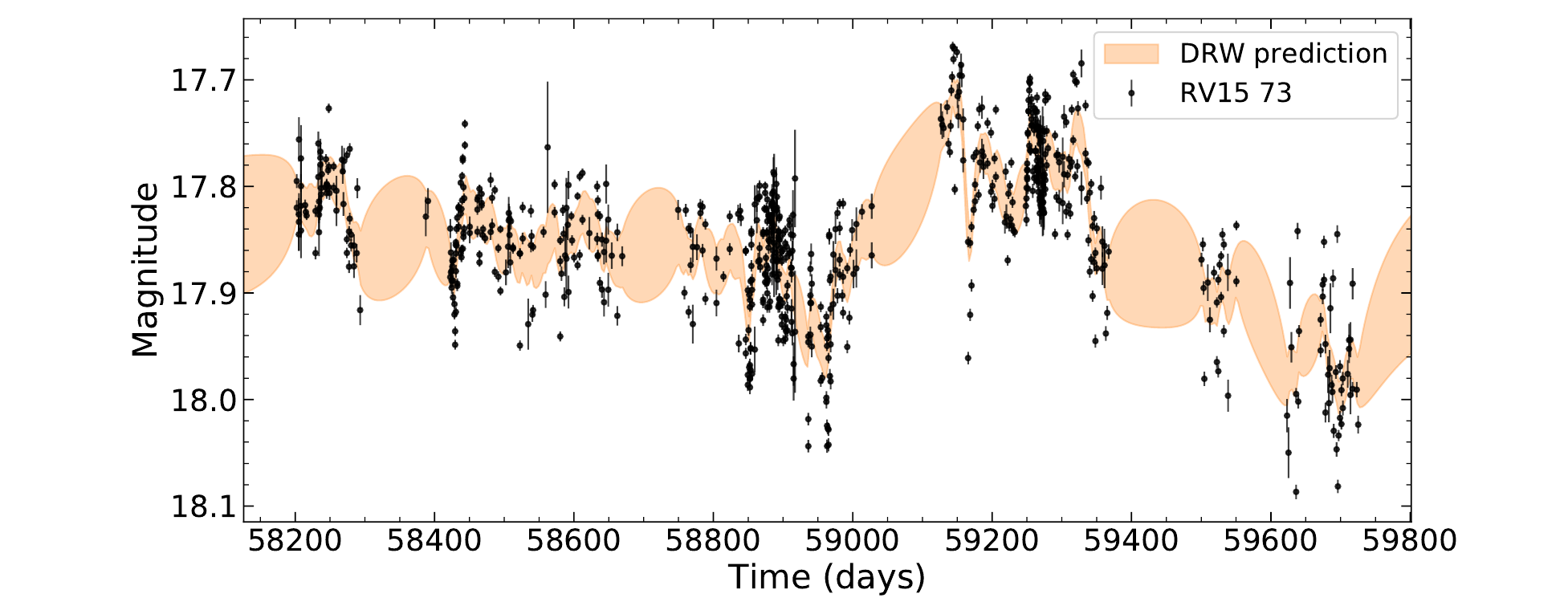}
\caption{Continuation of Figure~\ref{fig:lc_appx}.}
\end{figure*}

\begin{figure*}
\includegraphics[width=0.48\textwidth]{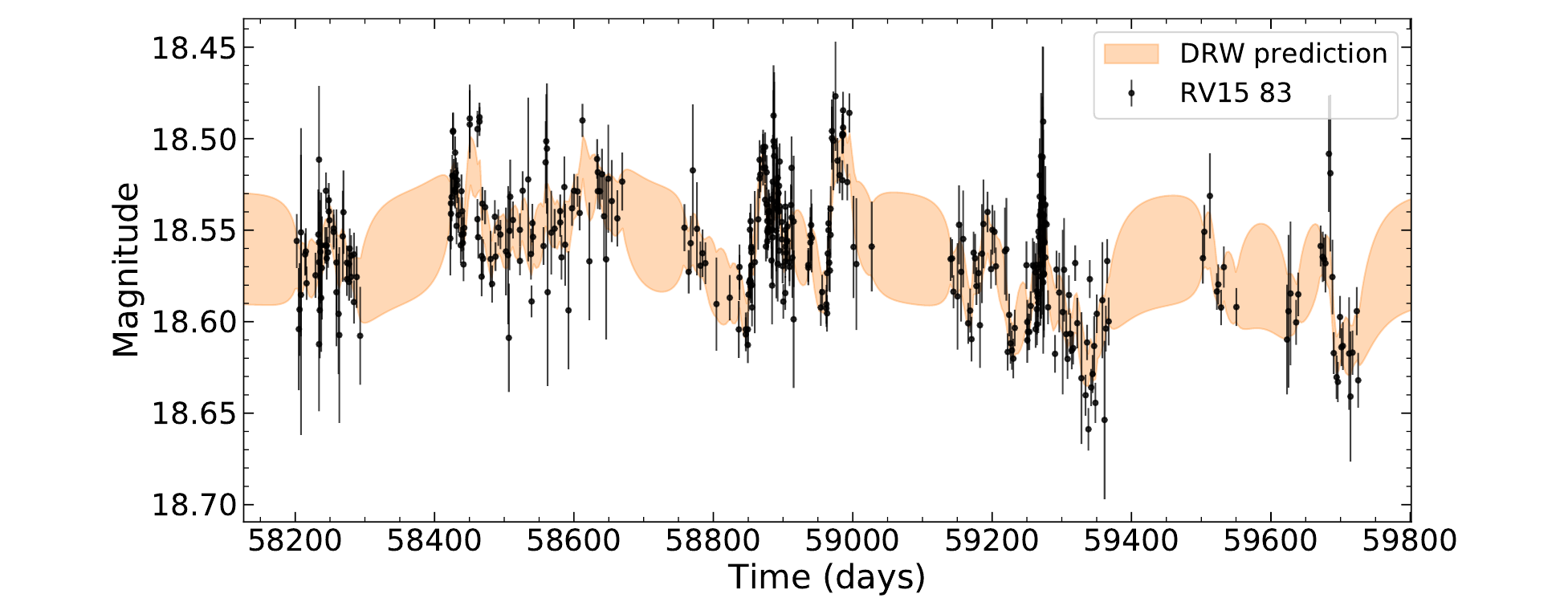}
\includegraphics[width=0.48\textwidth]{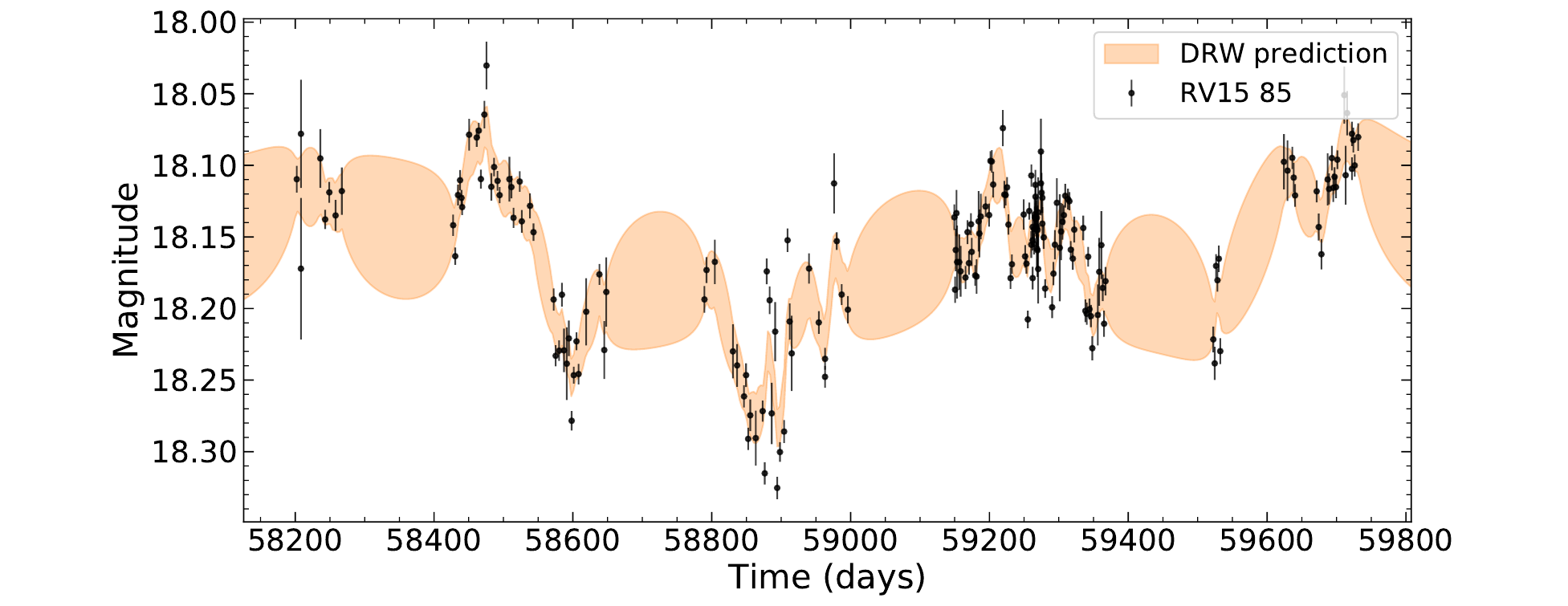}
\includegraphics[width=0.48\textwidth]{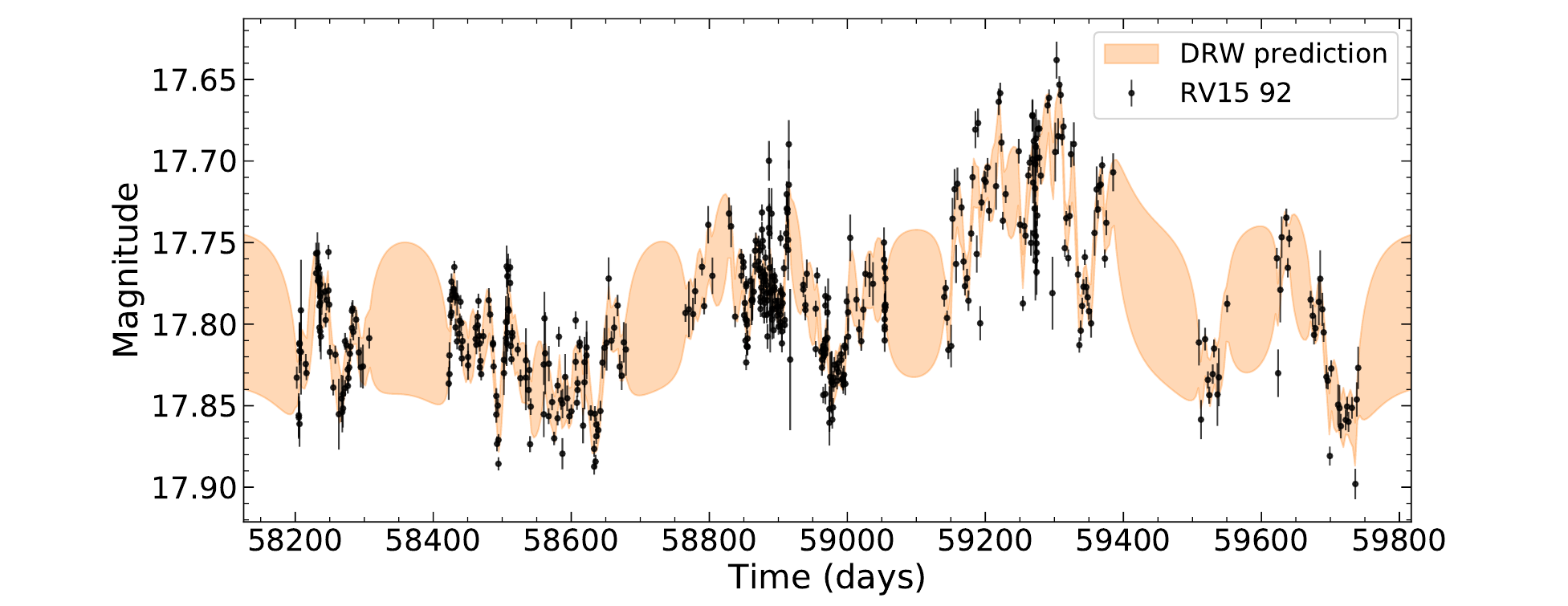}
\includegraphics[width=0.48\textwidth]{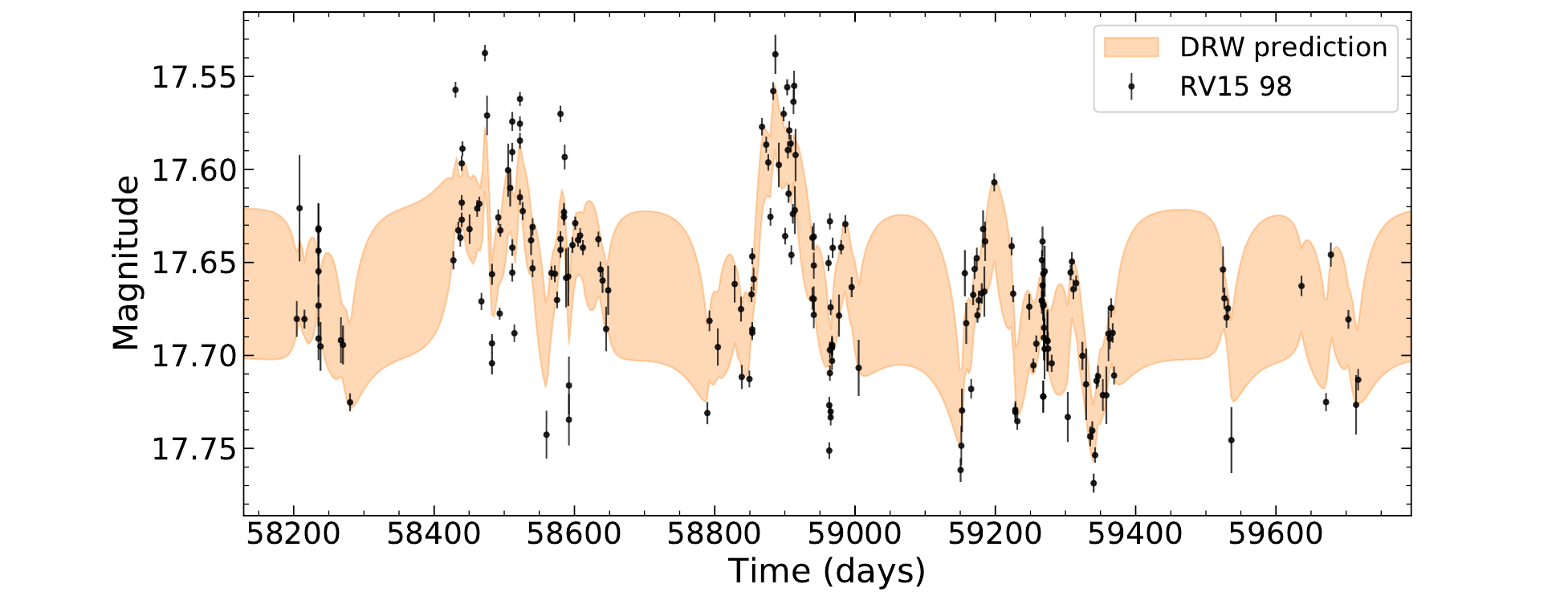}
\includegraphics[width=0.48\textwidth]{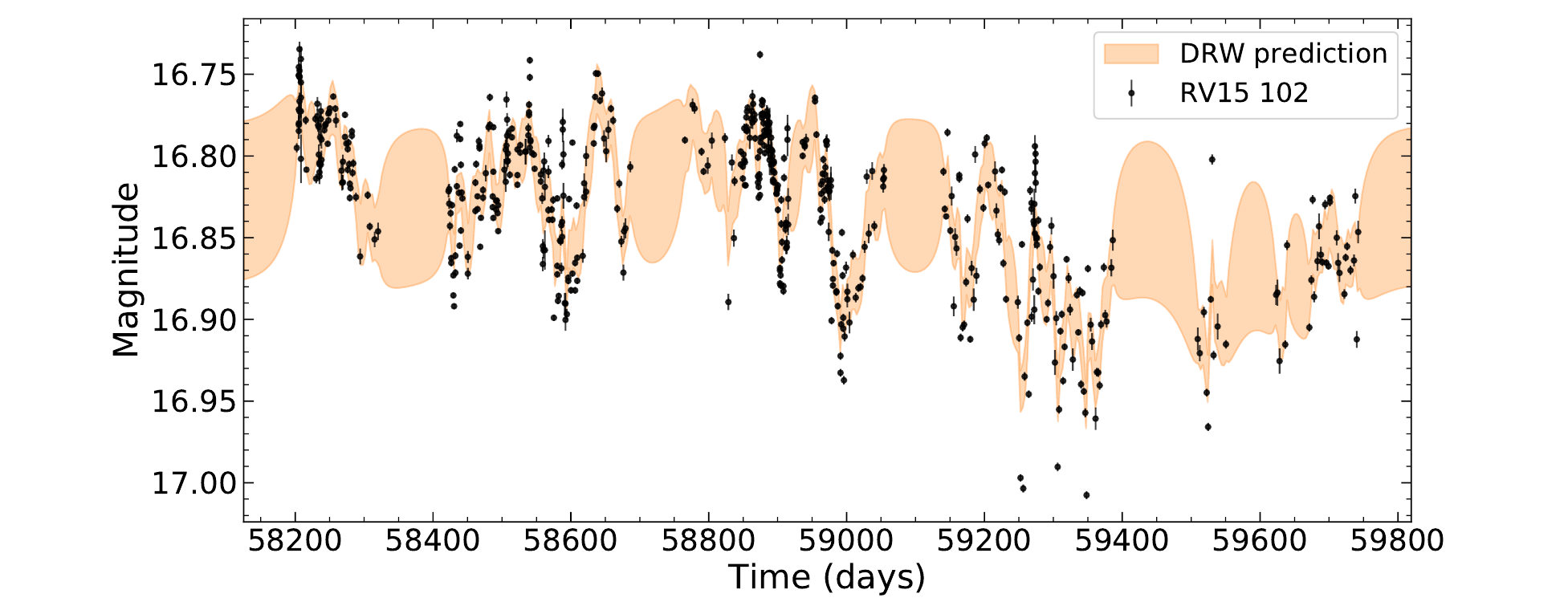}
\includegraphics[width=0.48\textwidth]{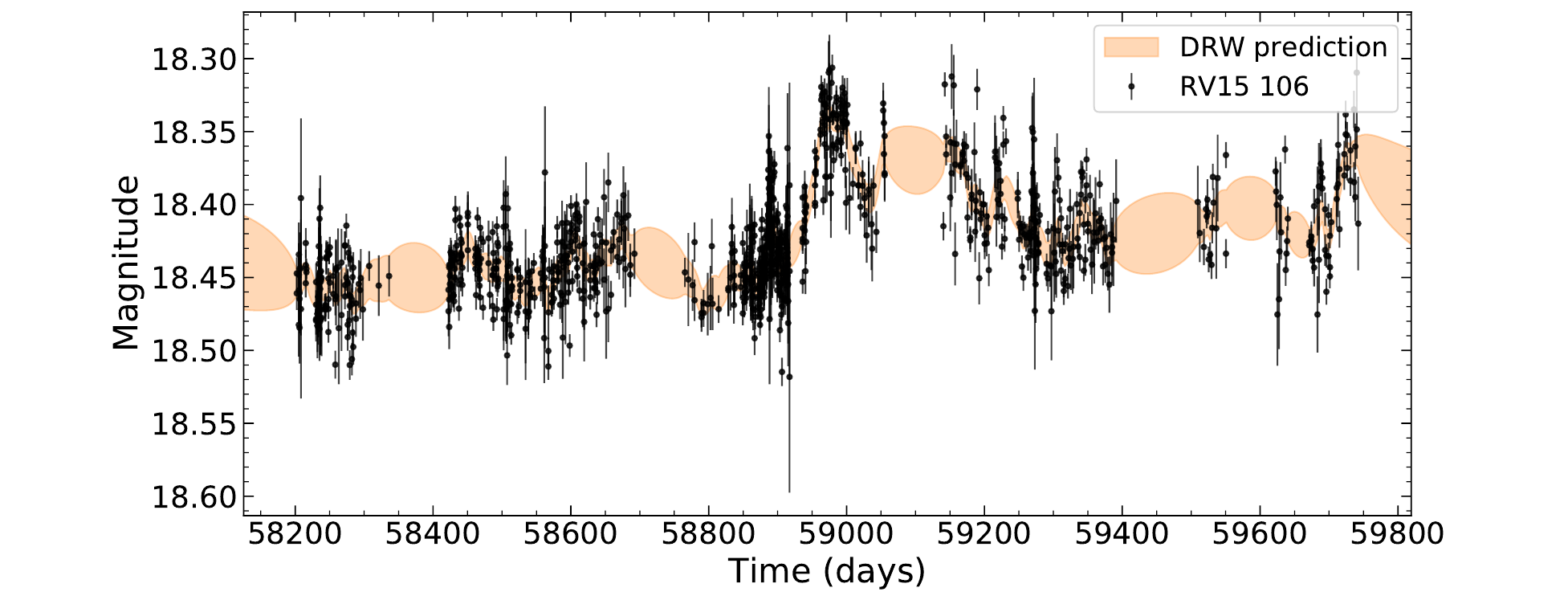}
\includegraphics[width=0.48\textwidth]{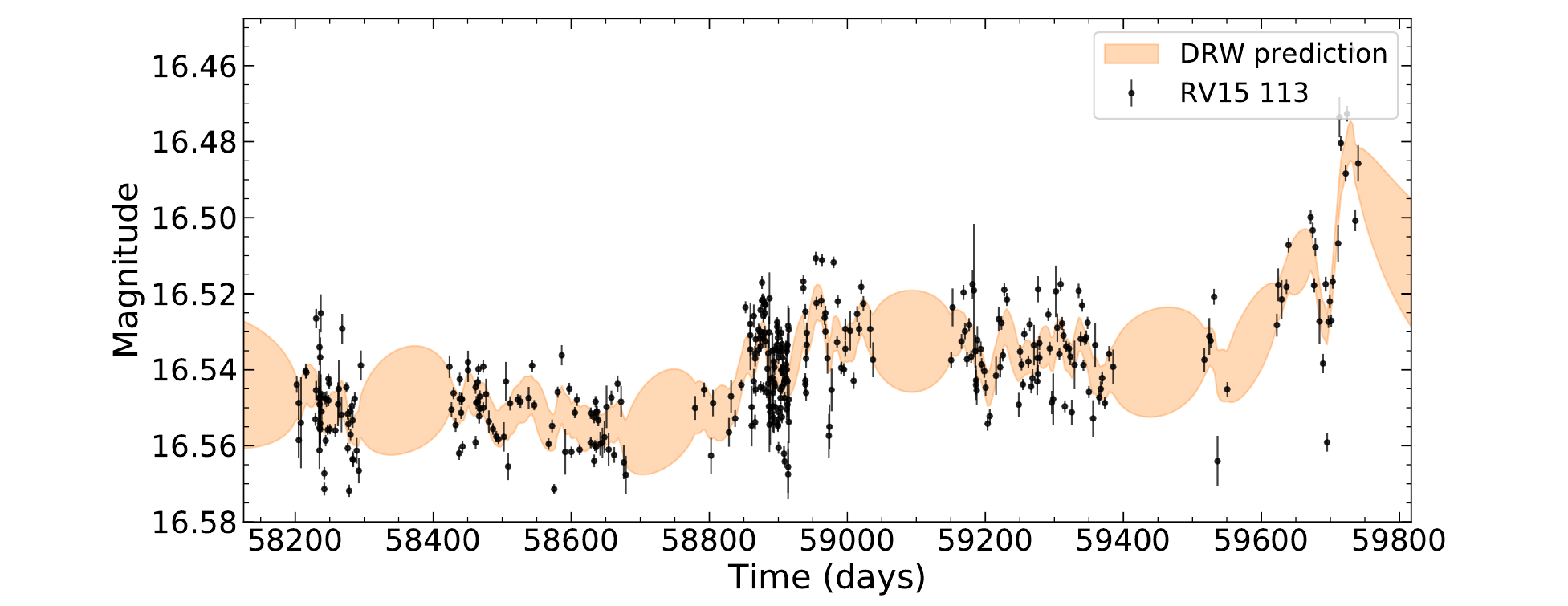}
\includegraphics[width=0.48\textwidth]{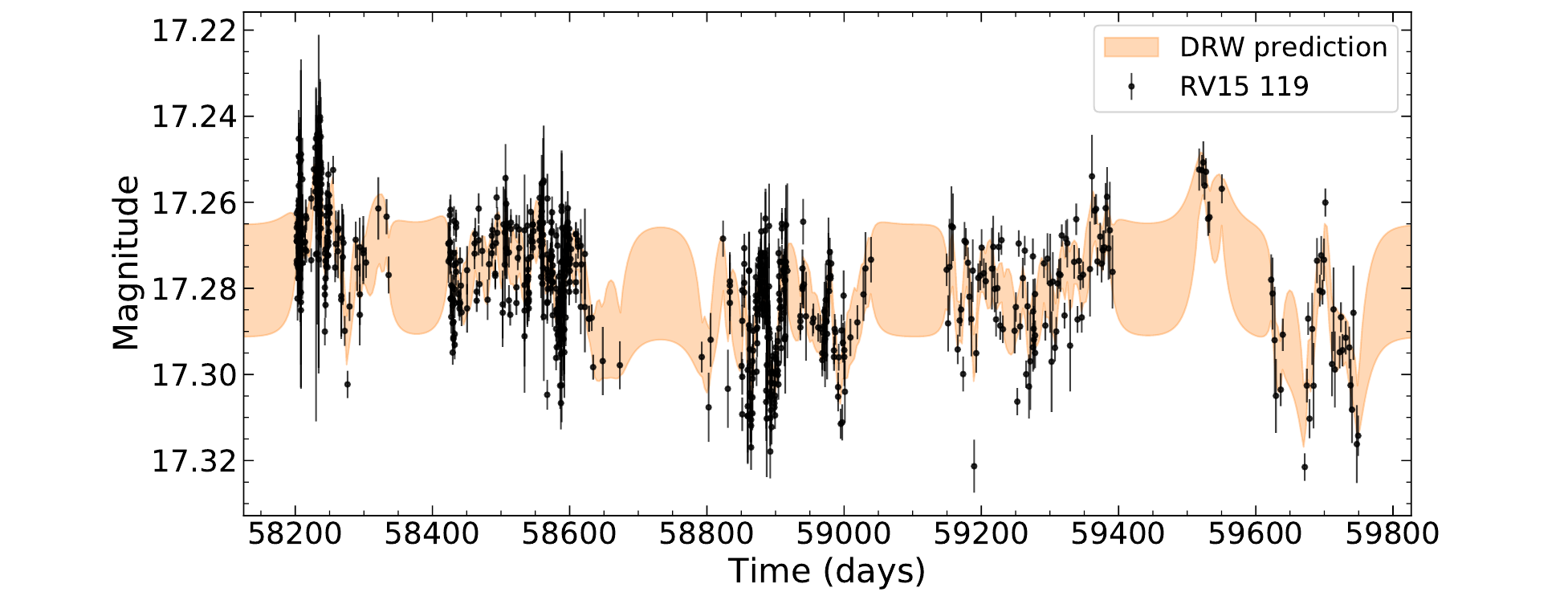}
\includegraphics[width=0.48\textwidth]{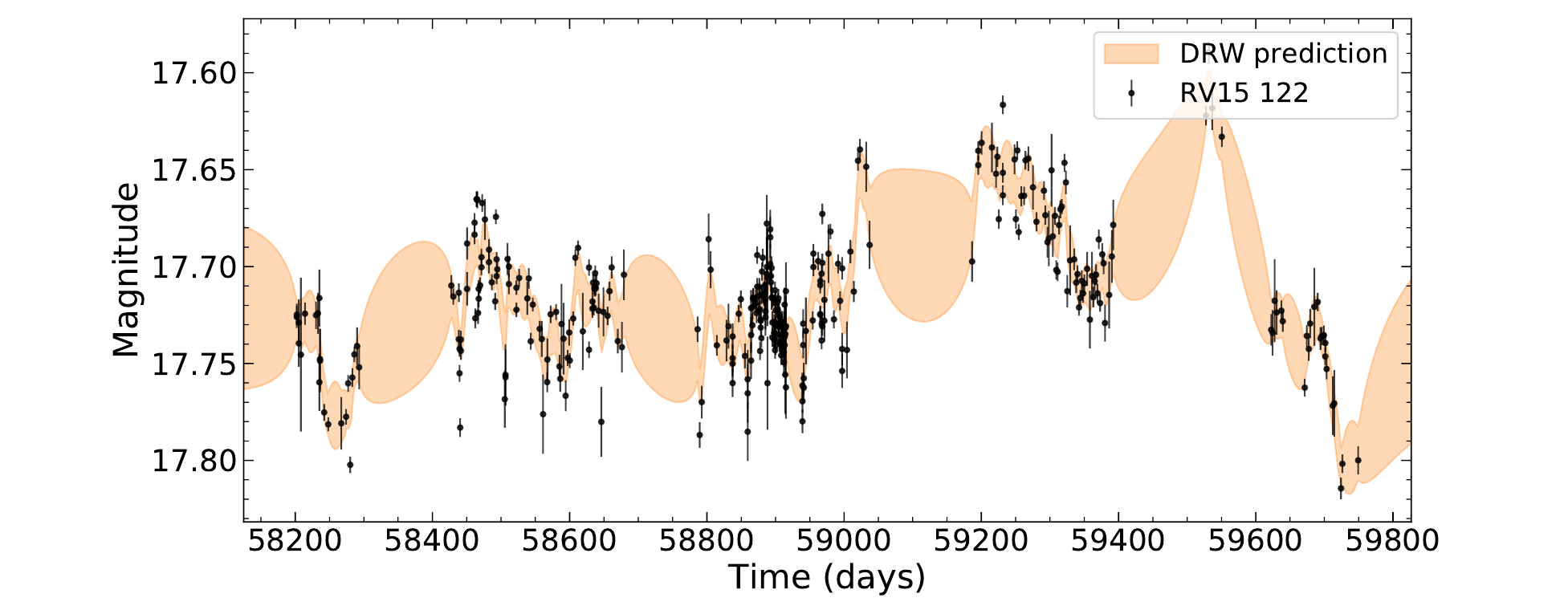}
\includegraphics[width=0.48\textwidth]{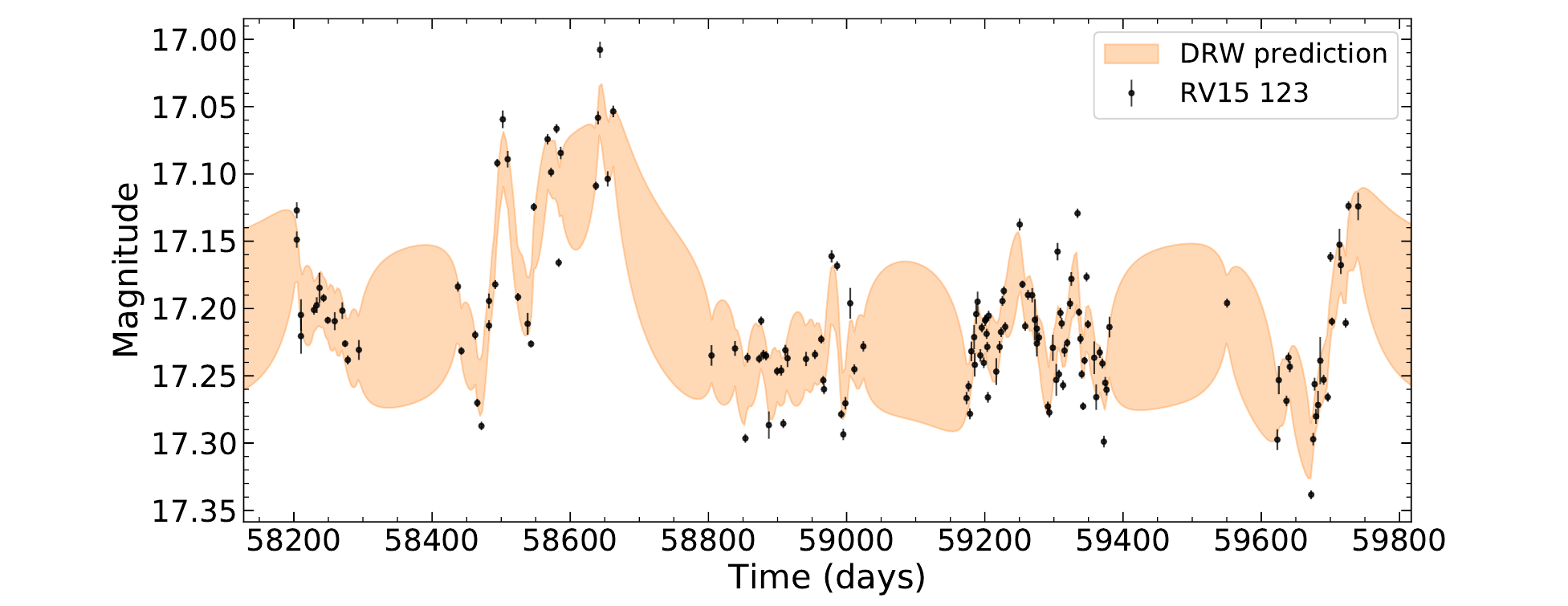}
\includegraphics[width=0.48\textwidth]{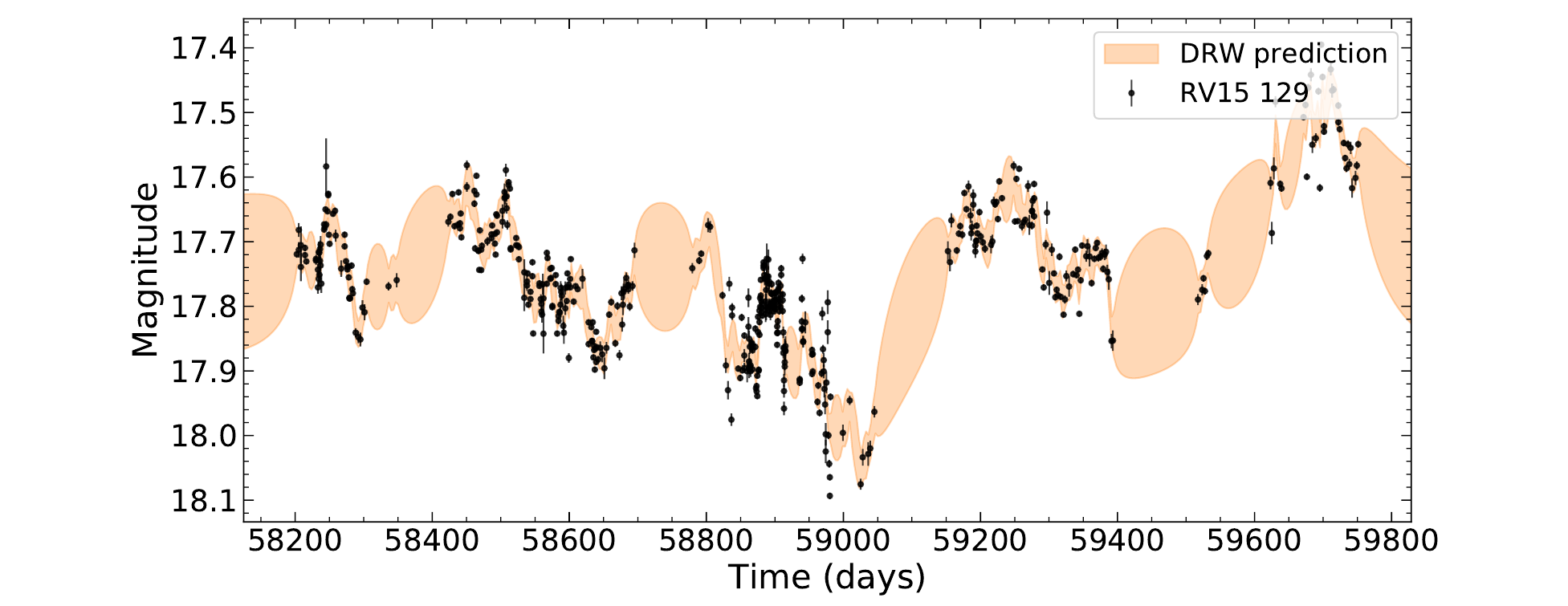}
\includegraphics[width=0.48\textwidth]{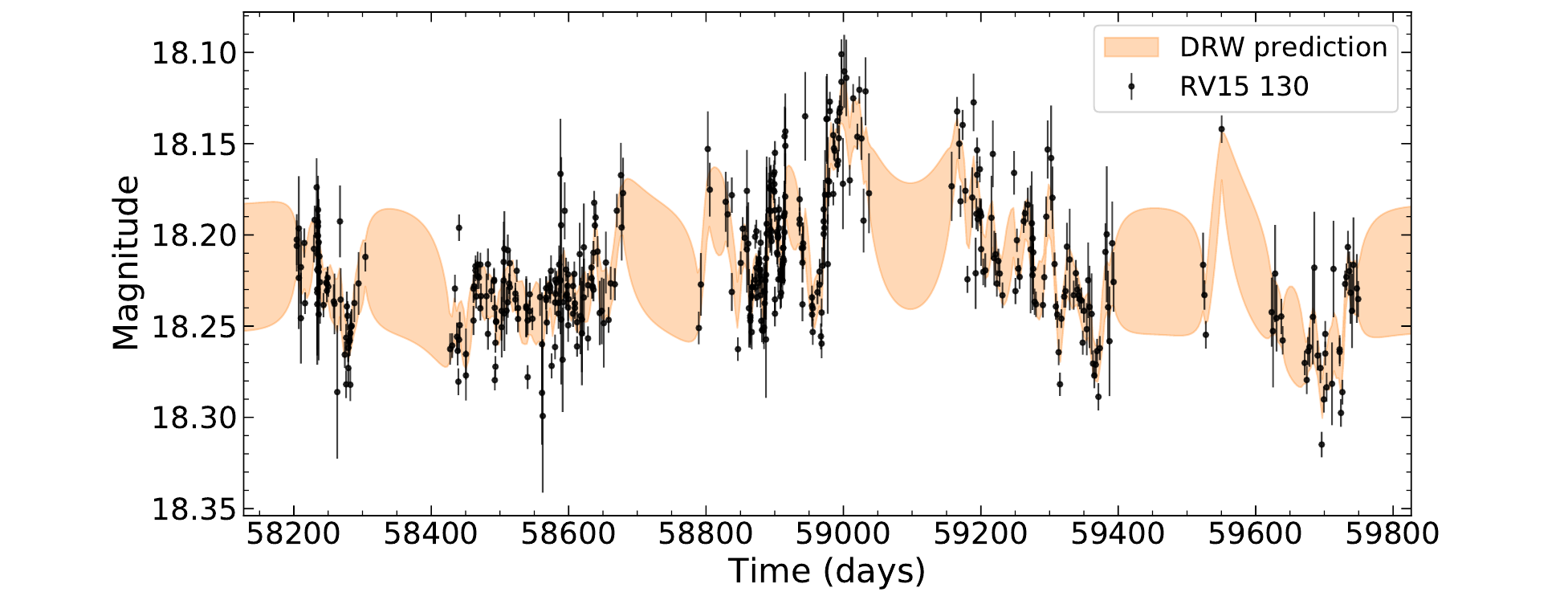}
\includegraphics[width=0.48\textwidth]{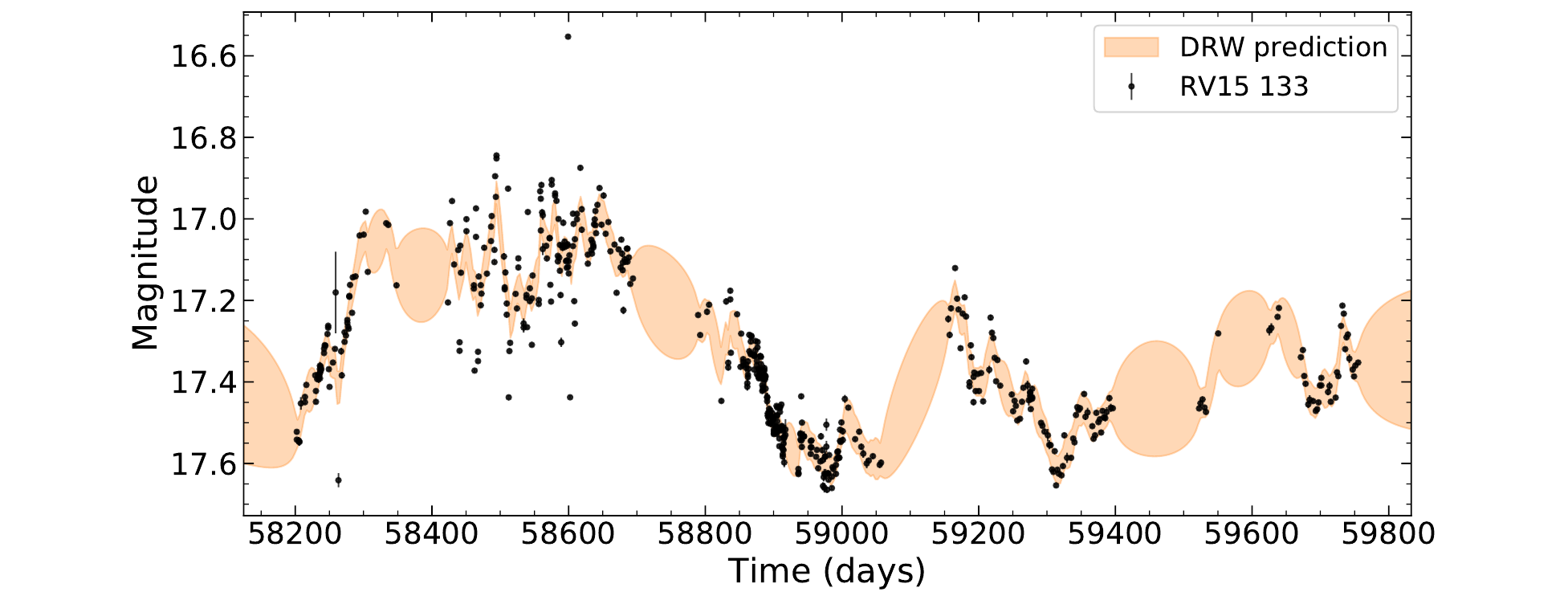}
\includegraphics[width=0.48\textwidth]{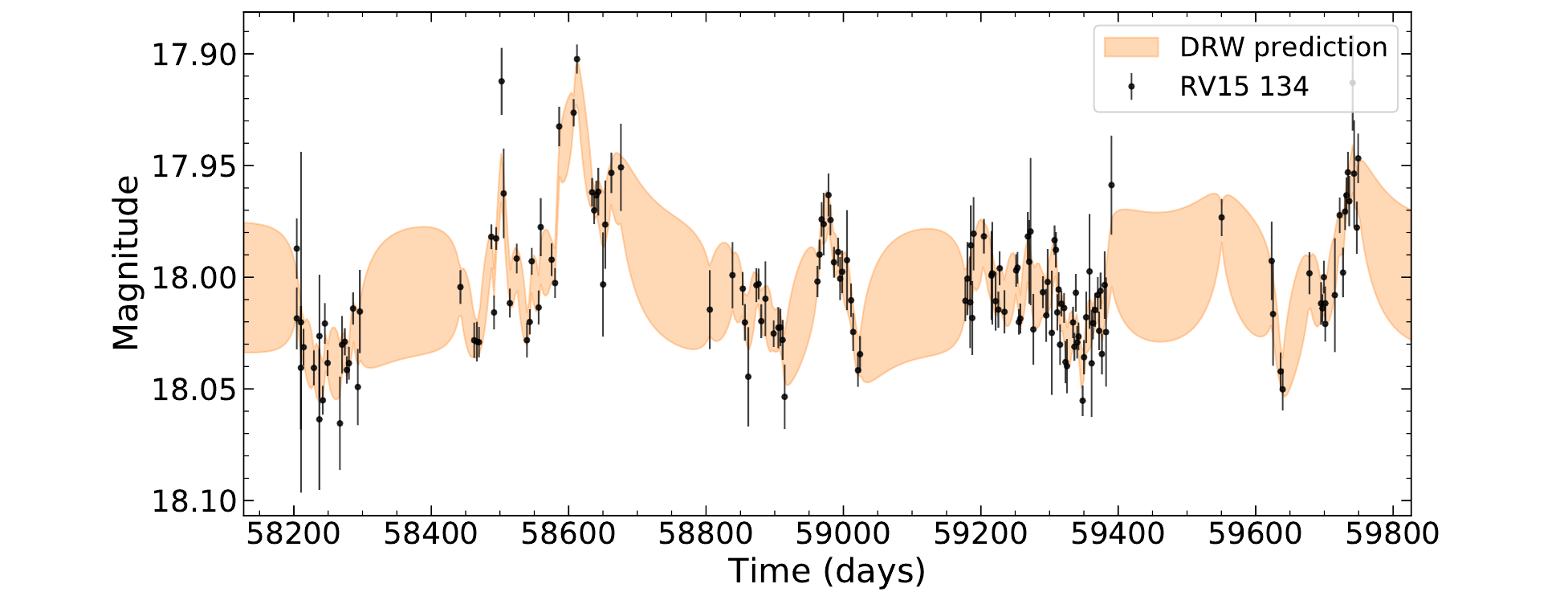}
\caption{Continuation of Figure~\ref{fig:lc_appx}.}
\end{figure*}

\begin{figure*}
\includegraphics[width=0.48\textwidth]{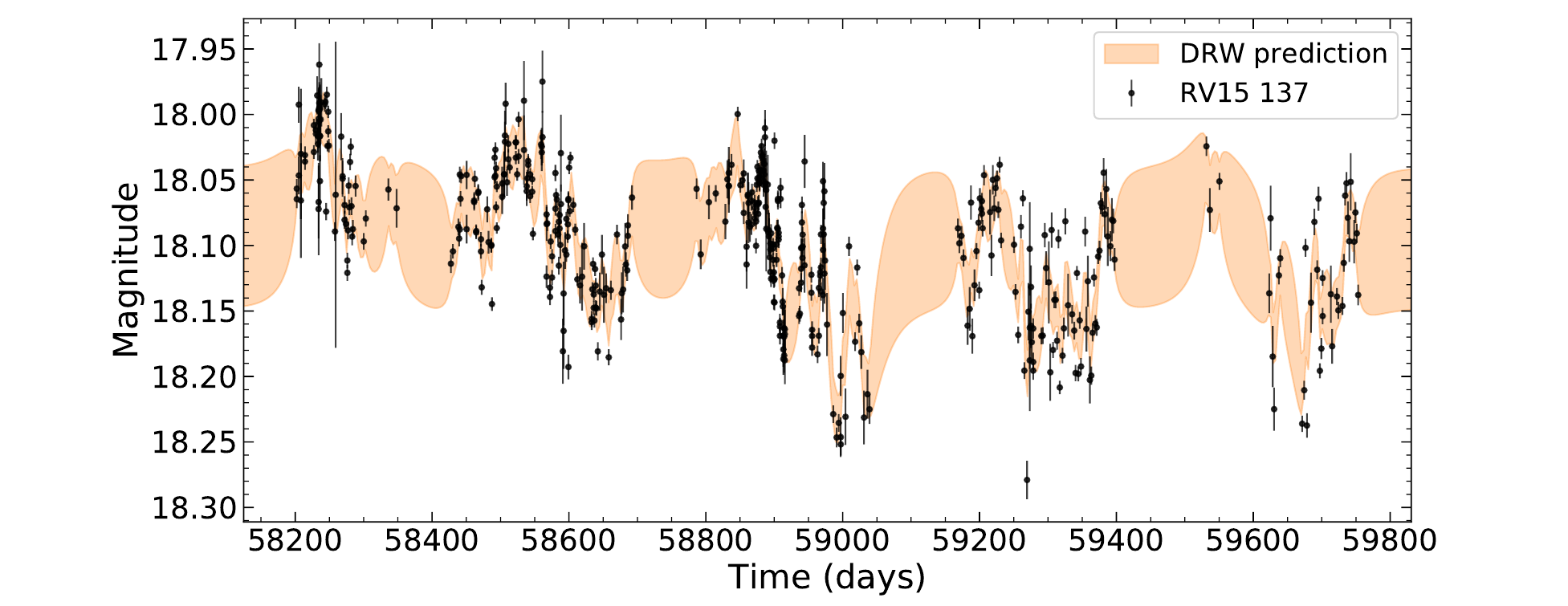}
\includegraphics[width=0.48\textwidth]{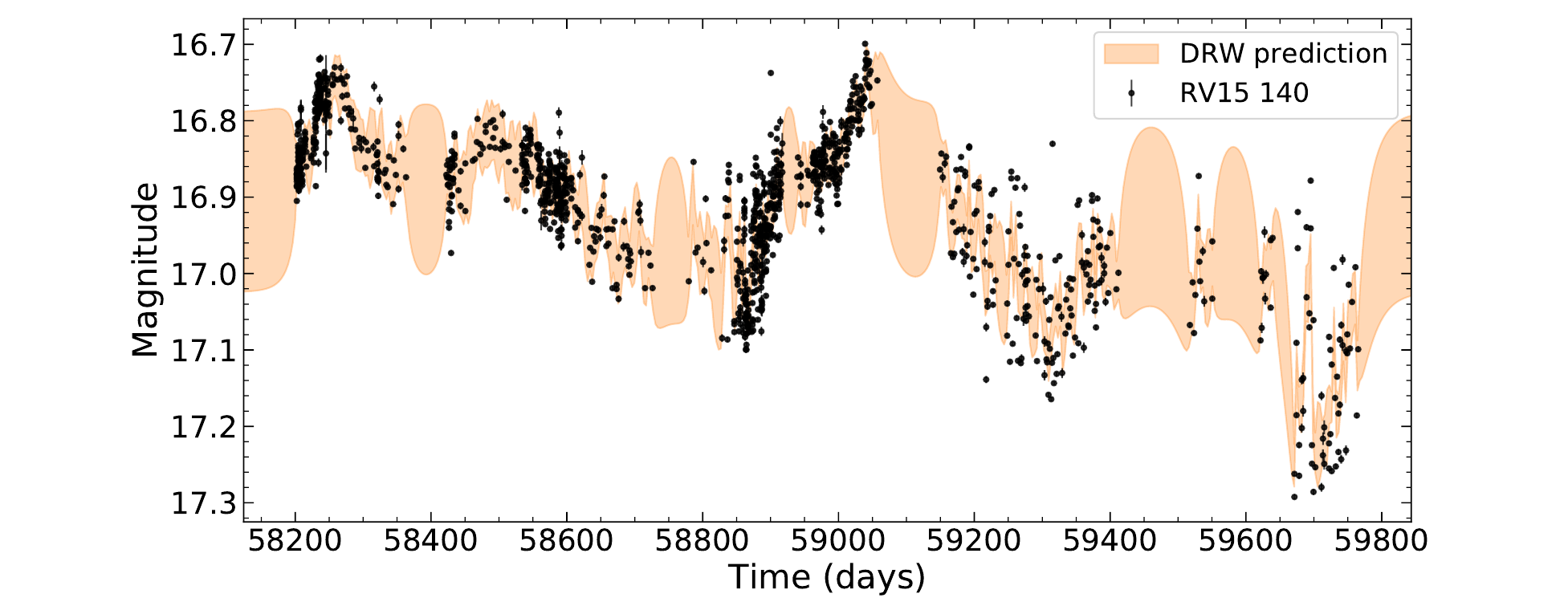}
\includegraphics[width=0.48\textwidth]{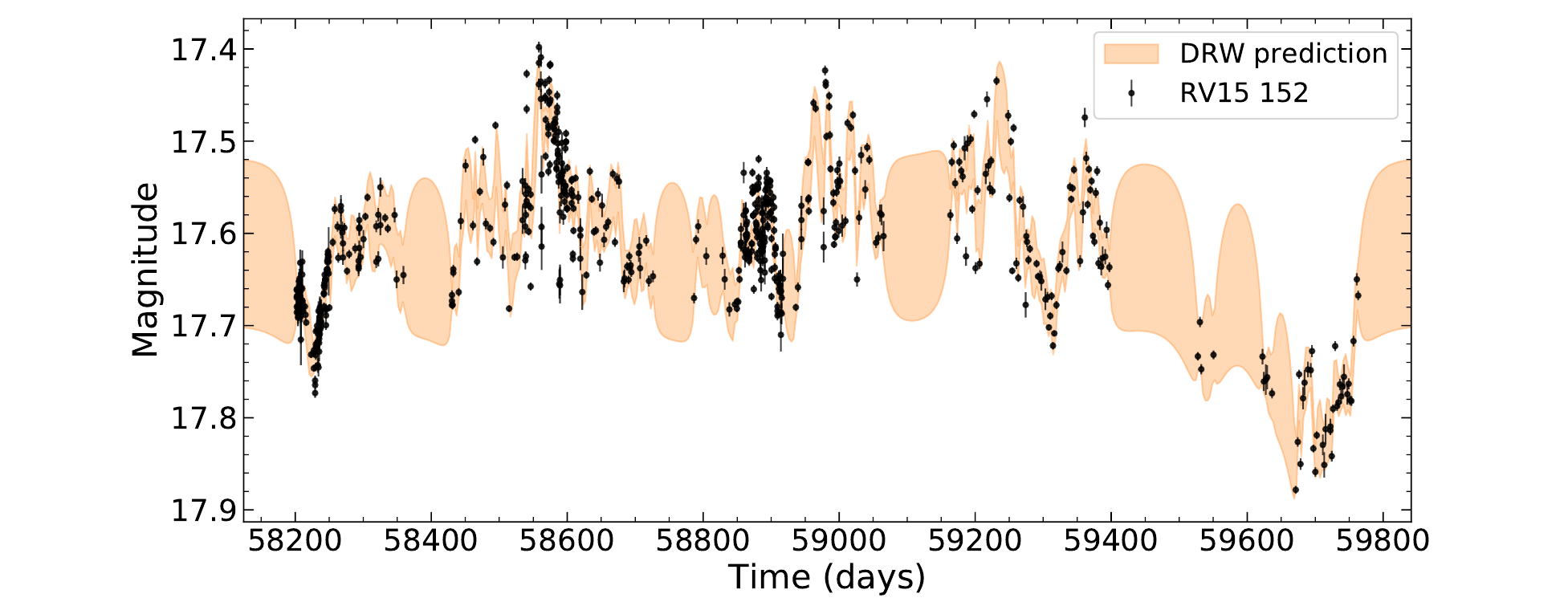}
\includegraphics[width=0.48\textwidth]{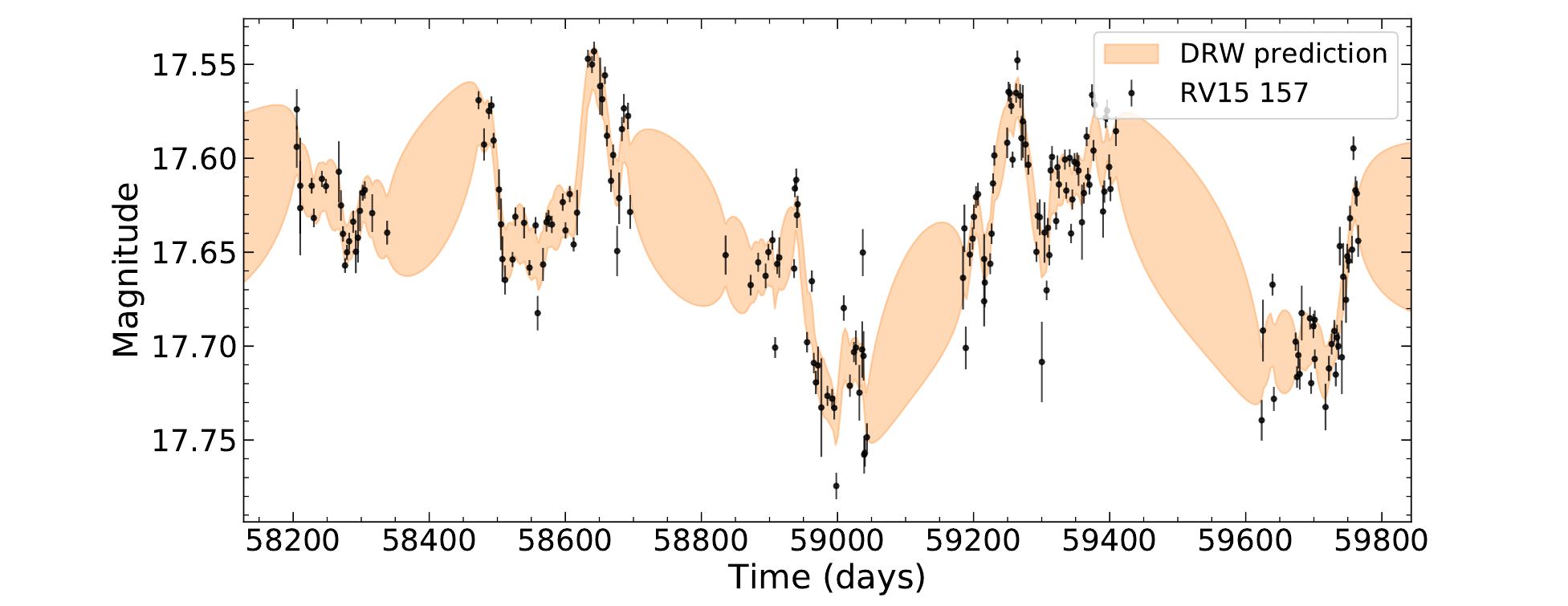}
\includegraphics[width=0.48\textwidth]{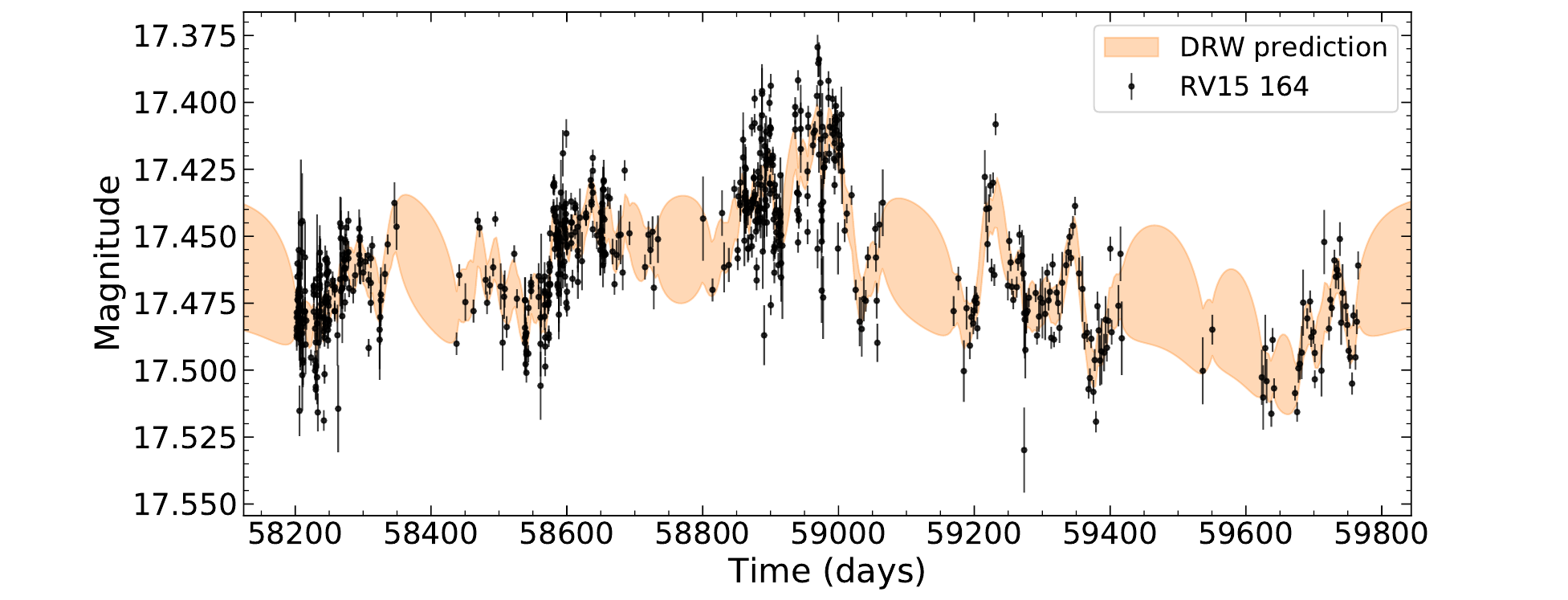}
\includegraphics[width=0.48\textwidth]{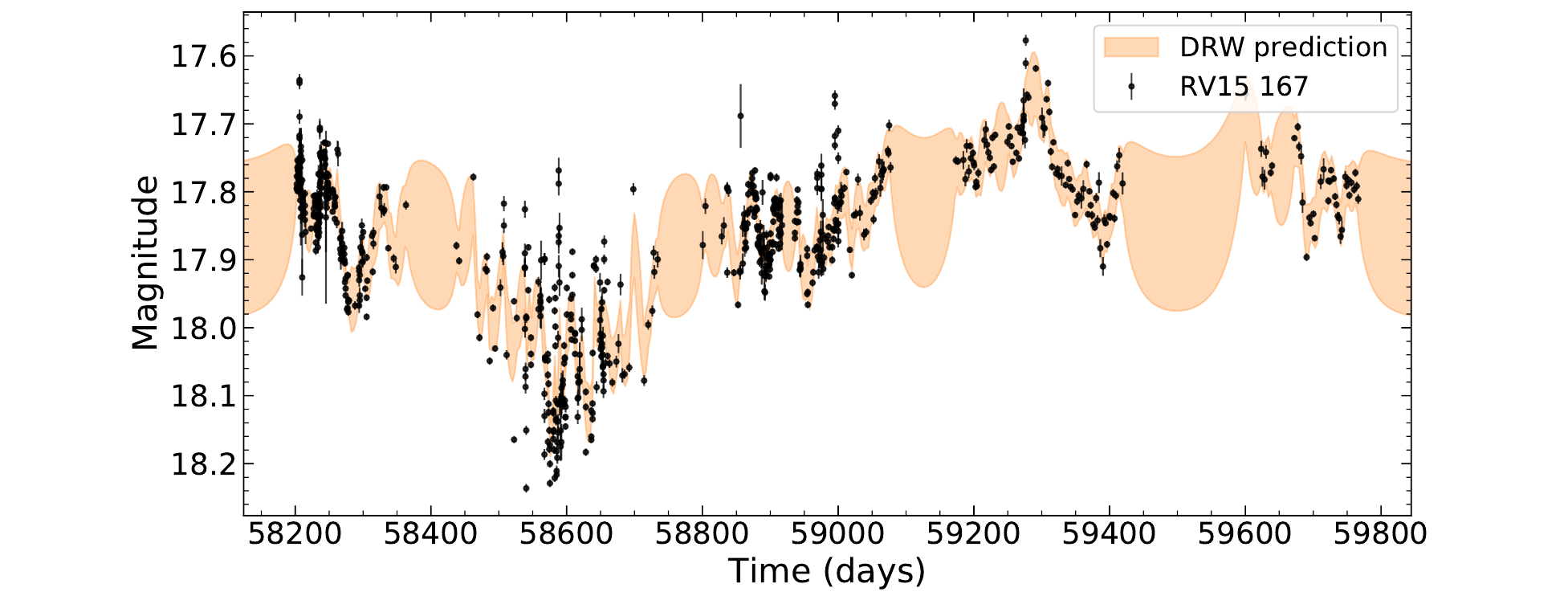}
\includegraphics[width=0.48\textwidth]{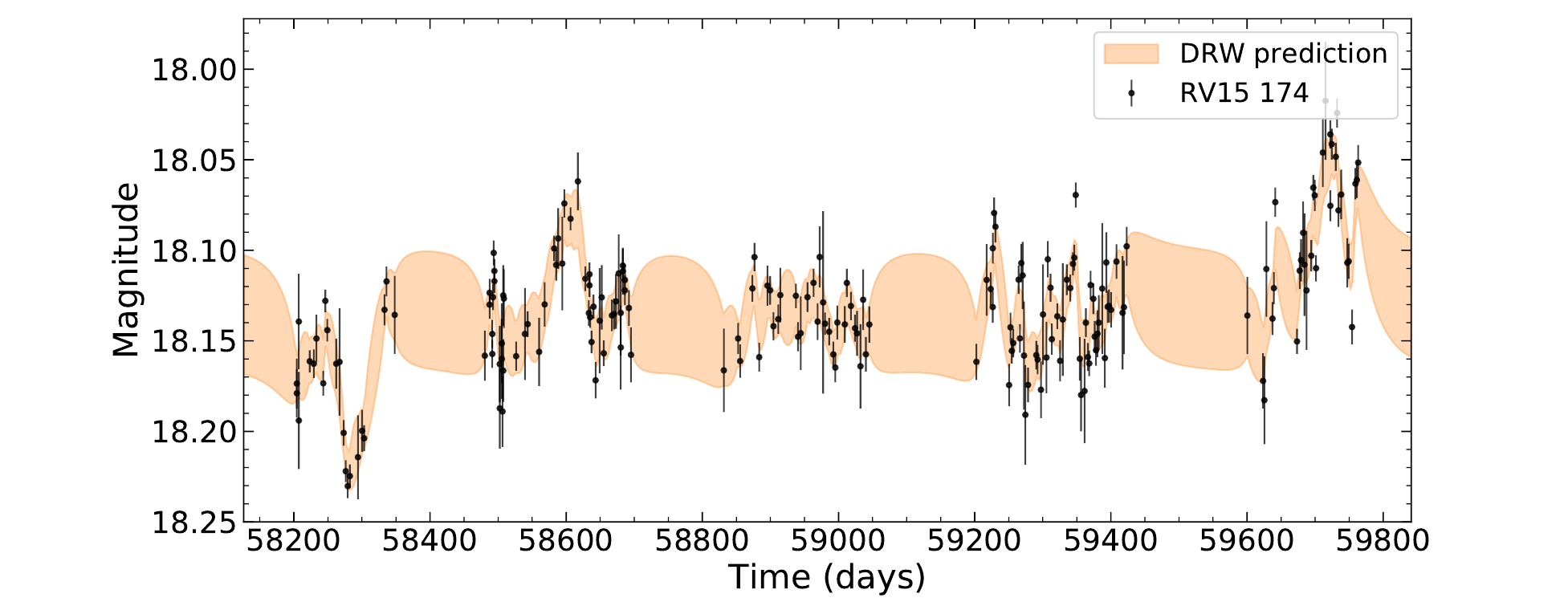}
\includegraphics[width=0.48\textwidth]{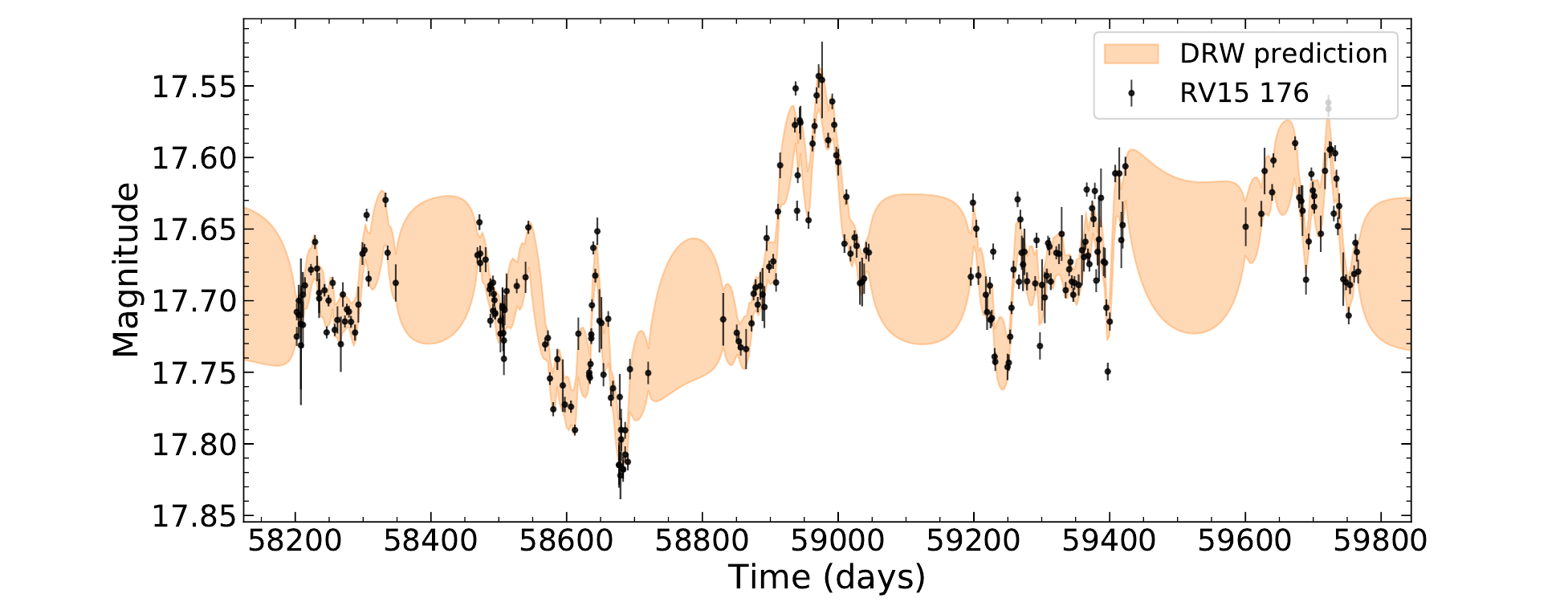}
\includegraphics[width=0.48\textwidth]{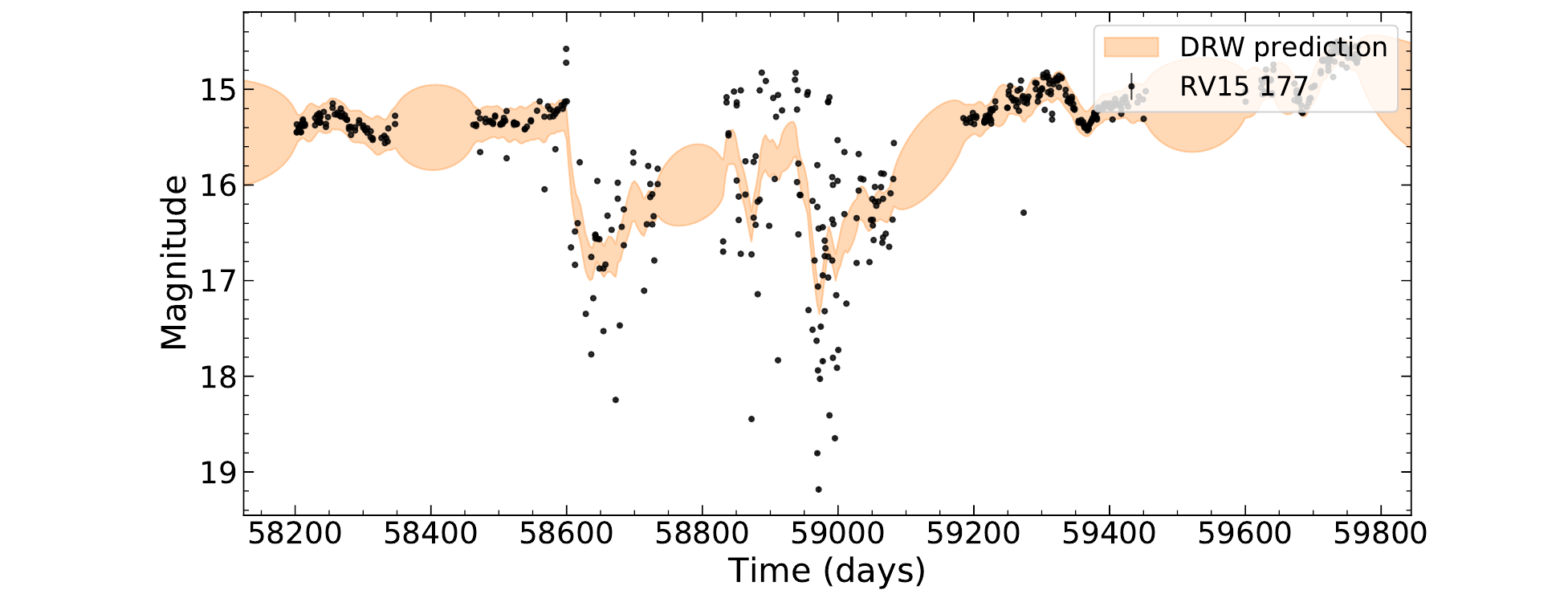}
\includegraphics[width=0.48\textwidth]{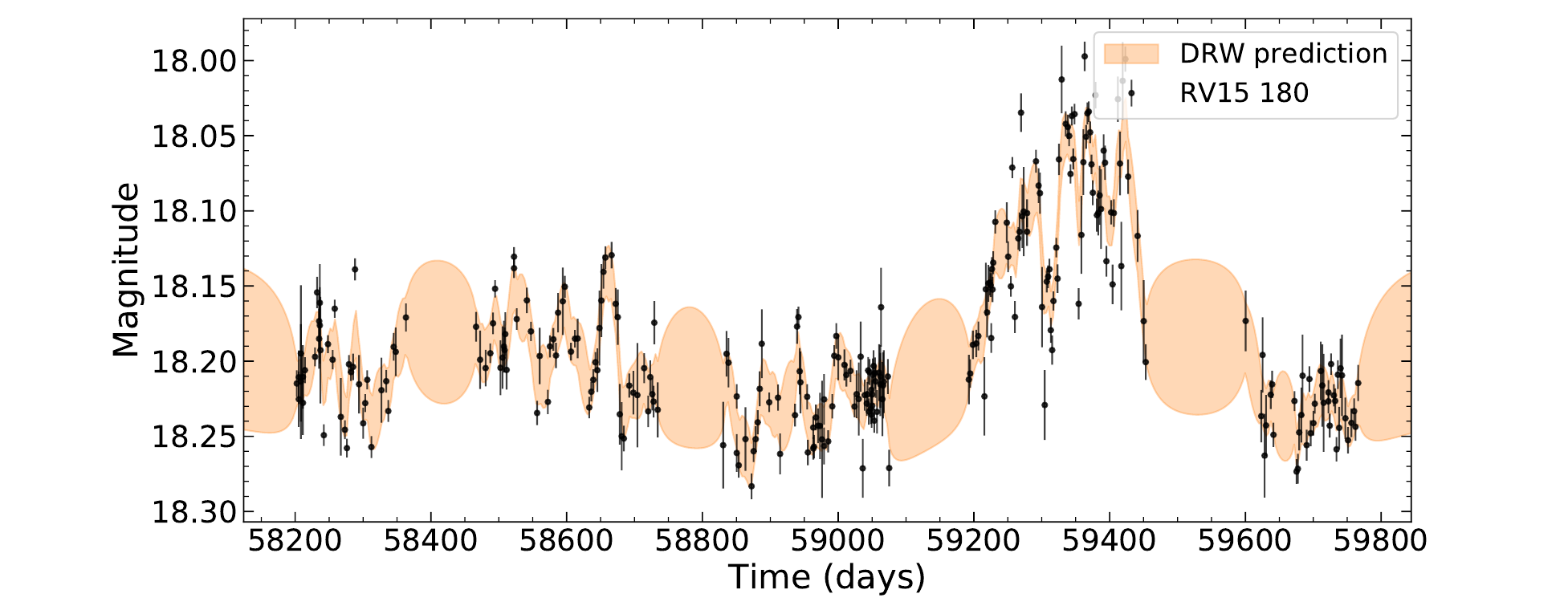}
\includegraphics[width=0.48\textwidth]{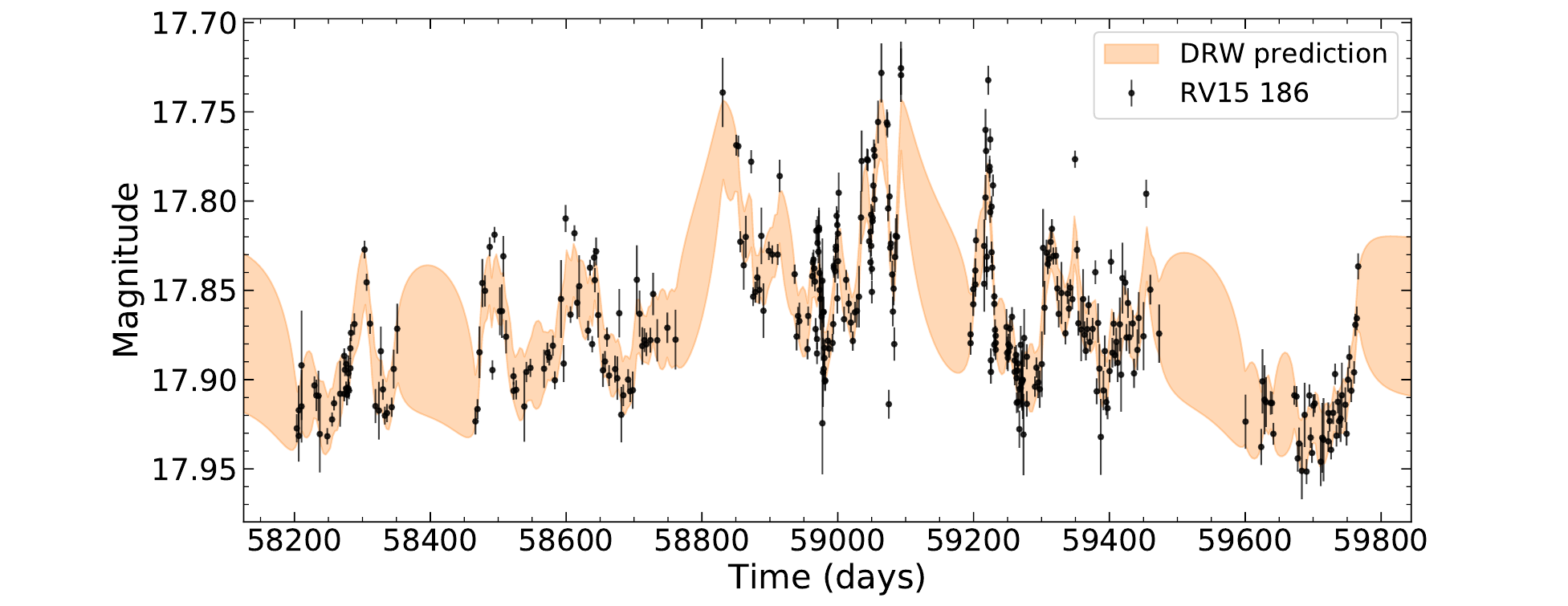}
\includegraphics[width=0.48\textwidth]{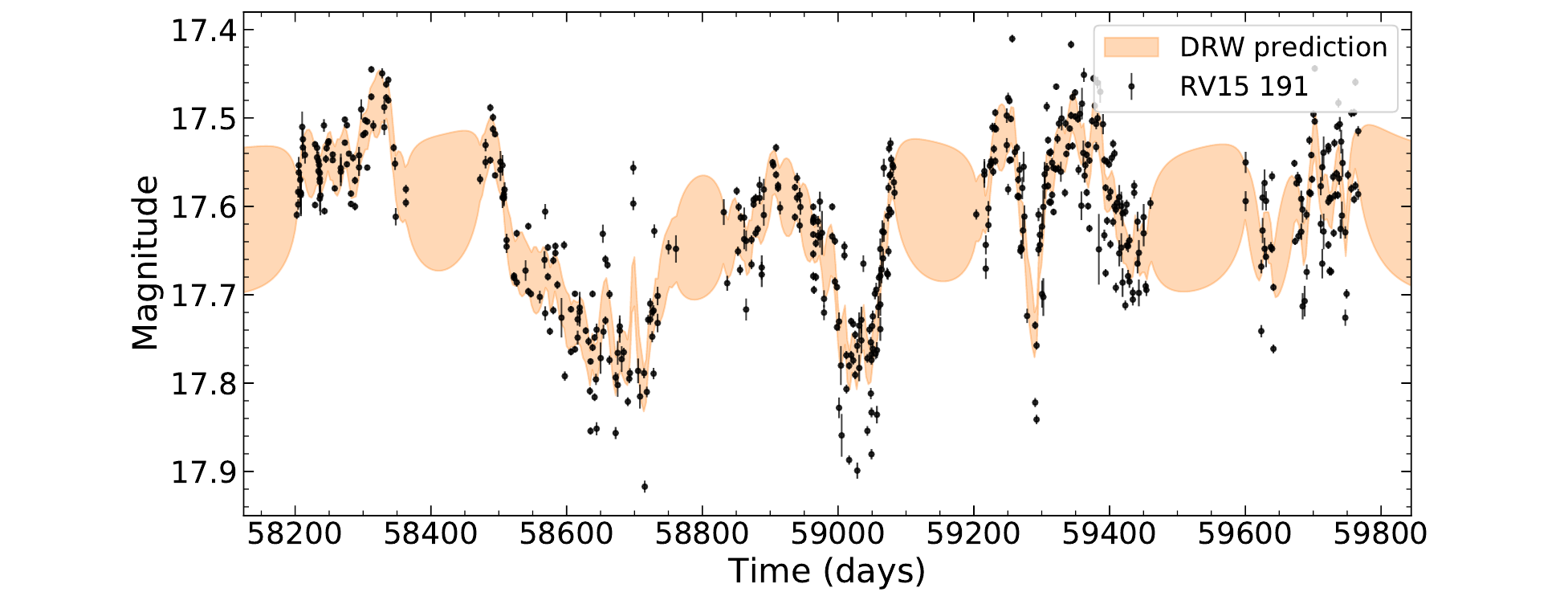}
\includegraphics[width=0.48\textwidth]{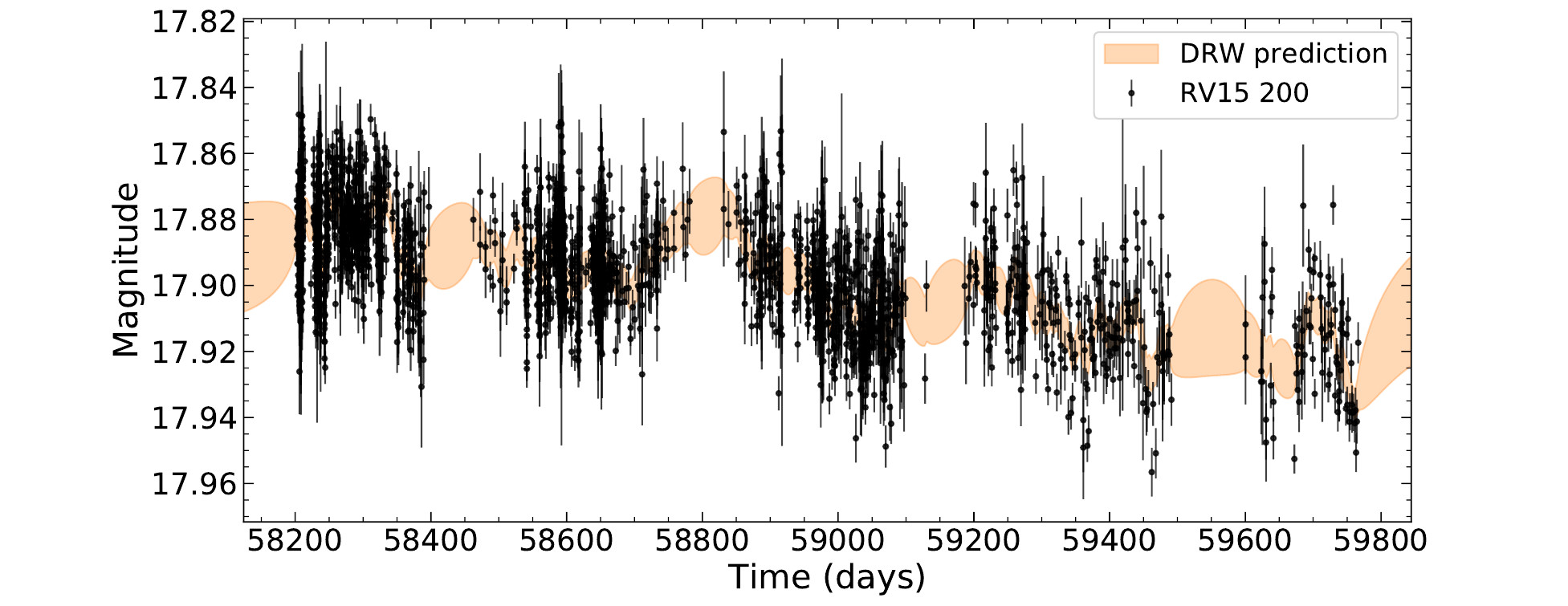}
\includegraphics[width=0.48\textwidth]{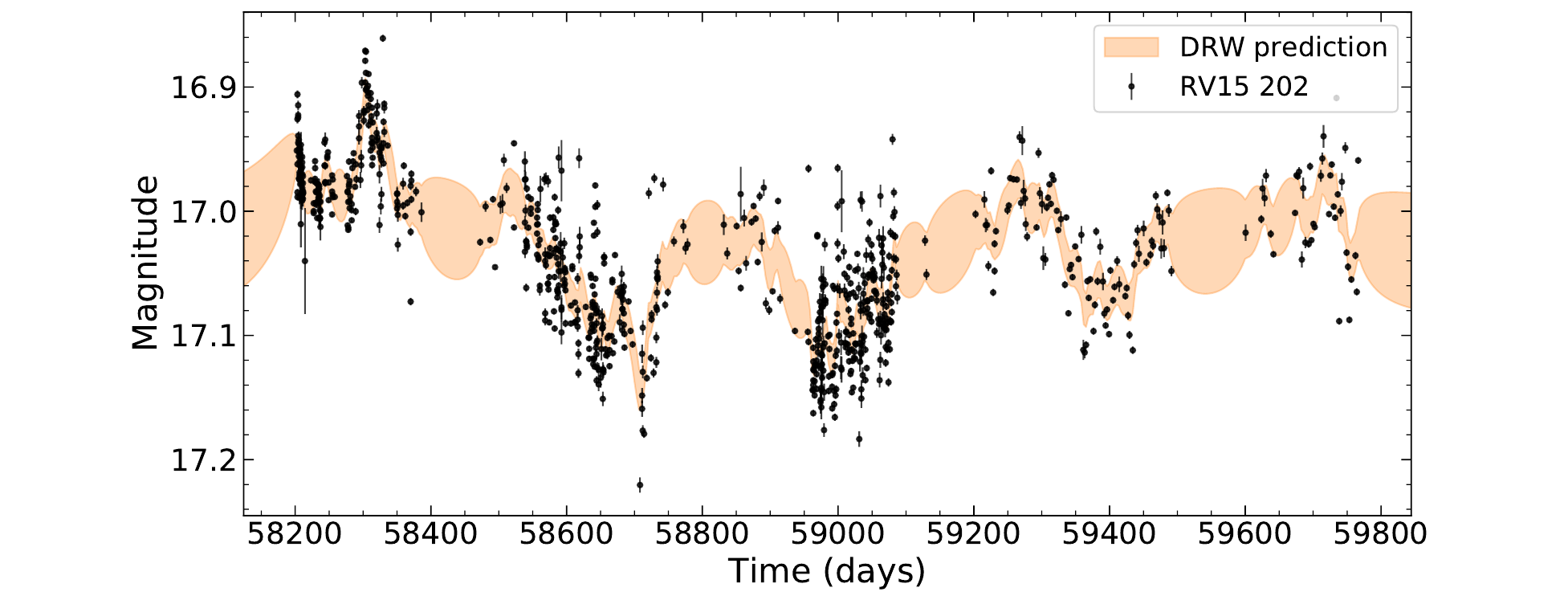}
\caption{Continuation of Figure~\ref{fig:lc_appx}.}
\end{figure*}

\begin{figure*}
\includegraphics[width=0.48\textwidth]{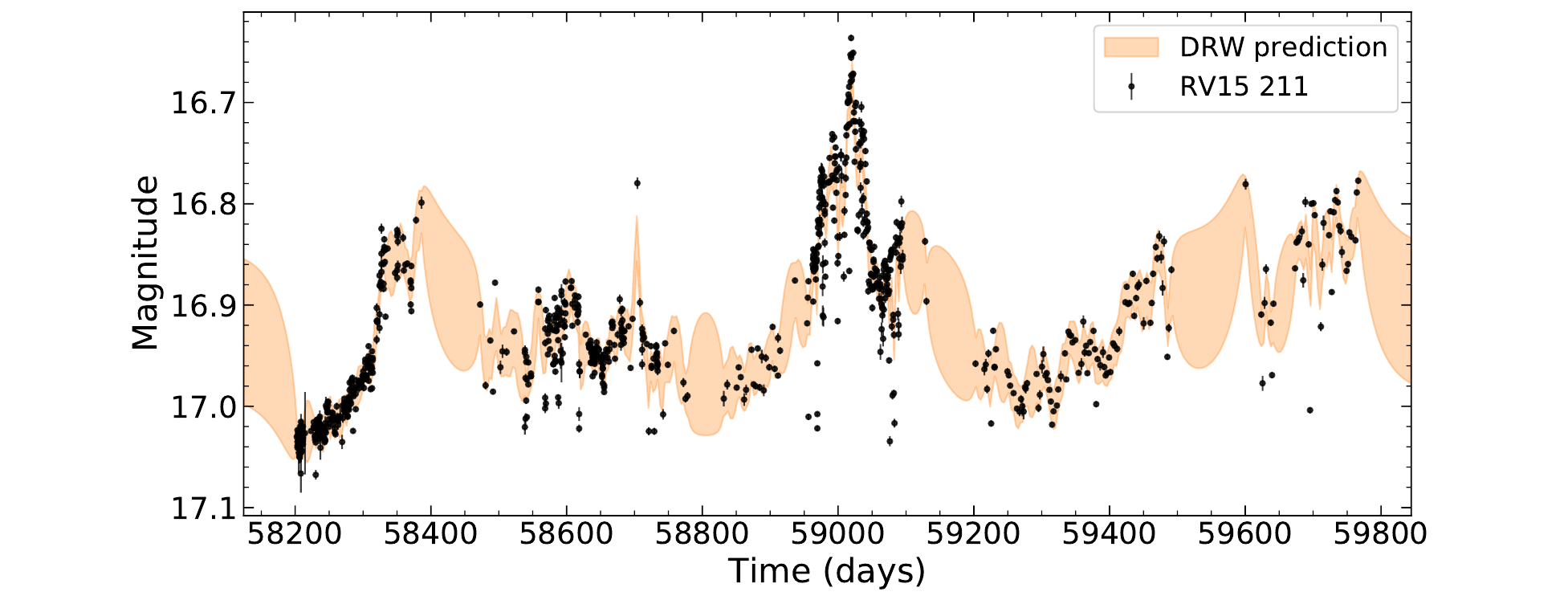}
\includegraphics[width=0.48\textwidth]{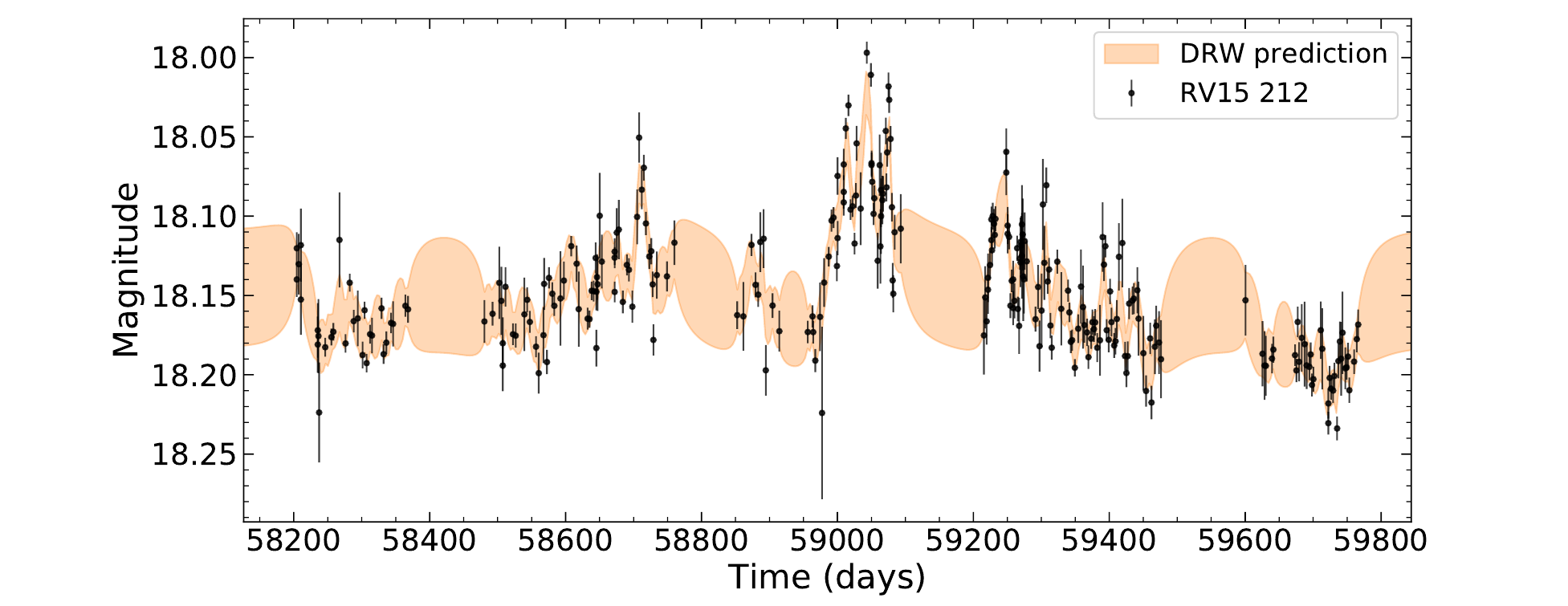}
\includegraphics[width=0.48\textwidth]{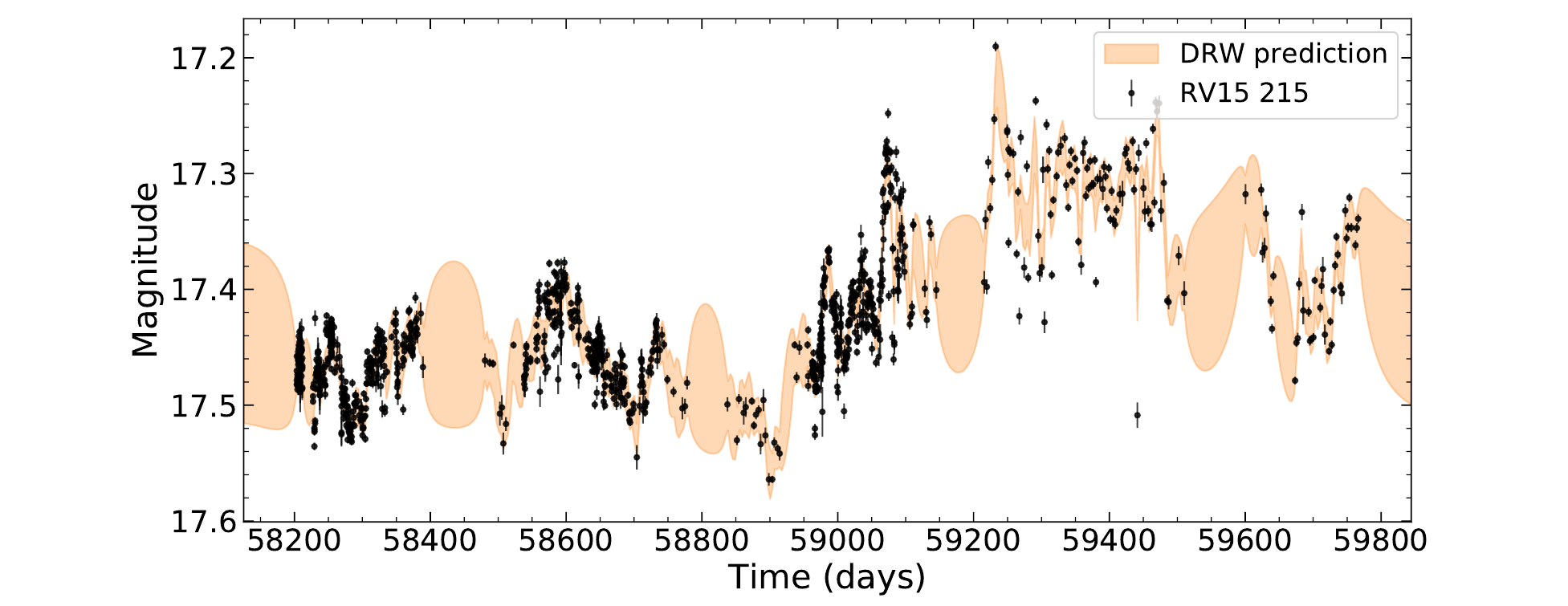}
\includegraphics[width=0.48\textwidth]{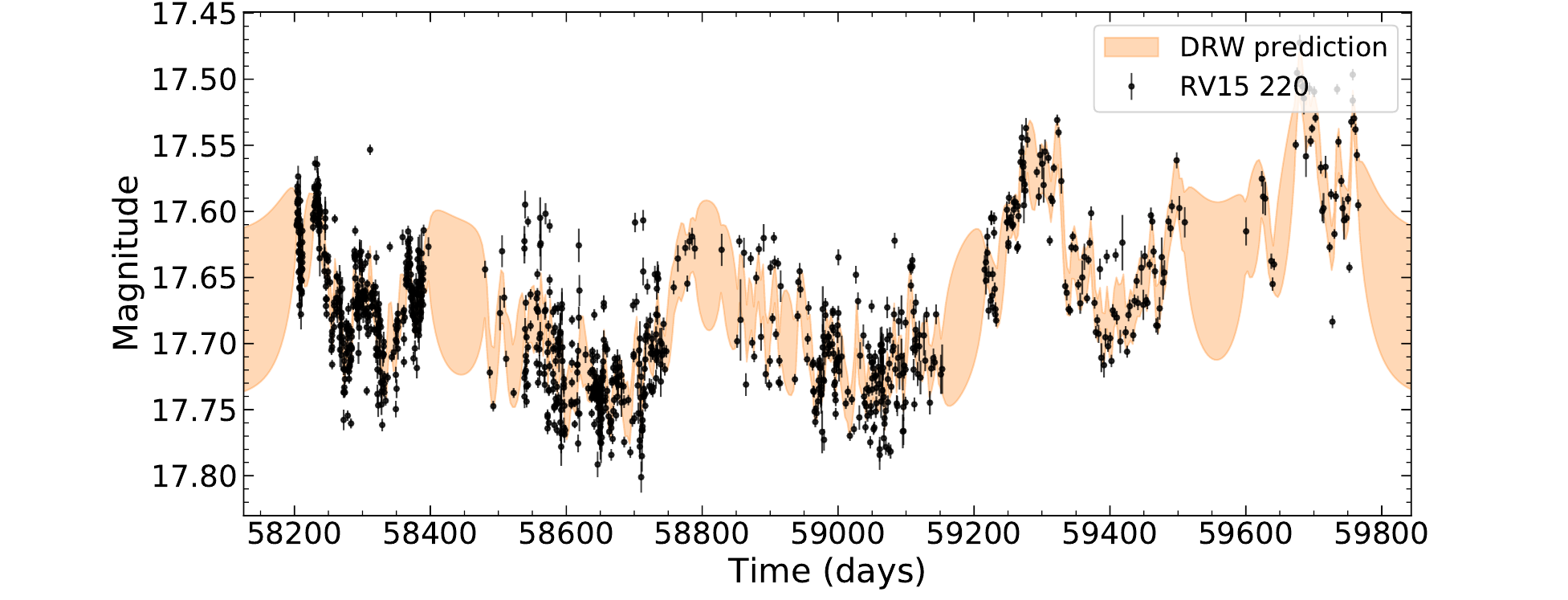}
\includegraphics[width=0.48\textwidth]{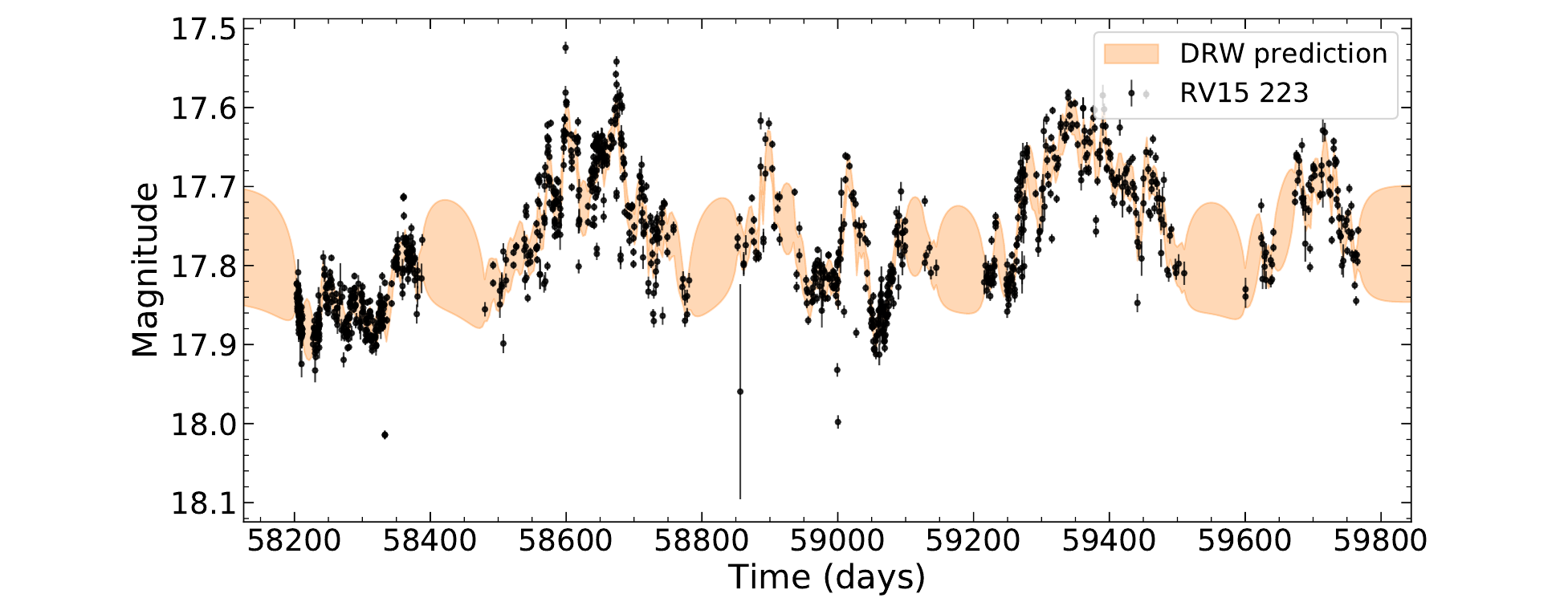}
\includegraphics[width=0.48\textwidth]{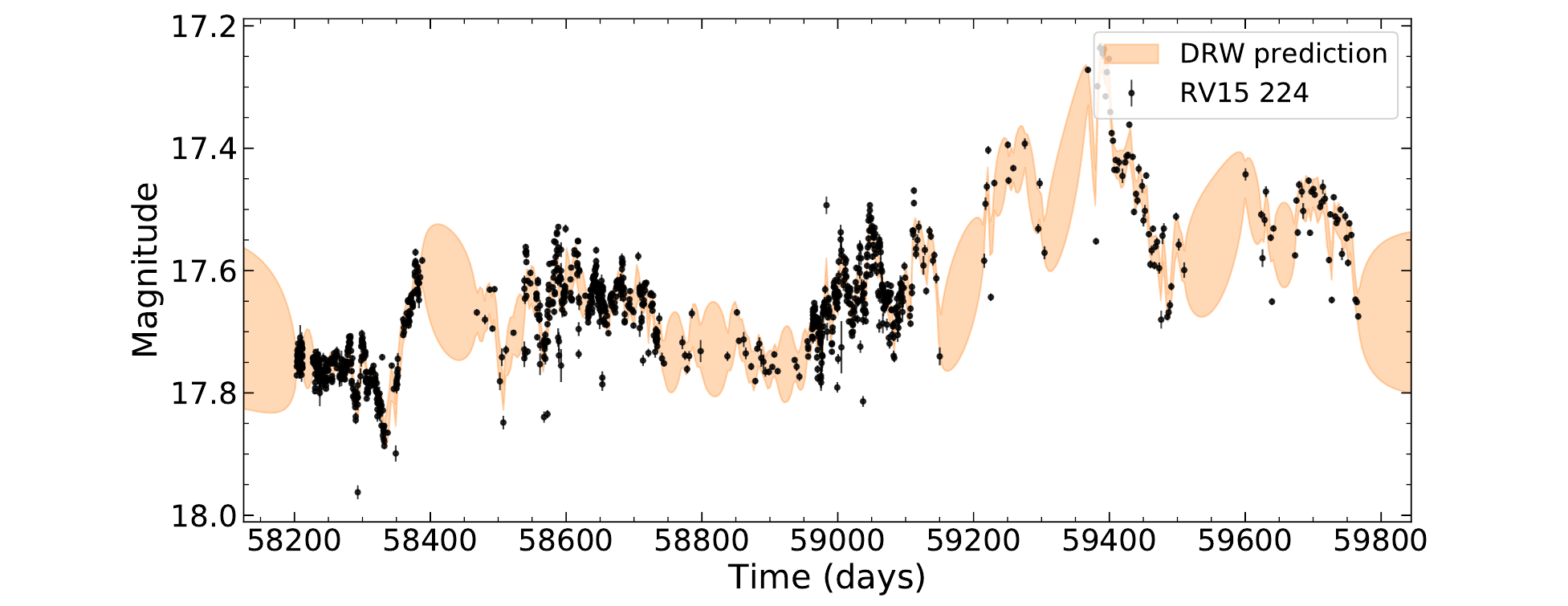}
\includegraphics[width=0.48\textwidth]{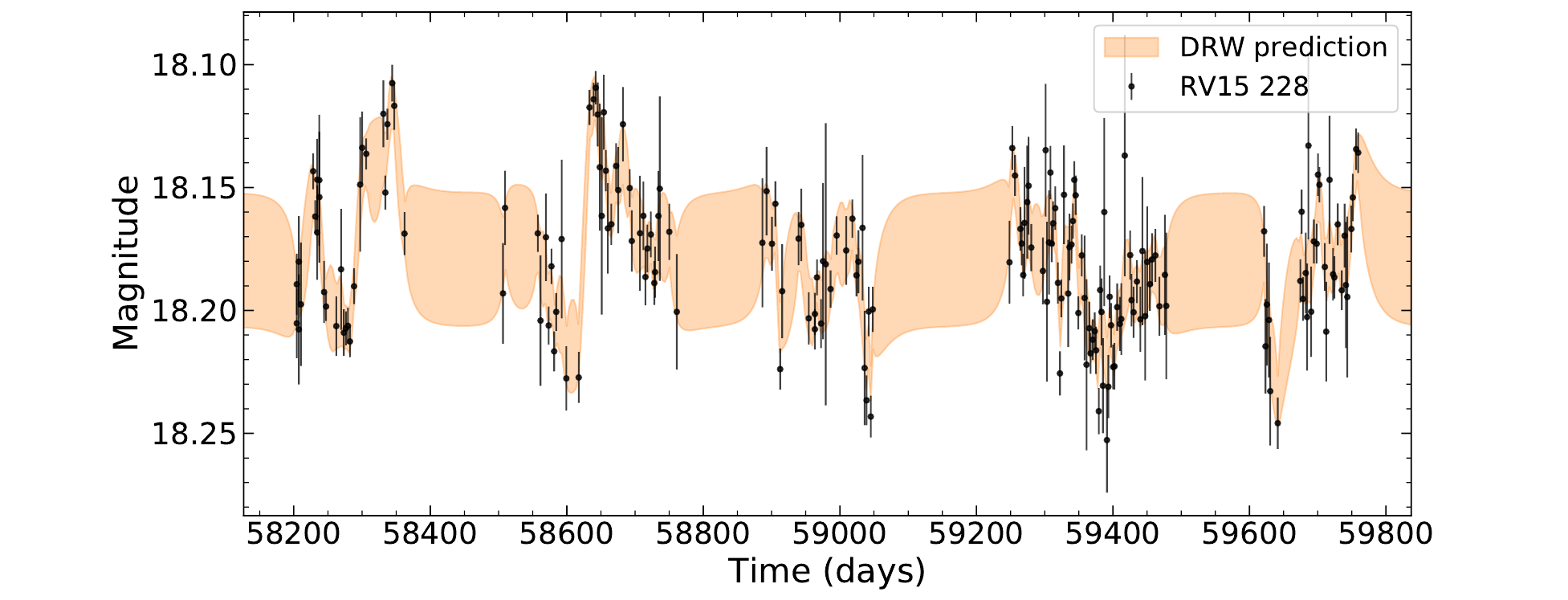}
\includegraphics[width=0.48\textwidth]{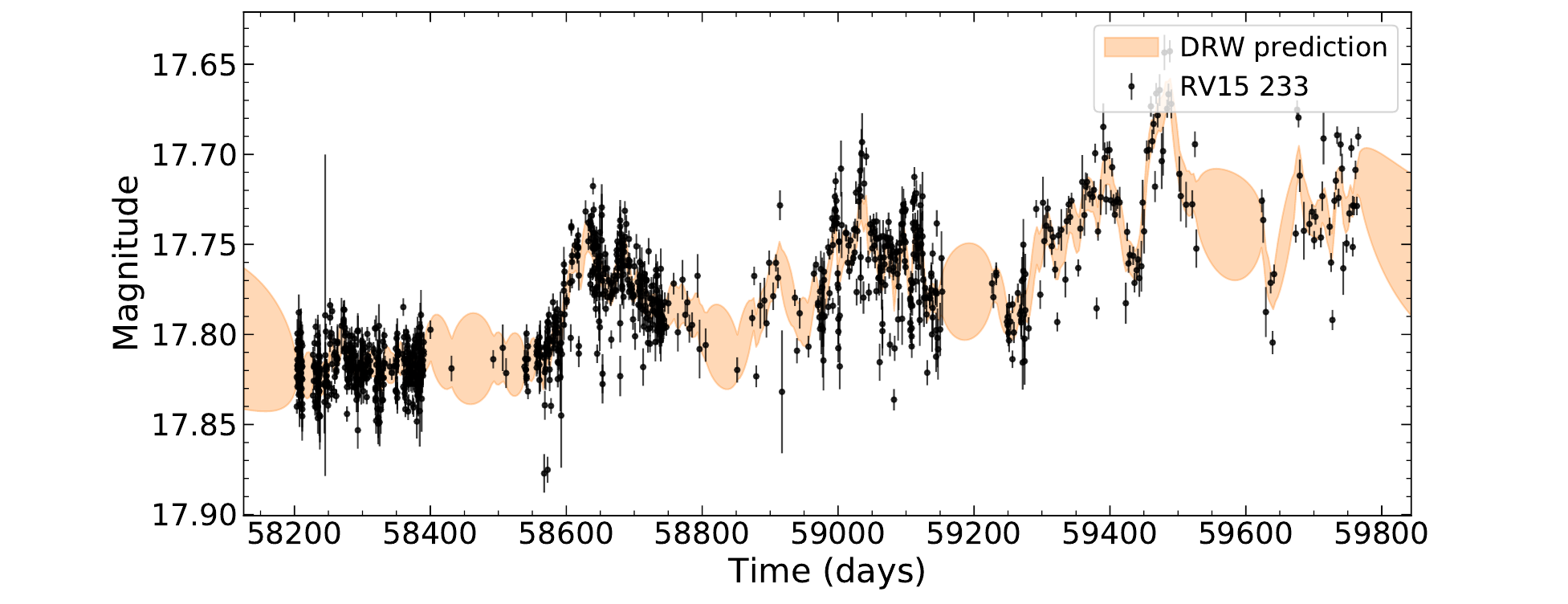}
\includegraphics[width=0.48\textwidth]{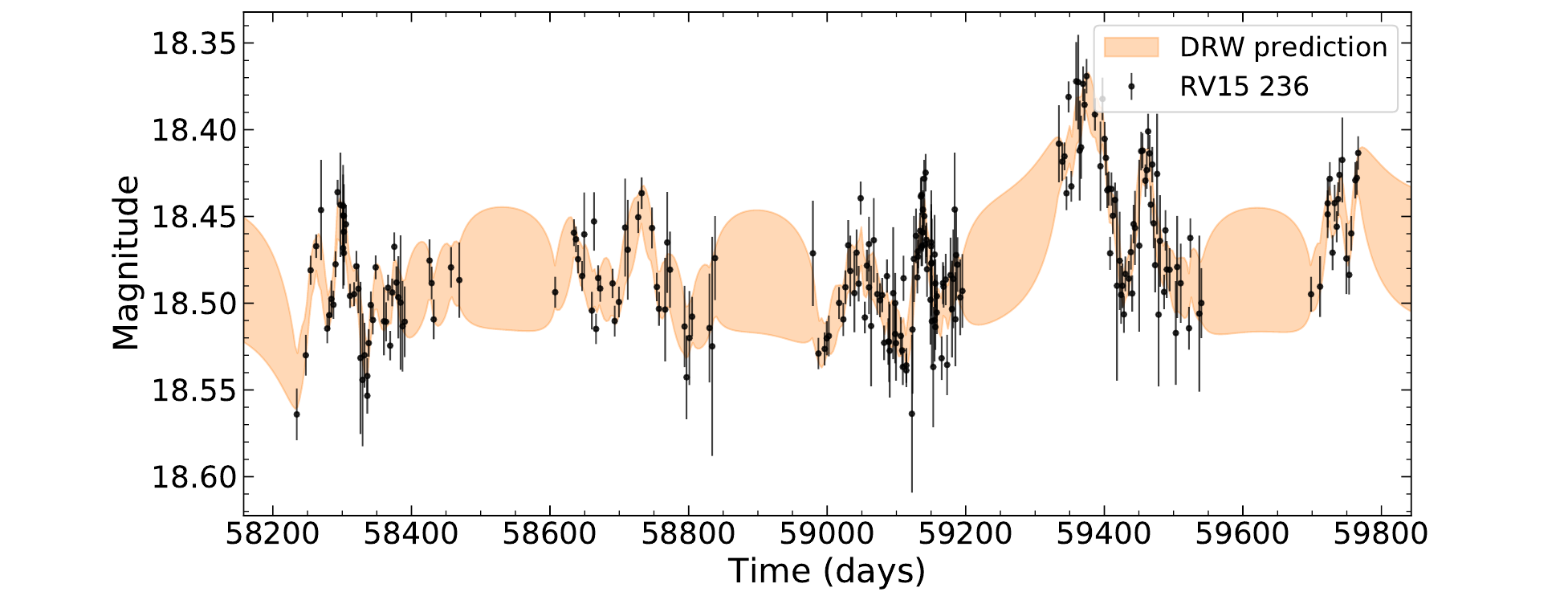}
\includegraphics[width=0.48\textwidth]{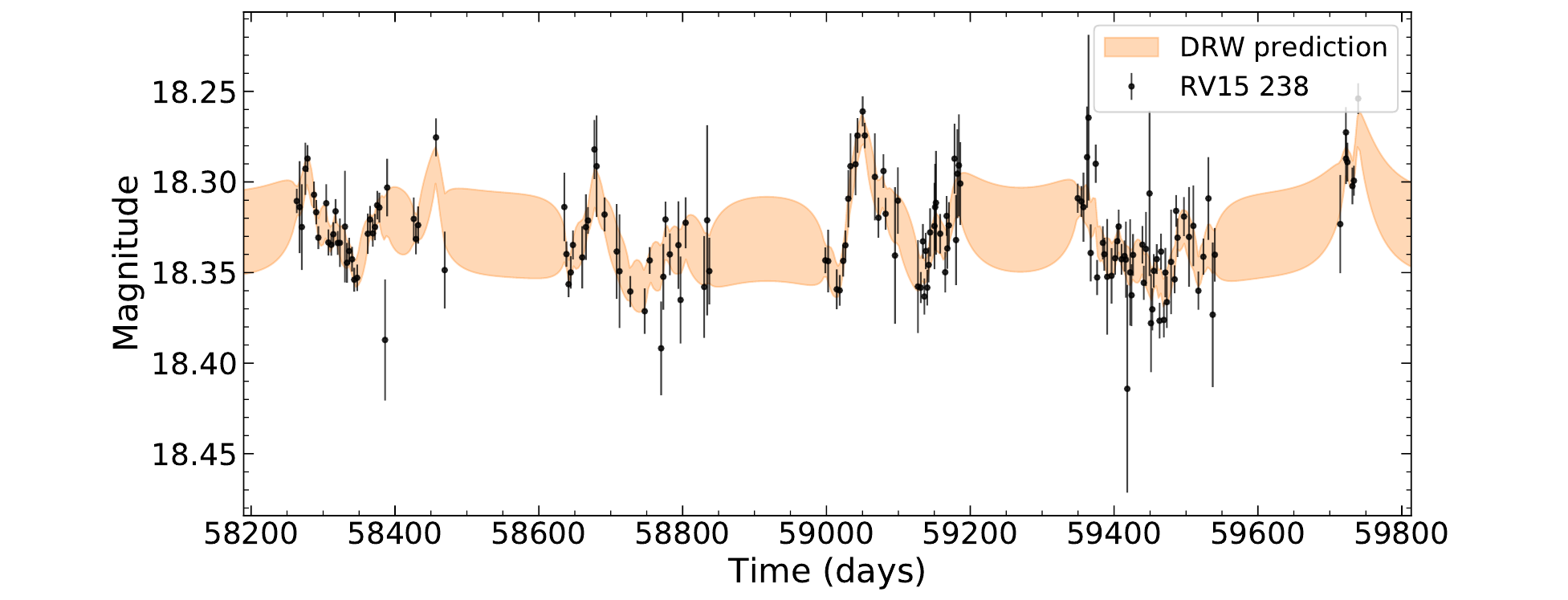}
\includegraphics[width=0.48\textwidth]{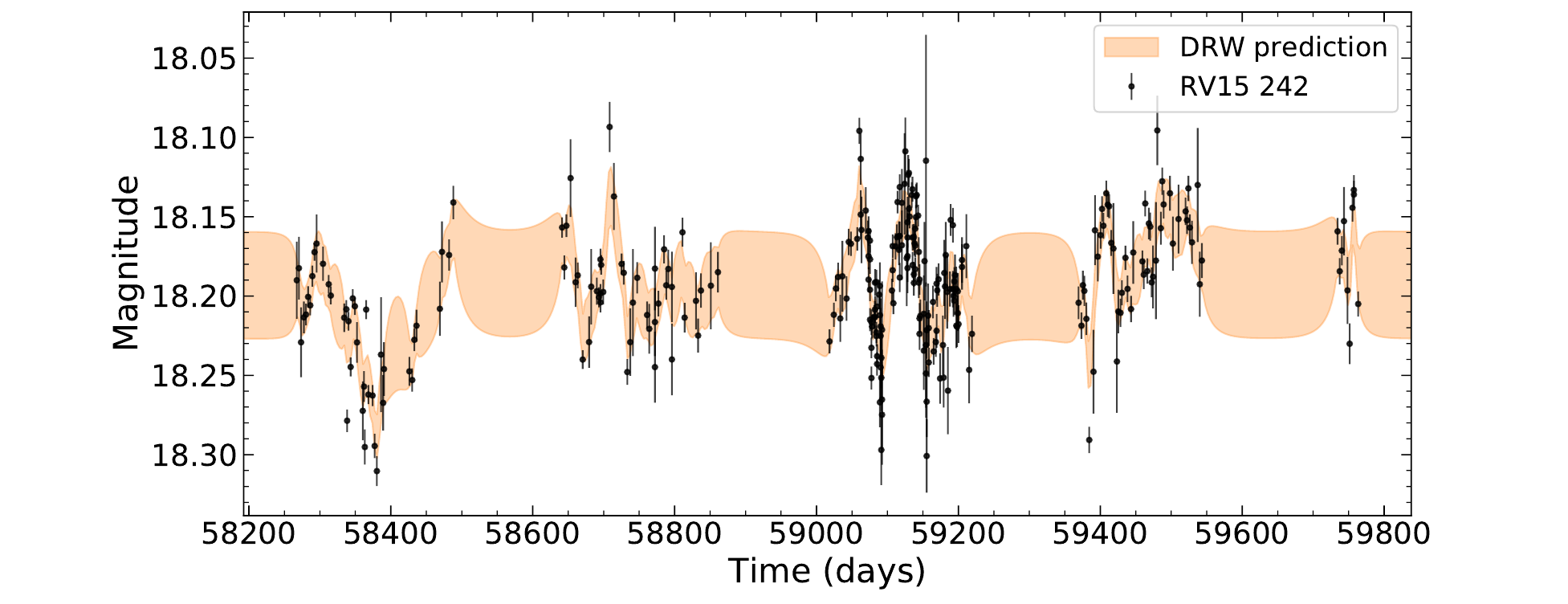}
\caption{Continuation of Figure~\ref{fig:lc_appx}.}
\end{figure*}

We show the difference imaging light curves and DRW modeling for all sources in our final, high-quality sample in Figure~\ref{fig:lc_appx}.



\bsp	
\label{lastpage}
\end{document}